\documentclass[a4paper,english, thm-restate]{lipics-v2021}
\hideLIPIcs

\usepackage[utf8]{inputenc}
\usepackage{xspace}
\usepackage{amsmath}
\usepackage{amsthm}
\usepackage{amsfonts}
\usepackage{MnSymbol}
\usepackage{mathtools}
\usepackage{textcomp}
\usepackage{varwidth}
\usepackage{graphicx}
\usepackage{mathrsfs} 
\usepackage{subfiles}
\usepackage{hyperref}
\usepackage{caption}
\usepackage{subcaption}
\usepackage{enumerate}
\usepackage{xcolor}
\usepackage[noadjust]{cite}

\usepackage{tikz}
\usetikzlibrary{automata, graphs,positioning,chains,arrows,calc, backgrounds}

\newcommand{\defining}[1]{\emph{#1}}

\DeclarePairedDelimiter\set{\lbrace}{\rbrace}

\DeclarePairedDelimiterX\setcond[2]{\{}{\}}{\mathchoice{\,}{}{}{}#1 \;\delimsize\vert\; #2\mathchoice{\,}{}{}{}}

\newcommand{\regular}{\mathsf{regular}}
\newcommand{\blocking}{\mathsf{blocking}}

\renewcommand{\phi}{\varphi}

\newcommand{\inv}[1]{#1^{-1}}

\newcommand{\Vars}[1]{\mathsf{var}(#1)}

\newcommand{\comp}{\equiv}
\newcommand{\CFIsym}{\mathsf{CFI}}
\newcommand{\CFI}[1]{\CFIsym(#1)}

\newcommand{\compCFI}[2]{\CFI{#1}/_{#2}}
\newcommand{\precompCFI}[2]{{(\CFI{#1},#2)}}

\newcommand{\nat}{\mathbb{N}}
\newcommand{\FF}{\mathbb{F}}
\newcommand{\tup}[1]{\bar{#1}}
\newcommand{\iso}{\cong}

\newcommand{\klogic}[1]{\mathscr{L}_{#1}}
\newcommand{\kmlogic}[2]{\mathscr{L}_{#1, #2}}

\newcommand{\graphG}{\mathcal{G}}
\newcommand{\graphH}{\mathcal{H}}
\newcommand{\grid}{\mathcal{C}}
\newcommand{\vertexG}{V_{\graphG}}
\newcommand{\vertexH}{V_{\graphH}}
\newcommand{\vertexC}{V_{\mathcal{C}}}
\newcommand{\edgeG}{E_{\graphG}}
\newcommand{\edgeH}{E_{\graphH}}
\newcommand{\edgeC}{E_{\mathcal{C}}}
\newcommand{\coloringG}{\chi_{\graphG}}
\newcommand{\coloringH}{\chi_{\graphH}}

\newcommand{\dom}{\mathsf{dom}}

\newcommand{\graphXG}{\mathcal{X}(\mathcal{G})}
\newcommand{\graphXH}{\mathcal{X}(\mathcal{H})}
\newcommand{\vertexXG}{V_{\graphXG}}
\newcommand{\vertexXH}{V_{\graphXH}}
\newcommand{\edgeXG}{E_{\graphXG}}
\newcommand{\edgeXH}{E_{\graphXH}}

\newcommand{\ISO}{\mathsf{ISO}(\graphG, \graphH)}
\newcommand{\XISO}{\mathsf{ISO}(\graphXG, \graphXH)}

\newcommand{\kbequivr}[2]{\simeq_{\mathcal{B}}^{#1,#2}}
\newcommand{\notkbequivr}[2]{\not\simeq_{\mathcal{B}}^{#1,#2}}
\newcommand{\kbequiv}[1]{\simeq_{\mathcal{B}}^{#1}}
\newcommand{\kequiv}[1]{\simeq^{#1}}
\newcommand{\notkequiv}[1]{\not\simeq^{#1}}
\newcommand{\kequivr}[2]{\simeq^{#1,#2}}
\newcommand{\notkequivr}[2]{\not\simeq^{#1,#2}}

\newcommand{\X}{\mathcal{X}}
\newcommand{\XG}{\X_{\graphG}}
\newcommand{\XH}{\X_{\graphH}}

\newcommand{\XHI}{\X^{-1}_{\graphH}}
\newcommand{\XI}{\X^{-1}}

\definecolor{rwth}   {RGB}{  0  84 159}
\definecolor{rwth-75}{RGB}{ 64 127 183}
\definecolor{rwth-50}{RGB}{142 186 229}
\definecolor{rwth-25}{RGB}{199 221 242}
\definecolor{rwth-10}{RGB}{232 241 250}

\definecolor{darkblue}{rgb}{0,0,0.8}
\definecolor{darkgreen}{rgb}{0,0.7,0}
\definecolor{darkred}{rgb}{0.5,0,0}
\definecolor{darkyellow}{rgb}{0.6,0.6,0}

\definecolor{mediumblue}{rgb}{0.2,0.2,1}
\definecolor{lightblue}{rgb}{0.5,0.5,1}
\definecolor{lightgreen}{rgb}{0.5,1,0.5}
\definecolor{lightred}{rgb}{1,0.5,0.5}
\definecolor{lightyellow}{rgb}{1,1,0.5}

\definecolor{colA}{RGB}{0,80,250}
\definecolor{colB}{RGB}{60,200,230}
\definecolor{colC}{RGB}{60,180,75}
\definecolor{colD}{RGB}{230,25,75}
\definecolor{colE}{RGB}{245,130,48}
\definecolor{colF}{RGB}{145,30,180}
\definecolor{colG}{RGB}{240,50,230}
\definecolor{colH}{RGB}{170,110,40}

\tikzstyle{vertex} = [circle, fill=black, inner sep=0mm, minimum size = 2mm]

\tikzstyle{basevertex} = [circle, fill=black, inner sep=0mm, minimum size = 6mm]

\NewDocumentCommand{\drawConnection}{ O{black} O{red} m m m }{
        \ifthenelse{#5=0}{
                \draw[#1, draw, thick]
                (#3-0) -- (#4-1)
                (#3-1) -- (#4-0);
        }
        {
                \draw[#2, draw, thick]
                (#3-0) -- (#4-0)
                (#3-1) -- (#4-1);
        }
}

\newcommand{\drawBaseGraph}[4]{
        \def\length{7}
        \def\height{1}
        \begin{scope}[scale = 2, shift = {(-2,1)}, yscale=-1]
                \foreach \i in {0,...,\length} {
                        \foreach \j in {0,...,\height} {
                                \node[baseVertex] (u-\i-\j) at (\i,\j) {};
                        }
                }
                
                \pgfmathtruncatemacro{\prevlen}{\length-1}
                \pgfmathtruncatemacro{\prevhei}{\height-1}
                \foreach \i in {0,...,\prevlen} {
                        \pgfmathtruncatemacro{\nexti}{1+\i};
                        \foreach \j in {0,...,\height} {
                                \ifthenelse{\equal{\i}{#1} \AND \equal{\j}{#2} \AND \equal{\nexti}{#3} \AND \equal{\j}{#4}}
                                {
                                        \path[twistedBaseEdge] (u-\i-\j) to node{\includegraphics[width=6mm]{robber.png}} (u-\nexti-\j);
                                }
                                {
                                        \draw[baseEdge] (u-\i-\j) -- (u-\nexti-\j);
                                }
                        }
                }
                \foreach \i in {0,...,\length} {
                        \foreach \j in {0,...,\prevhei} {
                                \pgfmathtruncatemacro{\nextj}{1+\j};
                                \ifthenelse{\equal{\i}{#1} \AND \equal{\j}{#2} \AND \equal{\i}{#3} \AND \equal{\nextj}{#4}}
                                {\path[twistedBaseEdge] (u-\i-\j) to node{\includegraphics[width=6mm]{robber.png}} (u-\i-\nextj);}
                                {\draw[baseEdge] (u-\i-\j) -- (u-\i-\nextj);}
                        }
                }
                
        \end{scope}
}

\tikzstyle{edgeVertex} = [circle, fill=black, inner sep=0, minimum size =1mm]
\tikzstyle{baseVertex} = [basevertex, gray]
\tikzstyle{baseEdge} = [draw, gray, ultra thick]
\tikzstyle{twistedBaseEdge} = [draw, red, ultra thick]

\newcommand{\id}[1]{#1}

\newcommand{\drawgadget}[5] {
        
        \draw[dashed, thick, gray] (#3,#4) circle (0.65);
        
        \ifnum #1=1
        \node[vertex,#5] (#2\id{0}) at (#3,#4){};
        \else \ifnum #1=2
        \node[vertex,#5] (#2\id{11}) at (#3+0.25,#4-0.25){};
        \node[vertex,#5] (#2\id{00}) at (#3-0.25,#4+0.25){};
        
        \else
        \foreach \x in {0,...,1}
        \foreach \y in {0,...,1}
        \foreach \z in {0,...,1}
        {
                \ifodd \numexpr \x + \y + \z \else
                \node[vertex, #5] (#2\x\y\z) at ({#3 + 0.5*( \z - 0.3*\x) -0.2}, {#4 + 0.5*(\x + 0.7*\z - 1.7*\y -\x*\z)+0.1}){};
                \fi
        }
        \fi
        \fi
        
}

\newcounter{memcount}
\newcommand{\countmem}[2]{%
        \setcounter{memcount}{0}%
        \foreach \i in #1{%
                \ifthenelse{\equal{\i}{#2}} {%
                        \stepcounter{memcount}%
                }{}%
        }%
        
}

\newcommand{\suffx}{x}
\newcommand{\suffy}{y}
\newcommand{\suffz}{z}
\newcommand{\suffsum}{sum}
\newcommand{\suffseq}{seq}
\newcommand{\suffrr}{rr}

\newcommand{\foreachdim}[4] {
        \ifnum #1=1
        \expandafter\foreach \csname#2\suffx\endcsname in {0,...,1} {
                \expandafter\def\csname#2\suffsum\endcsname{%
                        \csname#2\suffx\endcsname}
                \expandafter\def\csname#2\suffseq\endcsname{%
                        \csname#2\suffx\endcsname}
                \expandafter\def\csname#2\suffrr\endcsname{%
                        {\csname#2\suffx\endcsname}}
                #3
        }
        \else \ifnum #1=2
        \expandafter\foreach \csname#2\suffx\endcsname in {0,...,1} {
                \expandafter\foreach \csname#2\suffy\endcsname in {0,...,1} {
                        \expandafter\def\csname#2\suffsum\endcsname{%
                                \csname#2\suffx\endcsname + %
                                \csname#2\suffy\endcsname}
                        \expandafter\def\csname#2\suffseq\endcsname{%
                                \csname#2\suffx\endcsname\csname#2\suffy\endcsname}
                        \expandafter\def\csname#2\suffrr\endcsname{%
                                {\csname#2\suffx\endcsname,\csname#2\suffy\endcsname}}
                        #3
                }
        }
        \else
        \expandafter\foreach \csname#2\suffx\endcsname in {0,...,1} {
                \expandafter\foreach \csname#2\suffy\endcsname in {0,...,1} {
                        \expandafter\foreach \csname#2\suffz\endcsname in {0,...,1} {
                                \expandafter\def\csname#2\suffsum\endcsname{%
                                        \csname#2\suffx\endcsname + %
                                        \csname#2\suffy\endcsname +%
                                        \csname#2\suffz\endcsname}
                                \expandafter\def\csname#2\suffseq\endcsname{%
                                        \csname#2\suffx\endcsname\csname#2\suffy\endcsname\csname#2\suffz\endcsname}
                                \expandafter\def\csname#2\suffrr\endcsname{%
                                        {\csname#2\suffx\endcsname,\csname#2\suffy\endcsname,\csname#2\suffz\endcsname}}
                                #3
                        }
                }
        }
        \fi\fi
        
}

\newcommand{\drawconnectgadget}[9] {
        
        \foreachdim{#1}{a} {
                \foreachdim{#2}{b}{
                        \pgfmathsetmacro\as{\arr[#5]}
                        \pgfmathsetmacro\bs{\brr[#6]}
                        \ifodd \numexpr \asum \else
                        \ifodd \numexpr \bsum \else
                        \ifthenelse{
                                \(\(\NOT \equal{#9}{twist}\) \AND \equal{\as}{\bs}\) \OR %
                                \(\equal{#9}{twist} \AND \(\NOT \equal{\as}{\bs}\)\)
                        }{
                                \countmem{{#7}}{\aseq\id{E}\bseq}
                                \ifthenelse{\value{memcount} > 0}{
                                        \path (#3\aseq) edge[bend right=5] (#4\bseq);
                                }{
                                        \countmem{{#8}}{\aseq\id{E}\bseq}
                                        \ifthenelse{\value{memcount} > 0}{
                                                \path (#3\aseq) edge[bend left=5] (#4\bseq);
                                        }{
                                                \path (#3\aseq) edge (#4\bseq);
                                        }
                                }
                        }{}
                        \fi\fi
                }
        }
        
}

\tikzstyle{basevertex} = [circle, fill=black, inner sep=0mm, minimum size = 6mm]

\colorlet{gcolA}{colH}
\colorlet{gcolB}{colB}
\colorlet{gcolC}{colC}
\colorlet{gcolD}{colD}
\colorlet{gcolE}{colF}
\colorlet{gcolF}{colE}
\colorlet{gcolG}{colG}
\colorlet{gcolH}{colA}

\title{Supercritical Size-Width Tree-Like Resolution Trade-Offs for Graph Isomorphism} 
\titlerunning{Supercritical Size-Width Tree-Like Resolution Trade-Offs for Graph Isomorphism}

\author{Christoph Berkholz}{Technische Universit\"at Ilmenau, Germany}{christoph.berkholz@tu-ilmenau.de}{https://orcid.org/0000-0002-3554-517X}{}
\author{Moritz Lichter}{RWTH Aachen University, Germany}{lichter@lics.rwth-aachen.de}{https://orcid.org/0000-0001-5437-8074}{
Funded by the European Union (ERC, SymSim, 101054974). Views and opinions
expressed are those of the author(s) and do not necessarily reflect those of the European
Union or~the ERC. Neither the European Union nor the granting authority
can be held responsible for~them.
}
\author{Harry Vinall-Smeeth}{Technische Universit\"at Ilmenau, Germany}{harry.vinall-smeeth@tu-ilmenau.de}{https://orcid.org/0000-0003-2422-9435}{}

\authorrunning{C.~Berkholz, M.~Lichter and H.~Vinall-Smeeth}
\Copyright{Christoph Berkholz, Moritz Lichter and Harry Vinall-Smeeth}



\funding{\emph{Christoph Berkholz and Harry Vinall-Smeeth}: Funded by the Deutsche Forschungsgemeinschaft (DFG, German Research Foundation) - project number 414325841.}

\nolinenumbers

\newcommand{\confORfull}[2]{#1}
\newcommand{\ifAppendix}[1]{#1}

\begin{document}
        
        \maketitle

\begin{abstract}
We study the refutation complexity of graph isomorphism in the
tree-like resolution calculus.
Torán and Wörz~\cite{DBLP:journals/tocl/ToranW23} showed that 
there is a resolution refutation of \emph{narrow width} $k$ for two graphs if and only if
they can be distinguished in $(k+1)$-variable first-order logic (FO$^{k+1}$)\confORfull{}{ and
hence by a count-free variant of the $k$-dimensional
Weisfeiler-Leman algorithm}.
While DAG-like narrow width $k$ resolution refutations have size at most $n^k$, tree-like refutations
may be much larger.
We show that there are
graphs of order $n$, whose isomorphism can be refuted in narrow width $k$ 
but only in tree-like size $2^{\Omega(n^{k/2})}$.
This is a \emph{supercritical} trade-off where bounding one
parameter (the narrow width) causes the other parameter (the size) to
grow above its worst case. The size lower bound is super-exponential
in the formula size and improves a related supercritical\confORfull{}{ width versus tree-like size} trade-off by Razborov~\cite{DBLP:journals/jacm/Razborov16}.\confORfull{}{\\}
To prove our result, we develop a new variant of the $k$-pebble
EF-game for FO$^k$ to reason about tree-like refutation
size in a similar way as the Prover-Delayer games in proof
complexity. We analyze this game on the compressed CFI graphs introduced by Grohe, Lichter,
Neuen, and Schweitzer \cite{DBLP:conf/focs/GroheLNS23}.
Using a recent improved
\emph{robust} compressed CFI construction of de Rezende, Fleming, Janett, Nordström, and Pang \cite{RFJN024},
we obtain a similar bound for \emph{width}~$k$ (instead of the
stronger but less common \emph{narrow width}) and make the
result more robust.
\end{abstract}         
        
        \section{Introduction}\label{sec:introduction}

A common theme in proof complexity is the difficulty of refuting a given CNF formula in a particular proof system.
There are many variants of this problem depending on the proof system and the notion of difficulty under investigation. We focus on (variants of) resolution,  perhaps the most-studied proof system. Here typical measures of difficulty include the minimum width, depth, space, and (tree-like) size over all refutations of the input formula. 

By analyzing the proof complexity of formulas that encode natural
combinatorial problems, we also gain insights about the inherent complexity of
these problems. In this paper, we focus on the graph isomorphism problem, the 
complexity status of which is still unknown~\cite{DBLP:conf/coco/Karp72}.
On the one hand, Babai~\cite{Babai2016} showed in a breakthrough result that graph isomorphism is solvable in quasi-polynomial time (see also~\cite{DBLP:conf/bcc/GroheN21}),
which makes it a rare natural candidate for a problem that might be neither NP-complete nor polynomial-time solvable.
On the other hand, only relatively weak complexity lower bounds are known~\cite{Toran04}.
This motivates the study of the hardness of graph isomorphism from other perspectives, such as proof complexity.

In this direction,
Torán~\cite{DBLP:conf/sat/Toran13} showed that graph isomorphism is hard for resolution:
There are non-isomorphic graphs on $n$ vertices such that every resolution refutation certifying non-isomorphism has size
$2^{\Omega(n)}$.
Graph isomorphism and the Weisfeiler-Leman algorithm---a well-known algorithm in the context of graph isomorphism---have both  been studied in different proof systems, including
 Sherali-Adams%
~\cite{DBLP:journals/siamcomp/AtseriasM13,
        DBLP:journals/jsyml/GroheO15},
(extended) polynomial
calculus
\cite{DBLP:conf/icalp/BerkholzG15,DBLP:conf/soda/BerkholzG17, Pago2023},
sum-of-squares %
\cite{DBLP:conf/soda/ODonnellWWZ14},
cutting planes%
~\cite{DBLP:conf/sat/ToranW23},
and in extensions of resolution with different symmetry rules~\cite{SchweitzerSeebach21, DBLP:conf/sat/ToranW23}.
We are interested in `tree-like' refutations, which intuitively means
that whenever we want to use a clause in a proof step we have to
re-derive it. Tree-like resolution corresponds exactly to Boolean
decision trees and is closely related to the DPLL
algorithm. We will be concerned with `trade-offs' between the \emph{width}---i.e. the maximum number of literals occurring in any clause---and the \emph{size} of tree-like refutations. 

Studying trade-offs---i.e.\! situations where in order for a refutation to be `easy' with respect to one measure, it has to be `hard' with respect to another---is a prominent theme in proof complexity. 
 For example, there are trade-offs between size and space~\cite{BeameBeckImpagliazzo2016,BenSassonNordstrom2011},
between width and depth, and between width and size~\cite{Berkholz2012, BerkholzNordstrom2020, Thapen2016}.
There are two senses in which a trade-off can be particularly strong. Firstly, a trade-off is \emph{robust}
if it not only shows that one measure~$A$ has to be large
when another measure~$B$ is small,
but also that~$B$ can be increased over some (hopefully) wide range
without that the bound on~$A$ decreasing.
Secondly, a trade-off is \emph{supercritical}
if, in case we restrict~$B$,
the measure~$A$ must be larger than the general upper bound on~$A$ (over all formulas) in the case that~$B$ is not restricted.

In fact,
one of
the first supercritical trade-offs for resolution---proved by Razborov~\cite{DBLP:journals/jacm/Razborov16}---concerns tree-like size and width:
For every $k=k(n)$,
there are $k$-CNF formulas over $n$ variables that can be refuted
in width $O(k)$, but such that every tree-like refutation of minimum width
requires depth $n^{\Omega(k)}$ and size
$2^{n^{\Omega(k)}}$.
When the width is unrestricted then, for each unsatisfiable formula over $n$ variables, there is a tree-like refutation of size at most $2^n$.
Hence, when bounding the width, the tree-like refutation size increases beyond its worst-case and gets super-exponential in the variable number.
This trade-off is not only supercritical but also robust:
The lower bounds also hold for
width-$k'$ refutations for larger $k' < n^{1-\varepsilon}/k$.

One drawback of Razborov's trade-off is that it is not supercritical if we measure size and width with respect to the formula size rather than the number of variables.
While the $k$-CNF formula of Razborov uses $n$ variables, it has size
about $N := n^{\Theta(k)}$.
Thus in terms of the formula size~$N$, the bound on tree-like resolution size is roughly $2^N$.
Our first main result concerns
\emph{narrow resolution}~\cite{DBLP:conf/sat/GalesiT05} which extends resolution by an additional rule
avoiding side effects caused by large width clauses in the input formula.
It is closely related to first-order logic: The  width
and the depth of
\emph{narrow} graph isomorphism refutations correspond to the number of variables and the quantifier depth respectively of first-order formulas distinguishing graphs~\cite{DBLP:conf/sat/ToranW23}.
We show a supercritical trade-off with respect to formula size between width and size for tree-like narrow resolution applied to graph isomorphism formulas. 

\begin{theorem}
        \label{thm: lowerbound-treelike-refutation-narrow-width-non-robust}
        For all integers $k \geq 3$ and $n \in \nat$,
        there are two non-isomorphic colored graphs~$\graphG$ and~$\graphH$ of order $\Theta(n)$ and color class size $16$ such that
        \begin{enumerate}
                \item there is a width-$k$ narrow resolution refutation of $\ISO$, and
                \item every width-$k$ tree-like narrow resolution refutation of $\ISO$ has size $2^{\Omega(n^{k/2})}$.
        \end{enumerate}
\end{theorem}

\noindent Narrow resolution is stronger than resolution; in particular, the minimal width of a narrow resolution refutation is
at most the minimal width of a (plain) resolution refutation. The
formula $\ISO$, which encodes graph isomorphism of $\graphG$ and
$\graphH$, has size
$O(n^4)$
for graphs of order~$n$.
Thus, Theorem~\ref{thm: lowerbound-treelike-refutation-narrow-width-non-robust} indeed
yields a supercritical trade-off between width and tree-like size for narrow resolution \emph{with respect to the formula size}.
Building upon very recent work due to de Rezende, 
                   Fleming, Janett, Nordström, and Pang~\cite{RFJN024}, who provided a not only supercritical but also robust variant of the trade-off in~\cite{DBLP:conf/focs/GroheLNS23}, 
we can show that Theorem~\ref{thm: lowerbound-treelike-refutation-narrow-width-non-robust}
also applies to usual (non-narrow) resolution %
and
the trade-off can be made somewhat robust.

\begin{theorem}
        \label{thm: lowerbound-treelike-refutation-width}
        For all integers $k \geq 3$, $1\leq t \leq \frac{2}{5}k-1$, and $n \in \nat$, there are two colored graphs~$\graphG$ and~$\graphH$
        of order $\Theta(n)$ and color class size $16$
        such that 
        \begin{enumerate}
                \item \label{itm:resolution-refutable} there is a width-$(k+16)$ resolution refutation of $\ISO$, and
                \item \label{itm:resolution-robust} every width-$(k+t-1)$ tree-like resolution refutation of $\ISO$ has size $2^{\Omega(n^{k/(t+1)})}$.
        \end{enumerate}
\end{theorem}

\noindent
With this theorem, we address
Razborov's call for supercritical bounds in terms of formula size~\cite{DBLP:journals/jacm/Razborov16}. Moreover, our trade-off applies to formulas encoding a natural combinatorial problem and is somewhat robust for $t > 16$ and sufficiently large $k$. Since the maximum size of a width $k$ tree-like refutation is $2^{O(n^k)}$, our lower bounds are almost optimal in this range. Our proof utilizes machinery from \emph{finite model theory}:
We introduce a new Ehrenfeucht–Fraïssé style game played on two graphs 
and show that lower bounds for this game imply lower bounds
on the tree-like size of narrow resolution refutations
of the corresponding graph isomorphism formula. %

\subparagraph*{Razborov's Trade-Off and Weisfeiler-Leman.}

To prove his trade-off, Razborov used a compression technique, known as \emph{hardness condensation}~\cite{DBLP:journals/jacm/Razborov16, BerkholzNordstrom16}, that is based on \emph{xorification} and variable reuse
and converts large but hard formulas into smaller ones that are still hard.
Xorification is a well-known technique which replaces every variable in a formula by an XOR of fresh variables.
Xorification, or variable substitution in general, has found many applications in proof complexity 
(see e.g.~\cite{DBLP:conf/stoc/Ben-SassonW99,Ben-SassonNordstrom2008, BenSassonNordstrom2011,BeckNT2013}).
Razborov's compression technique was adapted to the Weisfeiler-Leman (WL) algorithm---an
important algorithm in the field of graph isomorphism~\cite{Grohe2012, Babai2016, GroheNeuen2024}.
The algorithm, parameterized by a dimension~$k$, is a graph isomorphism
heuristic, that is, whenever it distinguishes two graphs they are not isomorphic
but, for every $k$, it fails to distinguish all non-isomorphic graphs~\cite{CaiFuererImmerman1992}.
Of particular interest is the number of iterations needed by the $k$-dimensional WL-algorithm to distinguish two graphs; this 
almost corresponds to the quantifier depth needed in $(k+1)$-variable first-order logic with counting to distinguish them.
Berkholz and Nordström~\cite{BerkholzNordstrom16}
adapted Razborov's compression technique
to construct $k$-ary relational structures
for which the $k$-dimensional WL-algorithm requires
$n^{\Omega(k/\log k)}$ iterations; here the best known upper bound is $O(n^{k-1}/\log n)$~\cite{GroheLN23}.
From the perspective of trade-offs, 
first-order logic (without a bound on the variables)
requires at most quantifier depth $n$ to distinguish all non-isomorphic graphs, which means this trade-off is also supercritical.
The lower bound was recently improved to $n^{\Omega(k)}$~\cite{GroheLN23}.

The trade-offs described above have a common drawback:
They are supercritical with respect to the number of variables and the number of vertices of the structures, respectively,
but not with respect to the formula size or the size of the structure (in terms of the number of tuples in the $k$-ary relations). The common reason is that hardness condensation turns $3$-CNF formulas into $k$-CNF formulas and $3$-ary structures into $k$-ary ones.
But recently,  Grohe, Lichter, Neuen, and Schweitzer~\cite{DBLP:conf/focs/GroheLNS23} introduced a powerful new compression technique for the so-called Cai-Fürer-Immerman (CFI) graphs~\cite{CaiFuererImmerman1992} to prove a lower bound of $\Omega(n^{k/2})$
for the iteration number of the $k$-dimensional WL-algorithm on graphs of order $n$.
The bound not only improves the known ones, it is also a
bound
on graphs
and, as graphs have size $O(n^2)$, the lower bound is supercritical with
respect to the structure size. 
The inspiration for this paper was to see if this new technique yields analogous proof complexity results. 

\subparagraph*{Our Techniques and New Games.}
Tree-like refutations can be (almost) exponentially larger than their non-tree-like counterparts~\cite{Ben-SassonIW2004}.
The usual tool to prove width-$k$ tree-like size lower bounds is the Prover-Delayer game~\cite{PudlakImpagliazzo2000}.
Prover maintains a partial assignment to at most $k$ variables.
In each round, Prover forgets one variable and asks Delayer for an assignment to another one. Delayer can either give such an assignment or allow Prover to set it; in the latter case Delayer scores a point.
Prover wants to find an inconsistent partial assignment and Delayer wants to gain as many points as possible. If Delayer has a strategy to score $p$ points, then every width-$k$ tree-like refutation has size at least $2^p$.
On xorified formulas, where each variable $v$ is replaced with an XOR of $v_0$ and $v_1$,
Delayer can always gain a point when Prover queries $v_0$ or $v_1$ so long as the other one is not already assigned. 
This leads to tree-like size lower bounds exponential in the depth of a refutation~\cite{Urquhart2011}.

However, the xorification of a graph isomorphism formula is not necessarily a graph isomorphism formula. Since we are interested in such formulas,
our idea is to instead apply xorification on the level of graphs. We show that twins in a graph, i.e., vertices with the same neighborhood, can play the role of XORs in a formula:
When we isomorphically map a pair of twins in one graph to a pair of twins in the other graph, the image of the first twin can be chosen arbitrary.
We consider \emph{twinned graphs}, where  every vertex is replaced by a pair of twins.
Ultimately, we want to show that the narrow tree-like refutation size  of a twinned graph is exponential in the refutation depth of the original graph.
Unfortunately, there seems to be no generic argument for this.
To show that this is indeed the case for the graphs we consider,
we introduce a variant of the Prover-Delayer game suited for narrow resolution.
Then we use techniques from finite model theory to show lower bounds for this game.
For other examples of 
finite-model-theoretic techniques in proof complexity see,
e.g.~\cite{Berkholz2012,DBLP:journals/jcss/AtseriasD08,GradelGPP2019}.

We cannot reuse the correspondence between width-$(k-1)$ narrow resolution and $k$\nobreakdash-variable
 first-order logic~\cite{DBLP:journals/tocl/ToranW23}, or equivalently the $k$-pebble game~\cite{DBLP:journals/jcss/Immerman82},
because we care about tree-like size, not only about width and depth.
The issue is that assigning a variable to one or to zero in graph isomorphism formulas is not symmetric: In terms of isomorphisms,
fixing the image of a vertex is usually more restrictive than forbidding a single vertex as the image of another vertex.
We introduce a new pebble game, called the \emph{$k$-pebble game with blocking}, which captures this difference between one and zero assignments.
Round lower bounds in the pebble game with blocking imply
exponential size lower bounds for tree-like resolution.

Another game is involved in this lower bound.
The hardness of uncompressed CFI graphs for $k$-variable first-order logic
is captured by the $k$-Cops and Robber game~\cite{SeymourThomas93, DBLP:conf/focs/GroheLNS23},
which forgets about the CFI construction and instead considers the simpler underlying \emph{base graphs}.
For compressed CFI graphs, this game was modified to the compressed $k$-Cops and Robber game~\cite{DBLP:conf/focs/GroheLNS23}.
To obtain lower bounds for the $k$-pebble game with blocking,
we have to introduce a blocking mechanism to the compressed $k$-Cops and Robber game. 
Via all these games, we obtain the 
$2^{\Omega(n^{k/2})}$ narrow width-$k$ tree-like size lower bound in Theorem~\ref{thm: lowerbound-treelike-refutation-narrow-width-non-robust}.

\subparagraph{From Narrow to Plain Resolution.}
We lift Theorem~\ref{thm: lowerbound-treelike-refutation-narrow-width-non-robust} to (non-narrow) resolution.
Since lower bounds for narrow resolution imply lower bounds for resolution,
transferring the lower bounds is trivial.
But it is unclear whether the relevant isomorphism formula can be refuted in (non-narrow) width $k$.
By increasing the width by the maximal color class size of these graphs (which is $16$),
we can simulate the narrow resolution refutation by a plain resolution refutation.
But now the lower bound from Theorem~\ref{thm: lowerbound-treelike-refutation-narrow-width-non-robust} does not apply anymore.
At this point, the aforementioned result from~\cite{RFJN024} comes to hand:
The compression of the CFI graphs get modified to obtain,
for every fixed $t < k$, graphs whose isomorphism formula can be refuted in narrow width $k$ but
every narrow width $k+t$ refutation has depth at least
$\Omega(n^{k/(t+1)})$.
So the lower bound is robust within the range
from $k$ to $k+t-1$.
This construction can be seen as an interpolation between the 
original compression~\cite{DBLP:conf/focs/GroheLNS23} with round lower bound $\Omega(n^{k/2})$
and the linear round lower bound $\Omega(n)$ by Fürer~\cite{Furer2001};
both appear as special cases for $t=1$ and $t=k$.
Our approach with twinned graphs also applies
to the improved construction implying a narrow width-$(k+t)$ tree-like size lower bound of $2^{\Omega(n^{k/(t-1)})}$, but we need to restrict the range of $t$ even further.
For $k$ large enough and $t>16$
we can actually refute isomorphism of the graphs in (non-narrow) width $k+t$
and finally obtain a supercritical trade-off between width and
tree-like size for resolution with respect to formula size (Theorem~\ref{thm: lowerbound-treelike-refutation-width}).

\subparagraph{Further Related Work.}

How the robust compressed CFI construction~\cite{RFJN024} yields  a supercritical width-depth trade-off for resolution was presented at the Oberwolfach workshop \emph{Proof Complexity and Beyond}~\cite{oberwolfach2024}.
The resulting preprint~\cite{RFJN024} also contains a trade-off for
tree-like resolution.
A key difference is that our trade-off applies to graph-isomorphism formulas. Also, different techniques are used.
We cannot apply hardness condensation techniques to graph isomorphism formulas but apply a form of xorification on the level of graphs and analyze them using model theoretic techniques. In contrast, the trade-off~of~\cite{RFJN024} is obtained via xorification; the parameters obtained are within a constant factor of one-another.

        \section{Preliminaries}\label{sec:preliminaries}

\subparagraph{Graphs.}
An \defining{(undirected) graph}~$\graphG$ is a tuple $(V,E)$ where $V$ is a finite set of \defining{vertices} 
and $E \subseteq \tbinom{V}{2}$ is a set of \defining{edges}.
The vertex set of~$\graphG$ is denoted by~$\vertexG$
and the edge set of~$\graphG$ by~$\edgeG$.
For $W \subseteq \vertexG$,
the \defining{subgraph of $\graphG$ induced by $W$} is denoted by $\graphG[W]$.
The \defining{distance} between two vertices $u,v \in \vertexG$
is the number of edges in a shortest path between~$u$ and~$v$ in~$\graphG$.
The distance between $U,W \subseteq \vertexG$
is the minimal distance of
all $u \in U$ and $v \in W$.
We will sometimes consider directed graphs,
where the set of edges is a subset of $\vertexG^2$,
but we mention this explicitly.
A directed graph is \defining{acyclic}, if it does not contain a (directed) cycle.
 A \defining{source} (or \defining{sink}) is a vertex without incoming (or outgoing) edges.
A \defining{colored} graph~$\graphG$ is a tuple $(V,E,\chi)$
such that $(V,E)$ is a graph and~$\chi$ is a map
$V \to \nat$.
We interpret~$\chi$ as a vertex coloring of~$\graphG$ and denote it by~$\coloringG$.
The \defining{color class} of $u \in \vertexG$
is the set $\inv{\coloringG}(\coloringG(u))$ of vertices of the same color as $u$.
The \defining{color class size} of~$\graphG$
is the maximal cardinality of its color classes.
The graph~$\graphG$ is \defining{ordered} if~$\chi$ is injective.
We can see every graph as a colored graph in which every vertex is colored~$0$.
An \defining{isomorphism} of colored graphs $\graphG$ and $\graphH$
is a bijection $f \colon \vertexG \to \vertexH$
such that,
for all $u,v\in\vertexG$, we have $\coloringG(u) = \coloringH(f(u))$, and $\set{u,v} \in \edgeG$
if and only if $\set{f(u),f(v)} \in \edgeH$.
If there is such an isomorphism, $\graphG$ and~$\graphH$ are \defining{isomorphic}.

\subparagraph{Resolution.}
A \defining{literal} is a proposition variable $x$ or its negation $\overline{x} := \neg x$. We set $\overline{\neg x} = x$.
A \defining{clause}~$C$ is a finite set of literals $\set{\lambda_1,\dots, \lambda_k}$.
We may write clauses as disjunctions, e.g., $C = (\lambda_1 \lor \dots \lor \lambda_k)$. %
A \defining{CNF formula}~$F$  is a finite set of clauses $\set{C_1,\dots,C_m}$,
which we may write as a conjunction $F = (C_1 \land \dots \land C_m)$.
The set of \defining{variables occurring} in a clause~$C$ is $\Vars{C}$
and for a CNF formula~$F$ it is $\Vars{F} := \bigcup_{C \in F} \Vars{C}$.
A \defining{(partial) assignment} for a CNF formula~$F$ is a (partial) map
$\sigma\colon \,\Vars{F} \to \set{0,1}$.
The \defining{domain} of~$\sigma$ is $\dom(\sigma)$.
The \defining{size} of~$\sigma$ is $|\dom(\sigma)|$.
The assignment $\sigma$ \defining{violates} a clause $C \in F$
if $\Vars{C} \subseteq \dom(\sigma)$ and~$\sigma$ satisfies no literal in~$C$.
For a variable $x \in \Vars{F}$ and a Boolean value $\delta \in \set{0,1}$,
let $\sigma[x\mapsto \delta]$ be the assignment with domain $\dom(\sigma) \cup \set{x}$ derived from $\sigma$ that sets $x$ to $\delta$, i.e., $\sigma[x\mapsto \delta](x) = \delta$ and $\sigma[x\mapsto \delta](y) = \sigma(y)$ for
all $y \in \dom(\sigma) \setminus \set{x}$.
For partial assignments~$\sigma$ and~$\sigma'$, we write $\sigma' \subseteq \sigma$ if $\dom(\sigma') \subseteq \dom(\sigma)$ and $\sigma'(x) = \sigma(x)$ for all $x\in \dom(\sigma')$. 
For $k\in \nat$, we write $[k]:= \{1, \dots, k\}$.
We now  introduce the proof systems studied in this paper.

\begin{definition}[Narrow Resolution~\cite{DBLP:conf/sat/GalesiT05}]
A \defining{narrow resolution derivation} $\pi$ of a clause~$D$ from a CNF formula~$F$ is a directed acyclic graph $\pi=(V,E)$
whose vertices are labeled with clauses,~$D$ is the label of a source of $\pi$, and all sinks of $\pi$ are labeled with clauses in $F$.
Moreover, for every vertex $v \in V$, its clause $C$ is derived from the clauses $C_1,\dots, C_\ell$ labeling the vertices, to which $v$ has an outgoing edge, by one of the following three rules.
\begin{enumerate}
\item {\bf Axiom Rule:} \label{itm:axiom} $\ell = 0$ and $C \in F$.
\item {\bf Resolution Rule:} \label{itm:resolution} $\ell=2$ and $C_1 = A \vee x$, $C_2 = B \vee \overline{x}$, and $C = A \vee B$.
\item {\bf Narrow Resolution Rule:} \label{itm:narrow-resolution} 
$\ell\geq 2$ and, up to reordering, $C_\ell =(A \vee \lambda_1 \vee \dots \vee \lambda_{\ell-1}) 
\in F$ is an axiom, $C = (A \vee A_1 \dots \vee A_{\ell-1})$, and
$C_{i} = (A_i \vee 
\overline{\lambda_i})$ for all $i \in [\ell-1]$.
\end{enumerate}
\end{definition}

\noindent A \defining{resolution derivation} is a narrow resolution derivation using only Rules~\ref{itm:axiom} and~\ref{itm:resolution}.
The derivation~$\pi$ is \defining{tree-like} if $\pi$ is a tree. In this case we call the unique source the \defining{root} and the sinks \defining{leaves}. 
The \defining{size}~$|\pi|$ of the derivation~$\pi$
is the number of vertices $|V|$.
The \defining{depth} of~$\pi$ is the length of the longest directed path in it.
The \defining{width of a clause}~$C$ is its number of literals.
The \defining{width of a derivation} $w(\pi)$ is the maximal width of all clauses in $\pi$.
The \defining{narrow width of a derivation} $w^{\ast}(\pi)$ is the maximal number of literals among all those clauses in $\pi$ that are not axioms.
A \defining{$k$-narrow} derivation is a narrow resolution derivation of narrow width at most~$k$.
A derivation of the empty clause from~$F$ is a \defining{refutation} of~$F$.

        \section{A Prover-Delayer Game for Tree-Like Narrow Resolution} \label{sec: Prover}

In this section, we introduce a game that allows us to prove lower bounds on the size of $k$-narrow tree-like refutations of \defining{graph isomorphism formulas}. We now introduce these formulas.
Let $\graphG$ and $\graphH$ be $n$-vertex colored graphs; following
\cite{DBLP:journals/tocl/ToranW23}, we define a CNF
formula $\ISO$ whose solutions correspond to isomorphisms $\graphG \to
\graphH$. 
For all vertices $u \in \vertexG$ and $v \in \vertexH$,
we add a propositional variable $x_{u,v}$. The variables $x_{u,v}$ have the intended meaning that~$u$ is mapped to~$v$.
The CNF formula $\ISO$ contains three types of clause.
\begin{itemize}
\item \defining{Color Clauses:}
for each vertex $u \in \vertexG$,
let $W_u := \inv{\coloringH}(\coloringG(u))$ be the vertices of $\graphH$ with the same color as $u$.
Add
 the clause $\bigvee_{v \in W_u} x_{u,v}$ to encode that $u$ is mapped to a vertex of the same color.
 For each $v \in \vertexH$,
 let $W_v := \inv{\coloringG}(\coloringH(v))$ and add the clause $\bigvee_{u \in W_v} x_{u,v} $. 
\item \defining{Bijection Clauses:} 
For all $u\in \vertexG$ and distinct $v,w \in \vertexH$, we add the clause $(\neg x_{u,v} \vee \neg x_{u,w})$ to encode that an isomorphism is a function.
For all distinct $u,v \in \vertexG$ and $w \in \vertexH$, we add the clause $(\neg x_{u,w} \vee \neg x_{v,w})$ to encode
injectivity of the desired isomorphism. 

\item \defining{Edge Clauses:} for all $u,u' \in \vertexG$ and $v, v' \in \vertexH$ with $u\neq u'$ such that $\{u,u'\} \in \edgeG$ if and only if $\{v, v' \} \not \in \edgeH$, we include the 
clause $\neg x_{u,v} \vee \neg x_{u',v'}$. This encodes the edge relation.
\end{itemize}
\noindent The formula $\ISO$ has $O(n^2)$ variables, $O(n^4)$ clauses, width equal  to the maximal color class size of $\graphG$ and $\graphH$ (unless every vertex gets a unique color; in this case the width is two),
and is satisfiable if and only if $\graphG$ is isomorphic to $\graphH$.

\subparagraph*{The \texorpdfstring{$k$}{k}-Narrow Prover-Delayer Game.}

Let $\graphG$ and $\graphH$ be non-isomorphic colored graphs.
The \defining{$k$-narrow Prover-Delayer game} on $\graphG, \graphH$ is played 
by two players, Prover and Delayer, who construct  partial assignments for $\ISO$ as follows. The game begins with the empty assignment $\sigma_0 = 
\varnothing$.
Let $\sigma_{t-1}$ be the assignment after the $(t-1)$-th round.
In round $t$, Prover chooses $\sigma \subseteq \sigma_{t-1}$ with $|\dom(\sigma)| \le k-1$ and makes one of the following kinds of moves.
\begin{enumerate}
\item \defining{Resolution Move:} Prover chooses a variable $x \not \in \dom(\sigma)$. Delayer chooses a response. 
\begin{enumerate}
\item \defining{Committal Response:} Delayer responds with $\delta \in \{0,1\}$ and sets $\sigma_t := \sigma [x \mapsto \delta ]$.
\item  \defining{Point Response:} Delayer gets a point; Prover picks $\delta \in \{0,1\}$ and sets $\sigma_t := \sigma [x \mapsto \delta ]$.
\end{enumerate}
\item  \defining{Narrow Move:} Prover chooses a color clause $C$ from $\ISO$. Again Delayer chooses one of two response types.
\begin{enumerate}
\item \defining{Committal Response:} Delayer chooses some $x \in C \setminus \sigma^{-1}(0)$ and sets $\sigma_t := \sigma [ x \mapsto 1 ]$.
\item  \defining{Point Response:} Delayer chooses distinct $x,y \in C \setminus \sigma^{-1}(0)$ and gets a point; Prover chooses $z \in \{x,y\}$ and sets $\sigma_t := \sigma [z \mapsto1 ]$. 
\end{enumerate}
\end{enumerate} 
\noindent If the assignment $\sigma_t$ violates a clause of $\ISO$,
the game ends and \defining{Prover wins}.  Otherwise, the game continues in round $t+1$. \defining{Prover has an $r$-point strategy} if, no matter how Delayer plays, Prover can always win the game while limiting Delayer to at most~$r$ points. If Prover does \emph{not} have an $r$-point strategy, then \defining{Delayer has an $(r+1)$-point strategy}.
It will be useful to start the $k$-narrow Prover-Delayer game on $\graphG, \graphH$ from assignments $\sigma \neq \emptyset$. In this case, the game starts at $\sigma_0 = \sigma$.
By constructing strategies for Delayer, the game can be used to show tree-like size lower bound for resolution refutations of graph isomorphism formulas\ifAppendix{; the full proof is given in Appendix~\ref{a: prover_delayer}}.

\begin{lemma} \label{lem: game}
For all $k\ge 1$, colored graphs $\graphG, \graphH$, and $k$-narrow tree-like refutations $\pi$ of $\ISO$, Prover has a $(\lceil \log(|\pi|)  \rceil)$-point strategy in the $(k+1)$-narrow Prover-Delayer game on $\graphG,\graphH$. 
\end{lemma}

\begin{proof}[Proof Sketch]
Prover follows $\pi$ starting at the empty clause at the root.
If a resolution rule is applied to a variable $x$, then Prover makes a resolution move for $x$. Similarly, Prover follows narrow resolution moves.
If Delayer makes a committal response, Prover moves to the corresponding child in $\pi$.
If Delayer makes a point response, Prover moves to the child with the smallest subtree `below it', at least halving the size of the subtree at the current position. Prover wins if a leaf is reached, so Delayer can score at most $(\lceil \log(|\pi|)  \rceil$  points.
\end{proof}

\subparagraph*{The \texorpdfstring{$k$}{k}-pebble Game and Narrow Resolution.}
We next recall the connection between
$(k-1)$-narrow resolution refutations 
and the $k$-variable fragment of first order logic~\cite{DBLP:journals/tocl/ToranW23}.
For an integer $k$, we write $\klogic{k}$ for the set of first order formulas using at most $k$ \emph{distinct} variables.
We denote the set of  $\klogic{k}$-formulas with quantifier depth at most $r$ by $\kmlogic{k}{r}$.
If two graphs $\graphG$ and $\graphH$ satisfy the same sentences of $\klogic{k}$ or $\kmlogic{k}{r}$,
the graphs are $\klogic{k}$-equivalent or $\kmlogic{k}{r}$-equivalent, respectively, and we write $\graphG \kequiv{k} \graphH$ or $\graphG \kequivr{k}{r} \graphH$, respectively.

These equivalences are characterized by the following game:
Let $\graphG$ and $\graphH$ be (colored) graphs and $k, r \in \mathbb{N}$. The \defining{$r$-round $k$-pebble game} on $\graphG, \graphH$ is played by two players, Spoiler and Duplicator.
A \defining{position} of the game is a pair $(\alpha,\beta)$ of partial assignments $\alpha \, \colon [k] \to \vertexG$ and $\beta \, \colon [k] \to \vertexH$ such that $\dom(\alpha) = \dom(\beta)$. 
These maps define positions of up to $k$ pebble pairs on~$\graphG$~and~$\graphH$.
Duplicator aims to show that~$\graphG$ and~$\graphH$ are isomorphic; Spoiler tries to show they are not.
Initially, no pebbles are placed. Let $(\alpha_t, \beta_t)$ be the position at the end of round $t < r$. At the beginning of round $t+1$, Spoiler picks one of the graphs, say $\graphG$, and $i \in [k]$. The $i$-th pebble pair is picked up and Spoiler places the $i$-th pebble for $\graphG$ on some $u \in \vertexG$ yielding the map $\alpha_{t+1}$.
Duplicator responds by similarly placing the $i$-th pebble for $\graphH$ on a vertex of $\graphH$ yielding $\beta_{t+1}$. 
If $(\alpha_{t+1}, \beta_{t+1})$ does \emph{not} \defining{induce a partial isomorphism}, meaning that $\alpha(i) \mapsto \beta(i)$ is not an isomorphism of the induced subgraphs $\graphG[\{\alpha(i) \mid i \in \dom(\alpha)\}]$ and $\graphH[\{\beta(i) \mid i \in \dom(\beta)\}]$,
then \defining{Spoiler wins}.
Otherwise, if $t+1 <r$, the play continues in the next round.
If $t+1=r$, then \defining{Duplicator wins}.
A player (Spoiler or Duplicator) has a \defining{winning strategy},
if they can win independently of the moves of the other player.

\begin{theorem}[\cite{DBLP:journals/jcss/Immerman82, DBLP:journals/tocl/ToranW23}] \label{thm: logic_equiv}
Let $k,r \in \nat$. The following are equivalent:
\begin{enumerate}
\item $\graphG \notkequivr{k}{r} \graphH$, i.e., $\graphG$ and $\graphH$ are \emph{not} $\kmlogic{k}{r}$-equivalent.
\item Spoiler has a winning strategy in the $r$-round $k$-pebble game on $\graphG, \graphH$.
\item There is a $(k-1)$-narrow resolution refutation of $\ISO$ of depth at most $r$.
\end{enumerate}
\end{theorem}

\noindent It will sometimes be convenient to start the game from position $(\alpha_0, \beta_0) \neq (\emptyset, \emptyset)$; nothing else in the rules of the games changes in this case. 
Similarly, we may sometimes not specify the number of rounds in advance; in this case the game only ends if Spoiler wins.

       \section{Twinned Graphs and Pebble Games} \label{sec: Clique}

In this section, we introduce the \emph{twinned graph} construction, which we described on a high-level in the introduction. This will allow us to transfer lower bounds on the $k$-pebble game with blocking to lower bounds on the $k$-narrow Prover-Delayer game (and therefore to lower bounds on $(k-1)$-narrow tree like refutation size). 

We use a colored graph~$\graphG$ to define a new colored graph as follows.
For each vertex $u \in \vertexG$, we set $\XG(u) := \{u_0, u_1\}$, where $u_0$ and $u_1$ are fresh vertices; intuitively these are copies of $u$. We define the \defining{twinned graph} $\graphXG$ with vertex set $\vertexXG := \bigcup_{u \in \vertexG} \XG(u)$ and edge set
\begin{align*}
  \edgeXG  &:= \setcond[\big]{ \{x,y\} }{ x \in \XG(u),\, y\in \XG(v), \{u,v\} \in \edgeG} \cup \setcond[\big]{\XG(u)}{ u\in \vertexG}.
\end{align*}  
We give $u_0$ and $u_1$ the same color in $\graphXG$ as $u$ has in $\graphG$. For notational convenience, we define $\hat{u}_0 := u_1$ and $\hat{u}_1 := u_0$. Moreover, we define $\XG^{-1}(u_i) := u$ for $i\in \{0,1\}$.

We first show that---under a mild condition---if there is a $k$-narrow refutation of $\ISO$,
then there is a $k$-narrow refutation of $\XISO$.
To state the condition, we need the following notion. Two distinct vertices $u,v \in \vertexG$ are \defining{twins} if for every $w \in \vertexG \setminus \set{u,v} $, we have that $\set{u,w} \in \edgeG$ if and only if $\set{v,w} \in \edgeG$. That is, the neighborhoods of $u$ and $v$ in $\graphG$ are, apart from $u$ and $v$ themselves, identical. Twins $u$ and $v$ are \defining{connected twins} if $\set{u,v} \in \edgeG$. Note that if $\graphG$ has no connected twins, then $\graphXG$ has exactly one pair of connected twins for each vertex in $\graphG$. This leads to the following observation. \ifAppendix{See Appendix~\ref{a: duplicator_to_delayer} for a proof.}

\begin{lemma} \label{lem: spoiler_win}
Let $k \ge 3$ and $\graphG$ and $\graphH$ be colored graphs that do not have connected twins.
If $\graphG \notkequivr{k}{r} \graphH$, then $\graphXG \notkequivr{k}{r+1} \! \graphXH$. 
\end{lemma}
\noindent By applying Theorem~\ref{thm: logic_equiv} we obtain the following corollary. 

\begin{corollary} \label{cor: spoiler_win}
Let $k \ge 2$ and $\graphG$ and $\graphH$ be colored graphs that have no connected twins.
If there exists a $k$-narrow refutation of $\ISO$ of depth $d$,
then there exists a $k$-narrow refutation of $\XISO$ of depth $d+1$.
\end{corollary}

\noindent It turns out that Prover-Delayer lower bounds on our twinned graphs are implied by round lower bounds for certain pebble games on the original graphs. The normal $k$-pebble game is the wrong tool for this task; intuitively,
the reason is the asymmetry between setting a variable of a graph isomorphism formula to zero or one.

\subparagraph*{The \texorpdfstring{$k$}{k}-Pebble Game with Blocking.}

Let $\graphG$ and $\graphH$ be colored graphs and $k, r \in \mathbb{N}$. We define the $r$-round $k$-pebble game with blocking on $\graphG$ and  $\graphH$ as follows. The game is played in rounds by Spoiler and Duplicator. A position in the game is a triple 
$(\alpha, \beta, c)$ of partial maps $\alpha\, \colon [k] \to \vertexG $, $\beta \, \colon [k] \to \vertexH$, and $c \,\colon [k] \to \{\regular, \blocking\}$
with $\dom(\alpha) = \dom(\beta) = \dom(c)$.
The first two maps give the positions of the pebbles and $c$ marks each pair of pebbles as either \defining{regular} or \defining{blocking}.
Regular pebbles (possibly) define partial isomorphisms as before, but blocking ones forbid certain ones as follows.

\begin{definition}[Partial Isomorphism with Blocking] \label{def: pi_block}
        Let $(\alpha, \beta, c)$ be a position in the $r$\nobreakdash-round $k$\nobreakdash-pebble game with blocking on $\graphG$ and  $\graphH$, $R := c^{-1}(\regular)$, and $B := c^{-1}(\blocking)$. Then $(\alpha, \beta, c)$ 
        \defining{induces a  partial isomorphism with blocking} if $(\left.\alpha\right|_R, \left.\beta\right|_R)$ induces a partial isomorphism and
        if every regular pebble respects every blocking pebble. Formally, this means that for every $p\in B$ and $q\in R$, 
        we have $(\alpha(p), \beta(p) ) \neq (\alpha(q), \beta(q))$.
\end{definition}
In the initial position $(\alpha_0, \beta_0, c_0)$,
all maps are empty.
Let $(\alpha_t, \beta_t, c_t)$ be the position after the $t$-th round.
At the beginning of the $(t+1)$-th round,
Spoiler can make either a \defining{regular move} or a \defining{blocking move}.
A regular move works in the same way as a move in the $k$-pebble game; the pebble pair moved in this turn is then marked $\mathsf{\regular}$.
For a blocking move, Spoiler picks $p\in [k]$ and places the $p$\nobreakdash-th pebble in $\graphG$ on some vertex $u\in \vertexG$ and in $\graphH$ on some vertex $v\in \vertexH$.
Duplicator next decides how to mark this pair.
If Duplicator chooses $\mathsf{\regular}$, then the round ends.
If instead Duplicator chooses $\blocking$, then the round continues and Spoiler can again choose to make either a regular or a blocking move. If $(\alpha_{t+1}, \beta_{t+1}, c_{t+1})$ does \emph{not} induce a partial isomorphism with blocking, then \defining{Spoiler wins} and the game ends. Otherwise if $t+1 < r$, the game continues in round $t+2$.
If $t+1 = r$, then \defining{Duplicator wins}.

We write $\graphG \kbequivr{k}{r} \graphH$ if Duplicator has a winning strategy in the $r$-round $k$-pebble game with blocking. If $\graphG \kbequivr{k}{r} \graphH$ for all $r \in \mathbb{N}$, then we write $\graphG \kbequiv{k} \graphH$. As for the $k$-pebble game, it will also be convenient to consider variants of the $k$-pebble game with blocking where we start from arbitrary positions or do not specify the number of rounds in advance. Note that while in the (non-blocking) $k$-pebble game it never makes sense for Spoiler to place a pebble on an already pebbled vertex, this is not the case in the $k$-pebble game with blocking.

\subparagraph*{Spoiler and Duplicator meet Prover and Delayer.}

We end the section by connecting the $k$-pebble game with blocking to the $k$-narrow Prover-Delayer game via the following lemma.

\begin{lemma}
        \label{lem:duplicator-blocking-to-delayer}
        Let $\graphG$ and $\graphH$ be colored graphs
        and $k\geq 2$ an integer. If $\graphG \kbequivr{k}{r} \graphH$, then Delayer has an $r$-point strategy in the $k$-narrow Prover-Delayer game on $\graphXG, \graphXH$.
\end{lemma}

	\begin{proof}[Proof Sketch]
\noindent In the $k$-narrow Prover-Delayer game on $\graphXG,\graphXH$,
Delayer  
simulates positions of the $k$-pebble game with blocking on $\graphG$, $\graphH$.
Intuitively, whenever a $\regular$ pebble pair is placed on vertices $u$ and $v$
(and there is not already a pebble pair on $u$ and $v$),
Delayer should score a point 
since it `does not matter' whether we map $u_0$ to $v_0$ or to $v_1$. As the round counter of the $k$-pebble game with blocking only advances when a pebble pair is marked as $\regular$, filling in the details is relatively straightforward. \ifAppendix{See Appendix~\ref{a: duplicator_to_delayer}.}
\end{proof}

\noindent Lemmas~\ref{lem: game} and~\ref{lem:duplicator-blocking-to-delayer} finally connect the pebble game with blocking to tree-like refutation size.

\begin{theorem} \label{cor: block_to_lb}
Let  $k \ge 1$, and $r \in \mathbb{N}$ and $\graphG$ and $\graphH$ be colored graphs.
If $\graphG \kbequivr{k+1}{r} \graphH$, then every $k$-narrow tree-like refutation of $\XISO$ has size at least $2^r$.
\end{theorem}

 \section{Compressing CFI Graphs}\label{sec:compressing-cfi}

By what we have seen so far (Corollary~\ref{cor: spoiler_win} and Theorem~\ref{cor: block_to_lb}), to prove Theorem~\ref{thm: lowerbound-treelike-refutation-narrow-width-non-robust} it suffices to show that Duplicator can survive a large number of rounds in the $k$-pebble game with blocking on suitably chosen colored graphs $\graphG, \graphH$. In this section, we describe a framework which allows us to construct such graphs.

Concretely, we recall a recent approach to construct pairs of graphs that
require quantifier depth $\Omega(n^{k/2})$
to be distinguished in
$k$-variable first order logic $\klogic{k}$ (and also with counting)~\cite{DBLP:conf/focs/GroheLNS23}.
The key idea is a novel compression technique of the so-called Cai-Fürer-Immerman (CFI) graphs~\cite{CaiFuererImmerman1992} and a concrete compression construction for CFI graphs over grids. Having introduced this construction, we give a method for proving lower bounds for the $k$-pebble game with blocking on compressed CFI graphs. To do this, we first recall a variant of the Cops and Robber game, which can be used to derive lower bounds on the $k$-pebble game on compressed CFI graphs, and then extend this game with an appropriate notion of blocking. 

\confORfull{}{\subsection{CFI Graphs, Compressions, and the Cops and Robber Game}}
\subparagraph*{CFI Graphs.}
Let $\graphG =(\vertexG, \edgeG)$ be a connected ordered graph, called a \defining{base graph},
and $f \colon \edgeG \to \FF_2$ be a function,
where $\FF_2$ is the two-element field.
From $\graphG$ and $f$
we derive the colored \defining{CFI graph} $\CFI{\graphG,f}$:
Vertices of~$\graphG$ are called \defining{base vertices}.
Every base vertex of~$\graphG$ is replaced by a \defining{CFI gadget}
and gadgets of adjacent vertices are connected.
The vertices of the CFI gadget for a degree~$d$ base vertex $u\in \vertexG$
are the pairs $(u,\tup{a})$ for all $d$-tuples $\tup{a} =(a_1, \dots, a_d) \in \FF_2^d$ with$\sum_{i=1}^d a_i = 0$.
The vertex $(u,\tup{a})$ has \defining{origin}~$u$. Vertices~inherit the color of their origin.
Since every vertex of the base graph has a unique color, the vertices of each gadget form a color class of the CFI graph.
For all adjacent base vertices $u,v \in \vertexG$,
we add the following edges between the gadgets for~$u$ and~$v$:
Let~$u$ be the $i$-th neighbor of~$v$
and~$v$ be the $j$-th neighbor of~$u$ according to the order on $\vertexG$.
There is an edge between vertices $(u, \tup{a})$ and $(v, \tup{b})$ if and only if $a_i + b_j = f(\set{u,v})$,
where $a_i$ is the $i$-th entry of $\tup{a}$ and $b_j$ is the $j$-th entry of $\tup{b}$. See Figure~\ref{fig:compressed-cfi} for an example.
We say that two functions $f,g \, \colon \edgeG \to \FF_2$
\defining{twist an edge} $e \in \edgeG$
or the edge $e\in \edgeG$ is \defining{twisted by $f$ and $g$}
if $f(e) \neq g(e)$.
It is well-known that $\CFI{\graphG,f} \not\iso\CFI{\graphG,g}$
if and only if~$f$ and~$g$ twist an odd number of edges~\cite{CaiFuererImmerman1992}.

\subparagraph*{Compressing CFI Graphs.}

CFI graphs are a well-studied tool to derive lower bounds for
$k$-variable logic with counting or other logics, see e.g.
\cite{CaiFuererImmerman1992, Furer2001, BerkholzNordstrom16,
DawarGraedelLichter22, Lichter2023b}. This construction and its
generalizations have also been used to derive proof complexity lower bounds on graph isomorphism in various
proof systems
\cite{DBLP:conf/sat/Toran13,DBLP:conf/icalp/BerkholzG15,DBLP:conf/soda/BerkholzG17,DBLP:conf/soda/ODonnellWWZ14,DBLP:conf/sat/ToranW23,SchweitzerSeebach21}.
We now discuss the method of \emph{compressing CFI graphs}~\cite{DBLP:conf/focs/GroheLNS23}.
The goal is to reduce the size of the resulting graph
while essentially preserving the hardness of it.
The main idea is to identify the gadgets of certain base vertices.
The hardness of the resulting compressed CFI graphs
heavily depends on which gadgets get identified
and can be analyzed using a variant of the Cops and Robber game.
We now present this framework.

\begin{definition}[Graph Compression]
        An equivalence relation~$\comp$ on $\vertexG$ is a \defining{$\graphG$-compression} if for all $u,u',v, v' \in \vertexG$ it satisfies the following two conditions:
        \begin{enumerate}
                \item If $u \comp v$, then
                $u$ and $v$ are non-adjacent and of the same degree.
                \item \label{itm:compression-order-agree} If $\set{u,v},\set{u',v'} \in \edgeG$, $u \comp u'$, $v \comp v'$, and $u$ is the $i$-th neighbor of $v$ (according to the order on~$\vertexG$), then $u'$ is the $i$-th neighbor of $v'$.
        \end{enumerate}
\end{definition}
 
\noindent Let ${\comp} \subseteq \vertexG^2$ be a $\graphG$-compression.
It induces an equivalence relation on $\CFI{\graphG,f}$ (independently of the function $f \colon E \to \FF_2$), which we also denote by $\comp$, via
$(u, \tup{a}) \comp (v, \tup{b})$
if and only if $u \comp v$ and $\tup{a} = \tup{b}$.
Contracting all $\comp$\nobreakdash-equivalence classes in $\CFI{\graphG,f}$ into a single vertex yields the colored graph $\compCFI{\graphG,f}{\comp}$.
Formally, the vertices of $\compCFI{\graphG,f}{\comp}$ are the $\comp$-equivalence classes $u/_{\comp} := \setcond{w \in V_{\CFI{\graphG,f}}}{w \comp u}$,
and $u/_\comp$ and $v/_\comp$ are adjacent
if there are $u'\comp u$ and $v'\comp v$ such that~$u'$ and~$v'$ are adjacent in $\CFI{\graphG,f}$.
Observe that $\compCFI{\graphG,f}{\comp}$ is loop-free by the condition on $\comp$ that equivalent vertices of~$\graphG$ are non-adjacent.
The color of a $\comp$\nobreakdash-equivalence class in $\compCFI{\graphG,f}{\comp}$ is the minimal color of one of its members in $\CFI{\graphG,f}$.
To obtain reasonable graphs, $f$ has to be compatible with the compression~$\comp$ in the following sense.

\begin{figure}
                \def\spacing{2}
                \begin{subfigure}[t]{0.48 \textwidth}
                        \centering
                \begin{tikzpicture}[scale=0.75]
                        \draw [draw=none, use as bounding box] (-1.7*\spacing, -0.35*\spacing+0.2) rectangle (2.7*\spacing, 1.35*\spacing);
                        \begin{scope}
                                \node[basevertex,gcolA] (z1) at (-\spacing,0){};
                                \node[basevertex,gcolE] (z2) at (-\spacing,\spacing){};
                                \node[basevertex,gcolB] (a1) at (0,0){};
                                \node[basevertex,gcolC] (b1) at (\spacing,0){};
                                \node[basevertex,gcolD] (c1) at (2*\spacing,0){};
                                \node[basevertex,gcolF] (a2) at (0,\spacing){};
                                \node[basevertex,gcolG] (b2) at (\spacing,\spacing){};
                                \node[basevertex,gcolH] (c2) at (2*\spacing,\spacing){};
                                
                                \path[draw, ultra thick]
                                (z1) -- (z2)
                                (a1) -- (a2)
                                (b1) -- (b2)
                                (c1) -- (c2)
                                (z1) -- (a1) -- (b1) -- (c1)
                                (z2) -- (a2) -- (b2) -- (c2);
                                \path[draw, dashed, ultra thick]
                                ($(z1.center)-(1.2,0)$) -- (z1)
                                ($(z2.center)-(1.2,0)$) -- (z2)
                                ($(c1.center)+(1.2,0)$) -- (c1)
                                ($(c2.center)+(1.2,0)$) -- (c2);
                                
                                \path[draw, magenta, line width=2mm]
                                (z1) to (b2);
                                
                        \end{scope}
                \end{tikzpicture}
                \caption {The base graph with the compression.}
        \end{subfigure}
        \begin{subfigure}[t]{0.48\textwidth}
                \centering
                \begin{tikzpicture}[scale=0.75]
                        \draw [draw=none, use as bounding box] (-0.7*\spacing, -0.35*\spacing+0.2) rectangle (3.7*\spacing, 1.35*\spacing);
                        \clip (-0.5*\spacing, -0.5*\spacing) rectangle (3.5*\spacing, 1.5*\spacing);
                        
                        \drawgadget{3}{z1}{-\spacing}{0}{black}
                        \drawgadget{3}{z2}{-\spacing}{\spacing}{black}
                        
                        \drawgadget{3}{a1}{0}{0}{gcolA}
                        \drawgadget{3}{b1}{\spacing}{0}{gcolB}
                        \drawgadget{3}{c1}{2*\spacing}{0}{gcolC}
                        \drawgadget{3}{d1}{3*\spacing}{0}{gcolD}
                        \drawgadget{3}{e1}{4*\spacing}{0}{black}
                        \drawgadget{3}{a2}{0}{\spacing}{gcolE}
                        \drawgadget{3}{b2}{\spacing}{\spacing}{gcolF}
                        \drawgadget{3}{c2}{2*\spacing}{\spacing}{gcolG}
                        \drawgadget{3}{d2}{3*\spacing}{\spacing}{gcolH}
                        \drawgadget{3}{e2}{4*\spacing}{\spacing}{black}

                        \drawconnectgadget{3}{3}{a1}{z1}{0}{1}{011E101}{011E000}{}
                        \drawconnectgadget{3}{3}{a2}{z2}{0}{1}{011E101}{011E000}{}
                        \drawconnectgadget{3}{3}{b1}{a1}{0}{1}{011E101}{011E000}{}
                        \drawconnectgadget{3}{3}{c1}{b1}{0}{1}{011E101}{011E000}{}
                        \drawconnectgadget{3}{3}{b2}{a2}{0}{1}{011E101}{011E000}{}
                                \drawconnectgadget{3}{3}{c2}{b2}{0}{1}{011E101}{011E000}{}
                                \drawconnectgadget{3}{3}{d2}{c2}{0}{1}{011E101}{011E000}{}
                        \drawconnectgadget{3}{3}{d1}{c1}{0}{1}{011E101}{011E000}{}
                        \drawconnectgadget{3}{3}{e2}{d2}{0}{1}{011E101}{011E000}{}
                        \drawconnectgadget{3}{3}{e1}{d1}{0}{1}{011E101}{011E000}{}

                        \drawconnectgadget{3}{3}{a1}{a2}{2}{2}{000E000,011E011}{101E101,110E110}{}
                        \drawconnectgadget{3}{3}{b1}{b2}{2}{2}{000E000,011E011}{101E101,110E110}{}
                                \drawconnectgadget{3}{3}{c1}{c2}{2}{2}{000E000,011E011}{101E101,110E110}{}
                                \drawconnectgadget{3}{3}{d1}{d2}{2}{2}{000E000,011E011}{101E101,110E110}{}

                        \end{tikzpicture}
                        \caption {The corresponding (uncompressed) CFI graph.}
                \end{subfigure}
                \begin{subfigure}[t]{0.48\textwidth}
                        \centering
                        \begin{tikzpicture}[scale=0.75]
                                \draw [draw=none, use as bounding box] (-0.7*\spacing, -0.35*\spacing-0.5) rectangle (3.7*\spacing, 1.35*\spacing+0.2);
                                \clip (-0.5*\spacing, -0.5*\spacing) rectangle (3.5*\spacing, 1.5*\spacing);
                                
                                \drawgadget{3}{z1}{-\spacing}{0}{black}
                                \drawgadget{3}{z2}{-\spacing}{\spacing}{black}
                                
                                \drawgadget{3}{a1}{0}{0}{gcolA}
                                \drawgadget{3}{b1}{\spacing}{0}{gcolB}
                                \drawgadget{3}{c1}{2*\spacing}{0}{gcolC}
                                \drawgadget{3}{d1}{3*\spacing}{0}{gcolD}
                                \drawgadget{3}{e1}{4*\spacing}{0}{black}
                                \drawgadget{3}{a2}{0}{\spacing}{gcolE}
                                \drawgadget{3}{b2}{\spacing}{\spacing}{gcolF}
                                \drawgadget{3}{c2}{2*\spacing}{\spacing}{gcolG}
                                \drawgadget{3}{d2}{3*\spacing}{\spacing}{gcolH}
                                \drawgadget{3}{e2}{4*\spacing}{\spacing}{black}

                                \drawconnectgadget{3}{3}{a1}{z1}{0}{1}{011E101}{011E000}{}
                                \drawconnectgadget{3}{3}{a2}{z2}{0}{1}{011E101}{011E000}{}
                                \drawconnectgadget{3}{3}{b1}{a1}{0}{1}{011E101}{011E000}{}
                                \drawconnectgadget{3}{3}{c1}{b1}{0}{1}{011E101}{011E000}{}
                                \drawconnectgadget{3}{3}{b2}{a2}{0}{1}{011E101}{011E000}{}
                                        \drawconnectgadget{3}{3}{c2}{b2}{0}{1}{011E101}{011E000}{}
                                        \drawconnectgadget{3}{3}{d2}{c2}{0}{1}{011E101}{011E000}{}
                                \drawconnectgadget{3}{3}{d1}{c1}{0}{1}{011E101}{011E000}{}
                                \drawconnectgadget{3}{3}{e2}{d2}{0}{1}{011E101}{011E000}{}
                                \drawconnectgadget{3}{3}{e1}{d1}{0}{1}{011E101}{011E000}{}

                                \drawconnectgadget{3}{3}{a1}{a2}{2}{2}{000E000,011E011}{101E101,110E110}{}
                                \drawconnectgadget{3}{3}{b1}{b2}{2}{2}{000E000,011E011}{101E101,110E110}{}
                                        \drawconnectgadget{3}{3}{c1}{c2}{2}{2}{000E000,011E011}{101E101,110E110}{}
                                        \drawconnectgadget{3}{3}{d1}{d2}{2}{2}{000E000,011E011}{101E101,110E110}{}

                                        \path[draw, ultra thick, gcolD]
                                        (a1000) to (c2000)
                                        (a1011) to (c2011)
                                        (a1110) to (c2110)
                                        (a1101) to (c2101);
                                \end{tikzpicture}
                        \caption {The precompressed CFI graph.}
                \end{subfigure}
                \begin{subfigure}[t]{0.48\textwidth}
                        \centering
                        \begin{tikzpicture}[scale=0.75]
                                \draw [draw=none, use as bounding box] (-0.7*\spacing, -0.35*\spacing-0.5) rectangle (3.7*\spacing, 1.35*\spacing+0.2);
                                \clip (-0.5*\spacing, -0.5*\spacing-0.7) rectangle (3.5*\spacing, 1.5*\spacing);
                                
                                \drawgadget{3}{z1}{-\spacing}{0}{black}
                                \drawgadget{3}{z2}{-\spacing}{\spacing}{black}
                                
                                \drawgadget{3}{a1}{0}{0}{gcolA}
                                \drawgadget{3}{b1}{\spacing}{0}{gcolB}
                                \drawgadget{3}{c1}{2*\spacing}{0}{gcolC}
                                \drawgadget{3}{d1}{3*\spacing}{0}{gcolD}
                                \drawgadget{3}{e1}{4*\spacing}{0}{black}
                                \drawgadget{3}{a2}{0}{\spacing}{gcolE}
                                \drawgadget{3}{b2}{\spacing}{\spacing}{gcolF}
                                \drawgadget{3}{d2}{3*\spacing}{\spacing}{gcolH}
                                \drawgadget{3}{e2}{4*\spacing}{\spacing}{black}

                                \drawconnectgadget{3}{3}{b1}{a1}{0}{1}{011E101}{011E000}{}
                                \drawconnectgadget{3}{3}{c1}{b1}{0}{1}{011E101}{011E000}{}
                                \drawconnectgadget{3}{3}{b2}{a2}{0}{1}{011E101}{011E000}{}
                                        \drawconnectgadget{3}{3}{a1}{b2}{0}{1}{011E101}{011E000}{}
                                        \drawconnectgadget{3}{3}{d2}{a1}{0}{1}{011E101}{011E000}{}
                                \drawconnectgadget{3}{3}{d1}{c1}{0}{1}{011E101}{011E000}{}
                                \drawconnectgadget{3}{3}{e2}{d2}{0}{1}{011E101}{011E000}{}
                                \drawconnectgadget{3}{3}{e1}{d1}{0}{1}{011E101}{011E000}{}

                                \drawconnectgadget{3}{3}{a1}{a2}{2}{2}{000E000,011E011}{101E101,110E110}{}
                                \drawconnectgadget{3}{3}{b1}{b2}{2}{2}{000E000,011E011}{101E101,110E110}{}
                                \begin{scope}[every edge/.style={draw, bend left = 55}]%
                                        \drawconnectgadget{3}{3}{c1}{a1}{2}{2}{}{}{}
                                \end{scope}
                                \drawconnectgadget{3}{3}{d1}{d2}{2}{2}{000E000,011E011}{101E101,110E110}{}

                                \drawconnectgadget{3}{3}{a1}{z1}{1}{0}{011E101}{011E000}{}
                                \drawconnectgadget{3}{4}{a2}{z2}{1}{0}{011E101}{011E000}{}
                        \end{tikzpicture}
                        \caption{The compressed CFI graph.}
                \end{subfigure}
                
                \caption{Compressed CFI graphs for a grid of    height $2$ as base graph,
                        a very simple compression,
                        which only identifies two base vertices,
                        and the function that assigns $0$
                        to all edges.
                        The compression on the base graph and the induced
                        one on the precompressed CFI graph is drawn in magenta.}
                \label{fig:compressed-cfi}
\end{figure}
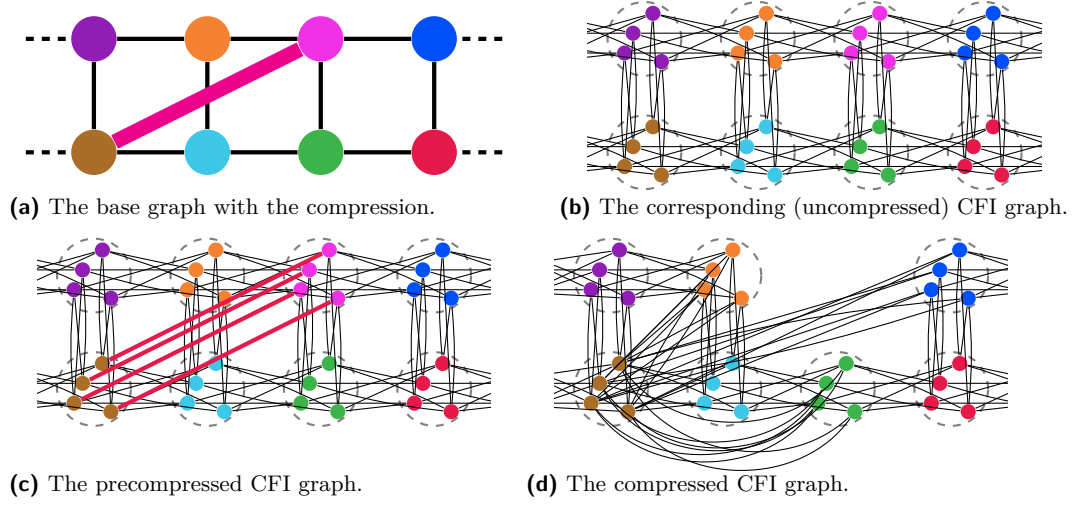

\begin{definition}[Compressible]
        A function $f\colon \edgeG \to \FF_2$ is \defining{$\comp$-compressible}
        if $f(\set{u,v}) = f(\set{u',v'})$ for all vertices $u,v,u',v' \in \vertexG$
        such that  $\set{u,v},
        \set{u',v'} \in \edgeG$,
        $u \comp u'$, and
        $v \comp v'$.
\end{definition}

\begin{definition}[Compressed CFI]
        For a $\graphG$-compression~$\comp$ and a $\comp$\nobreakdash-compressible function ${f \colon \edgeG \to \FF_2}$,
        the graph $\precompCFI{\graphG,f}{\comp}$ expanding the colored graph $\CFI{\graphG,f}$ with 
       $\comp$ is a \defining{precompressed CFI graph}, and
                the colored graph $\compCFI{\graphG,f}{\comp}$ is a \defining{compressed CFI graph}.
\end{definition}

\noindent Precompressed CFI graphs can also be seen as edge-colored graphs that use two colors for the edges---one the the regular edges and one for the equivalence relation. An example is shown in Figure~\ref{fig:compressed-cfi}.
The round number of the bijective $k$-pebble game (a variant of the $k$-pebble game that characterizes equivalence of $k$-variable first order logic with counting quantifiers) on precompressed and compressed CFI graphs are almost equal~\cite{DBLP:conf/focs/GroheLNS23}.
The corresponding statement for the $k$-pebble game with blocking is proved similarly. \ifAppendix{See Appendix~\ref{a: compress} for a proof.}
\begin{lemma}
        \label{lem:precomp-comp-same-number-of-round}
        Let $k\geq 3$, $r\in\nat$, $\comp$ be a $\graphG$-compression, and
         $f,g \, \colon \edgeG \to \FF_2$ be $\comp$\nobreakdash-compressible.
        \begin{enumerate}
                \item $\CFI{\graphG,f} \notkbequivr{k}{r} \CFI{\graphG,g}$
                implies $\precompCFI{\graphG,f}{\comp} \notkbequivr{k}{r} \precompCFI{\graphG,g}{\comp}$.
                \item \label{claim:precomp_implies_comp} $\precompCFI{\graphG,f}{\comp} \notkbequivr{k}{r} \precompCFI{\graphG,g}{\comp}$
                implies $\compCFI{\graphG,f}{\comp} \notkbequivr{k}{r} \compCFI{\graphG,g}{\comp}$.
                \item $\compCFI{\graphG,f}{\comp} \notkbequivr{k}{r} \compCFI{\graphG,g}{\comp}$ implies
                $\precompCFI{\graphG,f}{\comp} \notkbequivr{k}{r+2} \precompCFI{\graphG,g}{\comp}$.
        \end{enumerate}
\end{lemma}

\subparagraph*{The Compressed Cops and Robber Game.}

The ability of the bijective $k$-pebble game
to distinguish non-isomorphic CFI graphs
is captured by the $k$-Cops and Robber game~\cite{SeymourThomas93, DBLP:conf/focs/GroheLNS23}.
A variant of this game---the compressed $k$-Cops and Robber game---provides lower bounds for compressed CFI graphs.
To see this, we consider isomorphisms of CFI graphs. These always twist an even number of edges and can be described 
in terms of paths in the base graphs by twistings (defined below). Moreover, if these paths are compatible with the compression, they induce isomorphisms of  compressed CFI graphs.
For ordered base graphs, these twistings correspond one-to-one with isomorphisms of the (compressed) CFI graphs.

\begin{definition}[Twisting]
        A set $T \subseteq \setcond{(u,v)}{\set{u,v} \in E}$
        is called a \defining{$\graphG$-twisting}
        if, for every $u \in V$, the set $T \cap (\set{u} \times V)$
        is of even size.
        The twisting $T$
        \begin{itemize}
                \item \defining{twists} an edge $\set{u,v} \in E$
                if the set $T$
                contains exactly one of $(u,v)$ and $(v,u)$ and
                \item \defining{fixes} a vertex $u \in V$ if $T \cap (\set{u} \times V) = \emptyset$.
        \end{itemize}
\end{definition}

\noindent To obtain a reasonable notion of twistings for isomorphisms of  compressed CFI graphs,
the twistings have to be compatible with the compression.
For more details on (compressed) CFI graphs, their isomorphisms, and twistings,
we refer to the original paper~\cite{DBLP:conf/focs/GroheLNS23}.

\begin{definition}[Compressible Twisting]
        For a $\graphG$-compression~$\comp$,
        a $\graphG$-twisting~$T$ is called \linebreak \defining{$\comp$\nobreakdash-compressible} if the following holds for all $u,u' \in V$ with $u \comp u'$:
        Let~$u$ and~$u'$ be of degree~$d$.
        Then for every $i\in[d]$, we have
        $(u,v_i) \in T$ if and only if $(u',v'_i) \in T$,
        where~$v_i$ is the $i$-th neighbor of~$u$ and~$v'_i$ is the $i$-th neighbor of~$u'$ (according to the order on~$\graphG$).
\end{definition}

\noindent
The \defining{compressed $k$-Cops and Robber game}~\cite{DBLP:conf/focs/GroheLNS23} is played on a base graph~$\graphG$ and a $\graphG$\nobreakdash-com\-pression~$\comp$.
The Cops Player places cops on up to~$k$ $\comp$\nobreakdash-equivalence classes and the robber is placed on one edge of~$\graphG$.
Initially, only the robber is placed. The game proceeds in rounds:
\begin{enumerate}
        \item \label{itm:cop-picked-up-compressed} The Cops Player picks up a cop  and announces a new $\comp$-equivalence class~$C$ for this cop.
        \item The robber moves. To move from the current edge~$e_1$ to another edge~$e_2$, the robber has to provide a $\comp$\nobreakdash-compressible $\graphG$\nobreakdash-twisting
        that only twists the edges~$e_1$ and~$e_2$ and that fixes every vertex contained in a cop-occupied $\comp$\nobreakdash-equivalence class.
        \item  The cop that was picked up in Step~\ref{itm:cop-picked-up-compressed} is placed on $C$. The next round starts.
\end{enumerate}
The robber is \defining{caught} if the two endpoints of the robber-occupied edge are contained in cop-occupied $\comp$\nobreakdash-classes.
The cops have a winning strategy in~$r$~rounds,
if they can catch the robber in $r$~rounds independently of the moves of the robber.
Similarly, the robber has a \defining{strategy for the first  $r$~rounds}
if the robber can avoid being caught for $r$~rounds independently of the moves of the Cops Player.
The winner of the compressed game depends on the initial position of the robber.
This game yields lower bounds for the (bijective) $k$-pebble game:
\begin{lemma}[\cite{DBLP:conf/focs/GroheLNS23}]\label{lem:lowerbound-compressed-cops-and-robbers}
        Let $\comp$ be a $\graphG$-compression and 
        suppose $f,g \, \colon \edgeG \to \FF_2$ only twist a single edge $e$.
        If the robber, initially placed on the edge $e$, has a strategy for the first $r$ rounds in the compressed $k$-Cops and Robber game
        on $\graphG$ and $\comp$, then
        $\precompCFI{\graphG,f}{\comp} \kequivr{k}{r} \precompCFI{\graphG,g}{\comp}$.
\end{lemma}

\subparagraph*{Introducing Roadblocks for Cops.}
To obtain lower bounds for the $k$-pebble game with blocking,
we add `roadblocks' to the compressed Cops and Robber game
and prove a blocking  analogue of Lemma~\ref{lem:lowerbound-compressed-cops-and-robbers}.
Let $\graphG$ be an ordered graph.
A \defining{roadblock for a vertex} $u \in \vertexG$
is a nonempty set $N \subseteq \setcond{(u,v)}{\set{u,v} \in \edgeG}$ of (directed) edges incident to~$u$ of even size.
A $\graphG$-twisting~$T$ \defining{avoids a roadblock}~$N$ for a vertex~$u$
if $T\cap \set{u}\times \vertexG \neq N$.
In particular,~$T$ may contain a strict superset or subset of~$N$.
If~$T$ does not avoid~$N$, then~$T$ \defining{uses}~$N$. 
A \defining{roadblock for a $\comp$-equivalence class}~$C$ is a nonempty 
set~$N \subseteq [d]$ of even size, where~$d$ is the unique degree of the vertices in~$C$.
A $\graphG$-twisting~$T$ avoids the roadblock~$N$ on~$C$
if, for every vertex $u \in C$, 
the twisting~$T$ avoids the roadblock
$N_u := \setcond{(u,v_i)}{i \in N}$ for~$u$, where~$v_i$ denotes the $i$-th neighbor of~$u$.
If~$T$ is $\comp$-compressible and does not avoid~$N$,
then~$T$ uses~$N_u$ for every vertex $u \in C$;
we say that~$T$ uses~$N$.
Let $M \subseteq [d]$ be the set of all $i \in [d]$ such that $T$ contains the edge to the $i$-th neighbor of some and, since~$T$ is $\comp$-compressible, of every~$u \in C$.
We write $T(N)$ for the symmetric difference of~$N$ and~$M$.

The \defining{compressed and blocking $k$-Cops and Robber game}
is played on a base graph~$\graphG$ and a $\graphG$-compression~$\comp$.
The Cops Player controls cops and roadblocks.
The total number of cops and roadblocks is $k$ but the number of each may vary during the game.
Cops and roadblocks are placed on $\comp$-equivalence classes
and the robber is located on an edge.
Initially, only the robber is placed.
A round of the game proceeds as follows:
The Cops Player picks up a cop or a roadblock and can choose to play a cop move or a blocking move.

\begin{enumerate}
        \item A \defining{cop move} proceeds similarly to the non-blocking game.
        First, the Cops Player announces a $\comp$-equivalence class $C$.
        Next, the robber moves.
        To move from an edge~$e_1$ to another edge~$e_2$, the robber provides a $\comp$\nobreakdash-compressible $\graphG$\nobreakdash-twisting $T$ that
        only twists the edges~$e_1$ and~$e_2$,
        fixes every vertex contained in a cop-occupied $\comp$\nobreakdash-equivalence class, and
        avoids every roadblock.
        Afterwards, a cop is placed on the announced class $C$.
        \item For a \defining{blocking move}, 
        the Cops Player announces a $\comp$-equivalence class $C$ and a roadblock~$N$ for~$C$. 
        Next, the robber moves with a $\comp$\nobreakdash-compressible $\graphG$\nobreakdash-twisting~$T$ as in the cop move.
        If~$T$ uses~$N$, then a cop is placed on~$C$.
        Otherwise, the roadblock~$N$ is placed on~$C$.
        \item 
        The existing roadblocks are updated.
        If a roadblock~$N'$ is placed on a class~$C'$,
        then it is replaced by the roadblock $T(N')$ on~$C'$.
        Because in both a cop and a blocking move~$T$ avoids all roadblocks, $T(N')$ will always be a nonempty set.
        If a roadblock was placed in this move,
        the Cops Player can again choose to play either a cop or a blocking move 
        without increasing the round counter. 
\end{enumerate}
The notion of the robber being caught or having a strategy for the first $r$ round is the same as in the non-blocking game.
As in the non-blocking game, the starting edge of the robber is important.
The following lemma is proved similarly to Lemma~\ref{lem:lowerbound-compressed-cops-and-robbers}.
\ifAppendix{A full proof is given in Appendix~\ref{a: compress}.}

\begin{lemma}
        \label{lem:robber-to-duplictor-blocking}
        Suppose $\comp$ is a $\graphG$-compression and 
        $f,g \, \colon \edgeG \to \FF_2$ only twist a single edge $e$.
        If the robber, initially placed on the edge $e$, has a strategy for the first $r$ rounds in the compressed and blocking $k$-Cops and Robber game
        on $\graphG$ and $\comp$, then
        $\precompCFI{\graphG,f}{\comp} \kbequivr{k}{r} \precompCFI{\graphG,g}{\comp}$.
\end{lemma}
The $\comp$\nobreakdash-compressible twistings of the robber
induce isomorphisms of the compressed CFI graphs, which respect all currently placed pebbles.
These are used to move the twisted edge (`the robber') away from the pebbles.
Cops correspond to $\regular$ pebble pairs and roadblocks to $\blocking$ ones. The case distinction in Point~2 whether a cop roadblock is placed
ensures that in a blocking move in the pebble game with blocking the pebble pair gets marked as $\regular$ or $\blocking$ consistently with the current isomorphism.
Updating the roadblocks in Point~3 corresponds to applying the isomorphism induced by the twisting~$T$ to them.

\section{The Super-Linear Lower Bound with Roadblocks}
\label{sec:super-linear-lower-bound}
We now present and analyze the robust compressed CFI construction of~\cite{RFJN024}.
This work shows that the robber can survive for a large number of
rounds in the compressed Cops and Robber game for certain
compressions.
This section shows that the robber also has a strategy for a large number of rounds
in the game with roadblocks. By Lemma~\ref{lem:robber-to-duplictor-blocking} and Theorem~\ref{cor: block_to_lb}, such a result implies a lower bound on tree-like refutation size for graph isomorphism formulas.

\subsection{Compressing Cylindrical Grids}
\label{sec:compressing-cylindrical-grids}

Fix an integer $k\geq 3$ and a sufficiently large integer $w$.
Set $f(k) := 4k$. 
Let $p_1,\dots,p_k$
be pairwise coprime numbers such that
$\frac{w}{2} \leq p_i \leq w$
for every $i \in [k]$. For all sufficiently large $w$, such numbers exist~\cite{DBLP:conf/focs/GroheLNS23}.
Set $J := f(k)\cdot p_1\cdot\ldots \cdot p_k$.
Let $\grid$ be the $k \times J$ cylindrical grid,
that is, the $k\times J$ grid, in which we also connect the
top and bottom row.
The vertices of $\grid$ are pairs $(i,j)$ for all $i\in[k]$ and $j \in [J]$. They are ordered lexicographically.
We think of the first component as the row index and the second component as the column index.
We refer to the first $f(k)$ columns
as the \defining{left end} of~$\grid$, and
to the last $f(k)$ columns as the \defining{right end} of~$\grid$.
We use addition on row indices in a modulo-like manner.
For example, the $(k+1)$-th row is the first one and $p_{k+1} = p_1$.
For each  $t \in [k-1]$, we define the following equivalence $\comp^t$ via:
\[(i,j) \comp^{t} (i',j') \quad \Longleftrightarrow \quad i=i';\ f(k) < j,j' \leq J-f(k);\  \text{and } j-j'
\text { is divisible by } f(k)p_i\cdot\ldots \cdot p_{i+t}.\] 
These equivalences are $\grid$-compressions~\cite{RFJN024}.
Note that the vertices in the left or the right end of~$\grid$
are in singleton $\comp^t$-equivalence classes. The vertices in between are identified periodically, but the period is different in every row.
It is not hard to show that there are $\Theta(w^{t+1})$ $\comp^t$-equivalence classes. Together with the fact that CFI gadgets for degree $4$ base vertices have $8$ vertices, this implies the next lemma:

\begin{lemma}
        \label{lem:compression-size}
        For all $t \in [k-1]$ and $\comp^t$-compressible $f \colon \edgeC \to \FF_2$,
        the graph $\compCFI{\grid,f}{\comp^t}$ has order $\Theta(w^{t+1})$
        (where $k$ and $t$ are seen as constants) and color class size $8$.
\end{lemma}
\confORfull{}{
        \begin{proof}
                Let $t \in [k-1]$ and $i\in[k]$.
                In the $i$-th row, there are $f(k)p_i \cdot \ldots \cdot p_{i+t} + 2f(k) = \Theta(w^{t+1})$
                equivalence classes.
                Since only vertices in the same row are $\comp^t$-equivalence,
                there are $\Theta(w^{t+1})$ equivalence classes in total.
                Since all vertices in the cylindrical grid $\grid$ have degree at most $4$,
                for every $\comp^t$-equivalence class there are at most $8$ vertices added
                to $\compCFI{\grid,f}{\comp^t}$ (namely one CFI gadget of degree at most $4$).
                Hence, $\compCFI{\grid,f}{\comp^t}$ has order $\Theta(w^{t+1})$.
\end{proof}
}

\noindent While the order of the graphs is $\Theta(w^{t+1})$,
the robber has a strategy for  $\Omega(w^k)$ rounds:

\begin{theorem}[\cite{RFJN024}]
        \label{thm:compressed-lower-bound}
        For every $t \in [k-1]$, consider the compressed $(k+t)$-Cops and Robber game
        played on $\grid$ and $\comp^t$.
        If the robber is initially placed on an edge on the left or right end of~$\grid$,
        then the robber has a strategy for the first $\Omega(J)=\Omega(w^k)$ rounds.
\end{theorem}

\noindent Unfortunately, this theorem does not lift to the game with roadblocks; in order to lift it, we investigate the strategy of the robber in more detail: The robber is always located in either  the left or right end of the grid.
On uncompressed grids of height $k$, the optimal strategy of the Cops Player with at most $2k$ cops is to separate the left from the right end of grid using the cops and to move this separator slowly towards the robber (by at most a constant number of columns in each round).
So the robber can avoid getting caught for a number of rounds linear in the length of the grid: when a newly announced cop is about to form a separator, the robber moves to the end furthest from the separator.
In the compressed game, the strategy is similar.
However, the suitable notion of a separator and the analysis of the situations in which the robber can move from the one end of the grid to the other are more complicated.
We now describe them on an informal level to illustrate the central ideas.
For formal details and more explanations, we refer to \ifAppendix{Appendix~\ref{a: super-linear-lower-bound} and to }the original works~\cite{RFJN024,DBLP:conf/focs/GroheLNS23}.

To move the robber from one end of the grid to the other,
we use $\comp^t$\nobreakdash-compressible $\grid$\nobreakdash-twistings that twist exactly two edges, one in the first and one in the last column.
Such twistings are called \defining{$t$-end-to-end twistings} and are obtained from
\defining{$\ell$-periodic paths}~\cite{DBLP:conf/focs/GroheLNS23}.
Intuitively, these are paths from the first to the last column in the grid~$\grid$, which repeat every~$\ell$ columns. This means that an $\ell$-periodic path is defined by a path in columns~$0$ to $\ell-1$ repeating every~$\ell$ columns.
If~$\ell$ is the greatest common divisor of the compression periods of all the rows used by the path, then the path induces a $t$-end-to-end twisting:

\begin{lemma}[\cite{RFJN024,DBLP:conf/focs/GroheLNS23}]
        \label{lem:periodic-paths-to-twistings}
        Let $t \in [k-1]$, $\pi=(u_1,\dots,u_m)$ be an $\ell$-periodic path,
        and $I \subseteq[k]$ be the set of all rows
        of which $\pi$ contains vertices.
        If $\ell = \gcd \setcond{f(k)p_i\cdot \ldots \cdot p_{i+t}}{i \in I}$,
        then $\pi$ induces the $\comp^t$-compressible $\grid$-twisting
        $\setcond{(u_i, u_{i-1}), (u_i,u_{i+1})}{1<i<m}$.
        
\end{lemma}

\noindent We now turn to a suitable notion of `a separator' for the compressed grid.
Let $W$ be a set of $\comp^t$-equivalence classes.
A \defining{$t$-virtual cordon}~\cite{RFJN024} for $W$
is a separator $S \subseteq \vertexC$ that separates the left from the right end of the grid $\grid$ and satisfies additional conditions on the vertices that $S$ is allowed to contain. For example, if $W$ contains only one class of row $i$, then $S$ may only contain a single vertex from that class.
A set $W$ is \defining{$t$-critical}, if there is a $t$-virtual cordon for $W$
and there is no periodic path satisfying the conditions of Lemma~\ref{lem:periodic-paths-to-twistings} that avoids all vertices of the classes in $W$ (and actually even more).
Intuitively, for $t$-critical sets the robber cannot move between both ends. For non-$t$-critical sets of size at most $k+t-1$ (so in situations where at least one cop is picked up), this is always possible using periodic paths.
\begin{lemma}[\cite{RFJN024}]
        \label{lem:compressed-non-blocking-strategy}
        Let $t \in [k-1]$ and let $W$ be a set of at most $k+t-1$ many $\comp^t$-equivalence classes. 
        If $W$ is not $t$-critical,
        then there is a $t$-end-to-end-twisting avoiding all classes in $W$.
\end{lemma}
The minimal distance of the robber
to an inclusion-wise minimal $t$-virtual cordon for~$W$ measures the distance between the robber and the cops.
When an announced cop will make the position $t$-critical, the robber moves to the end to which this distance is larger.
This distance decreases by at most a constant in each round~\cite{RFJN024}.
So, the robber still has a strategy for a number of rounds linear in the grid length.
Since the compressed CFI graphs are much smaller,
we get a much better bound for the $k$-pebble game on the compressed CFI graphs. 

\subsection{Cops do not Benefit From Roadblocks}
We now show that for the $\grid$-compressions of the previous section,
the Cops Player does not benefit from roadblocks.
This means that, although blocking moves 
 possibly allow the Cops Player to make multiple moves
per round,
the number of rounds the robber can avoid getting caught does not 
change asymptotically compared to the game without roadblocks.
Note that converting a roadblock to a cop only makes it harder for the robber to move. To see this, observe that a roadblock prevents the robber from passing through a vertex (or class)
using a specified set of incident edges, while
 a cops prevents the robber from passing through the vertex (or class) at all.

\begin{lemma}
        \label{lem:no-separator-with-too-many-roadblocks}
        Let $t \in [k-1]$ and $c \leq \frac{2}{5}k-1$ be integers.
        Consider the compressed and blocking $(k + c)$-Cops and Robber
        game on $\grid$ and $\comp^t$
        and assume that cops are placed in at most $c$ rows. 
        Then there is a $t$\nobreakdash-end-to-end twisting
        that avoids all cop-occupied $\comp^t$\nobreakdash-equivalence classes
        and avoids all roadblocks.
        \end{lemma}
\begin{proof}[Proof Sketch]
        We call a roadblock \defining{horizontal} 
        if it blocks the use of exactly the two horizontal incident edges of a vertex or $\comp^t$-class.
        If there are no horizontal roadblocks in a \defining{cop-free} row,
        then the straight path through that row is a 
        $t$-end-to-end twisting and we are done. Assume for a contradiction, that no $t$-end-to-end twisting exists.
        A cop-free row is \defining{lonely},
        if it is sandwiched by cop-occupied rows.
        We show that there have to be additional (non-horizontal) roadblocks in each  non-lonely row.
        To do this we construct, for non-lonely rows $i$ and $i+1$, two $2$-periodic paths that only use vertices from rows $i$ and $i+1$,
        avoid all horizontal roadblocks, and
        do not share the same incident edges of any vertex in rows $i$ and $i+1$ (see Figure~\ref{fig:paths}).
        By Lemma~\ref{lem:periodic-paths-to-twistings}, if there are no non-horizontal roadblocks in rows $i$ and $i+1$, then
        these path would induce $t$-end-to-end twistings. 
        Because the two paths do not use common incident edges,
         an additional roadblock is required for each one to block the path.
         This allows us to lower bound the number of roadblocks in terms of $c$ and to contradict the assumption that
         $c \leq \frac{2}{5}k - 1$.
         Hence, the desired $t$-end-to-end twisting exists. \ifAppendix{See Appendix~\ref{a: super-linear-lower-bound} for the full proof.} \qedhere
        
        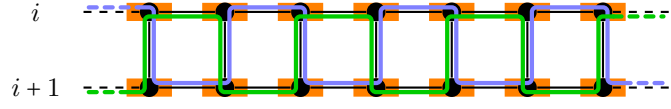
\begin{figure}
                \centering 
                \begin{tikzpicture}
                        \tikzstyle{vertex} = [circle, draw= none, fill=black, inner sep=2.5pt];
                        \def\rows{2}
                        \def\cols{7}
                        \def\colsminusone{6}
                        
                        \node at (-0.5, 1) {$i+1$};
                        \node at (-0.5, 2) {$i$};

                        \foreach \y in {1,...,\rows} {
                                \foreach \x in {1,...,\cols} {
                                        \node[vertex] (v\x\y) at (\x,\y) {};
                                }
                        }
                        \foreach \y in {1,...,\rows} {
                                \foreach \x in {1,...,\colsminusone} {
                                        \pgfmathtruncatemacro{\nextx}{int(\x + 1)}
                                        \draw[thick] (v\x\y) -- (v\nextx\y);
                                }
                        }
                        \def\dotteddist{0.9}
                        \foreach \y in {1,...,\rows} {
                                \draw[thick, dashed] (v1\y) -- ($(v1\y) -(\dotteddist,0)$);
                                \draw[thick, dashed] (v\cols\y) -- ($(v\cols\y) +(\dotteddist,0)$);
                        }
                        \foreach \x in {1,...,\cols} {
                                \draw[thick] (v\x1) -- (v\x2);
                        }
                        
                 \newcommand{\offset}{0.06}
                 \def\dotteddist{0.8}
                 \def\smalldist{0.2}
                 
                 \tikzstyle{periodicparth} = [ultra thick, rounded corners=0.6mm, line cap=round]
                        
                        \draw[blue!50!white, periodicparth]
                                ($(v12.center) -(\dotteddist+\offset,0) + (\offset,\offset)$)
                                edge[dashed]
                                ($(v12.center) -(\smalldist,0) + (\offset,\offset)$);
                        \draw[blue!50!white, periodicparth]
                                ($(v12.center) -(\smalldist,0) + (\offset,\offset)$) --
                                ($(v12.center) + (\offset,\offset)$)
                                \foreach [evaluate=\x as \nextx using int(\x+1), evaluate=\x as \nextnextx using int(\x+2)] \x in {1,3,5} {
                                        -- 
                                        ($(v\x1.center)+ (\offset,\offset)$) --
                                        ($(v\nextx1.center)+ (\offset,\offset)$) -- ($(v\nextx2.center)+ (\offset,\offset)$)--
                                        ($(v\nextnextx2.center)+ (\offset,\offset)$)
                        } -- ($(v\cols1.center)+ (\offset,\offset)$) --
                        ($(v\cols1.center)+  (\smalldist,0) + (\offset,\offset)$) 
                        edge[dashed]
                        ($(v\cols1.center)+  (\dotteddist-\offset,0) + (\offset,\offset)$) ;

                        \draw[green!80!black, periodicparth]
                        ($(v11.center) -(\dotteddist-\offset,0) - (\offset,\offset)$)
                        edge[dashed]
                        ($(v11.center) -(\smalldist,0) - (\offset,\offset)$);
                        \draw[green!80!black, ultra thick, rounded corners=0.6mm,  line cap=round]
                        ($(v11.center) -(\smalldist,0) - (\offset,\offset)$)--
                        ($(v11.center)- (\offset,\offset)$)
                        \foreach [evaluate=\x as \nextx using int(\x+1), evaluate=\x as \nextnextx using int(\x+2)] \x in {1,3,5} {
                                 -- 
                                 ($(v\x2.center)- (\offset,\offset)$) --
                                 ($(v\nextx2.center)- (\offset,\offset)$) -- 
                                 ($(v\nextx1.center)- (\offset,\offset)$) --
                                 ($(v\nextnextx1.center)- (\offset,\offset)$)
                        }
                        -- ($(v\cols2.center)- (\offset,\offset)$) --
                        ($(v\cols2.center)+  (\smalldist,0) - (\offset,\offset)$) 
                        edge[dashed]
                        ($(v\cols2.center)+  (\dotteddist+\offset,0) - (\offset,\offset)$) ;

                        \begin{scope}[on background layer] 
                                \foreach \y in {1,...,\rows} {
                                        \foreach \x in {1,...,\cols} {
                                        \draw[orange, line width=2.4mm]
                                        ($(v\x\y.center)-(0.3,0)$) -- ($(v\x\y.center)+(0.3,0)$);
                                        } 
                                }      
                        \end{scope}

                \end{tikzpicture}
                
                \caption {
                        \label{fig:paths}
                        The two $2$-periodic paths (in blue and in green)
                        constructed in the proof of Lemma~\ref{lem:no-separator-with-too-many-roadblocks}.
                        Edges in and between rows $i$ and $i+1$ in the cylindrical grid are drawn in black.
                        Both paths never use the same incident edges of any vertex.
                        Avoided horizontal roadblocks are drawn in orange.
                }
        \end{figure}
\end{proof}

\noindent Using the previous lemma, we are ready to prove the main technical result of this section.

\begin{lemma}
        \label{lem:compression-lower-bound-blocking}
        Let $1\leq t \leq \frac{2}{5}k -1$ be an integer.
        Then the robber, initially placed on the left or right end of~$\grid$,
        has a strategy for the first $\Omega(J)$
        rounds in the compressed and blocking $(k+t)$-Cops and Robber game
        on $\grid$ and $\comp^t$.
\end{lemma}
\begin{proof}[Proof Sketch] 
        We will convert all roadblocks into cops
        and make the game harder for the robber. 
        In this way, we use the notions of $t$-critical sets
        and $t$-virtual cordons for these positions.
        The robber always stays at one end of the grid:
         If the current position is not critical, the robber stays at the current end, until an announced cop or roadblock (seen as a cop)
                makes the position critical.
                Then using the $t$-end-to-end twisting from Lemma~\ref{lem:compressed-non-blocking-strategy} the robber moves
                to the end of the grid
                with larger distance to every minimal $t$-virtual cordon.
                This distance is in $\Omega(J)$~\cite{RFJN024}.
        If the current position is critical, we show that 
                blocking moves only allow the Cops Player to decrease this distance by $O(k)$, via
a case distinction on the number of cop-occupied rows.
                While this is at most $\frac{2}{5}k-1$,
                the robber can always use the $t$-end-to-end twisting given by Lemma~\ref{lem:no-separator-with-too-many-roadblocks} to switch ends.
                 Otherwise, the number of cop-occupied rows is at least $\frac{2}{5}k$. In this case, all intermediate positions between the blocking moves
                share at least one row in which only one and the same cop is placed, so
                by inductively applying \cite[Proposition 4.10]{RFJN024},
                we  show that all minimal  $t$-virtual cordon before and after the blocking moves are contained within $O(k)$ subsequent columns. 
                
        So, starting from a distance of $\Omega(J)$,
        the robber has a strategy such that this distance decreases by at most $O(k)$ in each round.
        Hence, the robber has a strategy for the first
      $\Omega(J/O(k))=\Omega(J)$ rounds (since $k$ is seen as a constant).
      \ifAppendix{The full proof is in Appendix~\ref{a: super-linear-lower-bound}.}
 \end{proof}

\noindent Finally, for sufficiently large $n$ and choosing $w = \lceil\sqrt[t+1]{n} \rceil$,
Lemmas~\ref{lem:compression-lower-bound-blocking},~\ref{lem:precomp-comp-same-number-of-round},~\ref{lem:compression-size}, and~\ref{lem:robber-to-duplictor-blocking}
together imply the desired round lower bound for 
the $(k+t)$-pebble game with blocking. \ifAppendix{See Appendix~\ref{a: super-linear-lower-bound} for details.}
\begin{theorem}
        \label{thm:compression-blocking-summary}
        For all integers $k\geq 3$, $1\leq t \leq \frac{2}{5}k-1$, and $n\in \nat$,
        there are  
        two colored graphs~$\graphG$ and~$\graphH$ of order $\Theta(n)$ and color class size~$8$
        such that
        \begin{enumerate}
                \item $\graphG \notkequiv{k+1} \graphH$, that is, Spoiler wins the $(k+1)$-pebble game on $\graphG, \graphH$,
                and 
                \item $\graphG \kbequivr{k+t}{\Omega(n^{k/(t+1)})} \graphH $,
                that is, Duplicator has a strategy for the first $\Omega(n^{k/(t+1)})$
                rounds in the $(k+t)$-pebble game with blocking on $\graphG, \graphH$.
        \end{enumerate}
\end{theorem}
\confORfull{}{
\begin{proof}
        Let $k \geq 3$, $1\leq t \leq  \frac{2}{5}k-1$, and $n \in \nat$.
        If $n$ is sufficiently large,
        for $w = \lceil\sqrt[t+1]{n} \rceil$ the required pairwise coprime numbers $p_1,\dots, p_k$ exist~\cite{DBLP:conf/focs/GroheLNS23}.
        Let $J = f(k) p_1\cdot\ldots \cdot p_k$
        and $\mathcal{C}$ be the $k\times J$ cylindrical grid.
        Also, let $\comp^t$ be the compression defined earlier with respect
        to $p_1,\dots, p_k$.
        By Lemma~\ref{lem:compression-lower-bound-blocking},
        the robber has a strategy for the first $\Omega( J) = \Omega(w^k) = \Omega(n^{k/(t+1)})$
        rounds in the compressed and blocking $(k+t)$-Cops and Robber game
        on $\mathcal{C}$ and $\comp^t$, when the robber is placed on the left or right end. 
        
        Now let $e$ be an edge in the left end,
        $f,g \, \colon \edgeC \to \FF_2$ only twist $e$,
        and set $\graphG :=\compCFI{\mathcal{C},f}{\comp^t}$ and $\graphH :=\compCFI{\mathcal{C},f}{\comp^t}$.
        By Lemmas~\ref{lem:precomp-comp-same-number-of-round}
        and~\ref{lem:robber-to-duplictor-blocking},
        Duplicator has a strategy for the first  $\Omega(n^{k/(t+1)})$
        rounds of the $(k+t)$-pebble game with blocking
        on $\graphG$ and $\graphH$.
        It is known that Spoiler wins the $(k+1)$-pebble game
        on CFI graphs over cylindrical grids and thus over the compressed ones
        by Lemma~\ref{lem:precomp-comp-same-number-of-round}.
        By Lemma~\ref{lem:compression-size},
        the graphs $\graphG$ and $\graphH$ have size $\Theta(n)$
        and are of color class size $8$.
\end{proof}}

        \section{Supercritical Width versus Tree-Like Size Trade-Offs}

We finally derive our main results; starting with narrow resolution.

\begin{theorem}\label{thm: lowerbound-treelike-refutation-narrow-width}
	For all integers $k \geq 3$,  $1\leq t \leq \frac{2}{5}k-1$, and $n\in \nat$, there are two colored graphs~$\graphG$ and~$\graphH$
	of order $\Theta(n)$ and color class size $16$
	such that 
	\begin{enumerate}
		\item there is a $k$-narrow resolution refutation of $\ISO$, and
		\item every $(k+t-1)$-narrow tree-like resolution refutation of $\ISO$ has size $2^{\Omega(n^{k/{(t+1)}})}$.
	\end{enumerate}
\end{theorem}	
\begin{proof}
	Let $k \geq 3$ and $1 \leq t \leq  \frac{2}{5}k-1$.
	By Theorem~\ref{thm:compression-blocking-summary}, for all $n \in \nat$, there are graphs~$\graphG$ and~$\graphH$ of color class size $8$ and order $\Theta(n)$
	such that $\graphG \notkequiv{k+1} \graphH$ 
	and $\graphG \kbequivr{k+t}{\Omega(n^{k/(t+2)})} \graphH$.
	It is easy to see that $\graphXG$ and $\graphXH$ have color class size $16$.
	By Theorem~\ref{thm: logic_equiv} and Lemma~\ref{lem: spoiler_win},
	there is a $k$-narrow resolution refutation for $\XISO$.
	Moreover, from Lemma~\ref{lem:duplicator-blocking-to-delayer} it follows that Delayer has a strategy to score $\Omega(n^{k/(t+1)})$ points in the $(k+t)$\nobreakdash-narrow Prover-Delayer game on $\graphXG, \graphXH$. Therefore, by Lemma~\ref{lem: game}, the result follows.
\end{proof}

\noindent \textbf{Theorem~\ref{thm: lowerbound-treelike-refutation-narrow-width-non-robust}}
is the case $t=1$ of Theorem~\ref{thm: lowerbound-treelike-refutation-narrow-width}. We now lift Theorem~\ref{thm: lowerbound-treelike-refutation-narrow-width} to usual resolution (without the narrow resolution rule).
First, if~$\graphG$ and~$\graphH$ have color class size $c$, then we can convert every $k$-narrow refutation of $\ISO$ into a (usual) refutation of $\ISO$ of width $k+c$.
Second, a width-$k$ refutation is in particular a width-$k$ narrow refutation.
\textbf{Theorem~\ref{thm: lowerbound-treelike-refutation-width}} follows immediately.
Note that while Assertion~\ref{itm:resolution-robust} of Theorem~\ref{thm: lowerbound-treelike-refutation-width} provides a lower bound for all $t \leq \frac{2}{5}k -1 $, Assertion~\ref{itm:resolution-refutable} only guarantees a refutation of width $k + 16$.
Therefore, the existing refutation must be large only for $k\geq 45$ and $17 \le t \leq \frac{2}{5}k -1$ .

\subparagraph{Conclusion and Open Questions.}
We established a new super-critical (narrow) width vs.~tree-like size trade-off on graph isomorphism formulas for resolution. The lower bound of $2^{\Omega(n^{k/2})}$ obtained for $t=1$ in Theorems~\ref{thm: lowerbound-treelike-refutation-width} and~\ref{thm: lowerbound-treelike-refutation-narrow-width} is close to the upper bound of $2^{n^k}$ for the tree-like size of resolution of (narrow) width $k$.
We exploited a compressed variant of the CFI graphs and round number lower bounds in the $k$-pebble game on them. However, we had to move from the $k$-pebble game to the $k$-pebble game with blocking, and reprove the round number lower bounds in this setting. This raises the question of whether there is a generic translation from round number lower bounds in the $k$-pebble game to tree-like size resolution lower bounds.
Another question is whether the decrease in the robustness in the trade-off from $2k$ in~\cite{RFJN024} to $\frac{7}{5}k$ in Theorem~\ref{thm: lowerbound-treelike-refutation-narrow-width} is necessary.
More broadly, we ask for a more robust compression or trade-off that can be applied to a much wider range than $2k$.

        \newpage

        \bibliography{../extras/biblio}

\begin{thebibliography}{10}

\bibitem{DBLP:journals/jcss/AtseriasD08}
Albert Atserias and V{\'{\i}}ctor Dalmau.
\newblock A combinatorial characterization of resolution width.
\newblock {\em J. Comput. Syst. Sci.}, 74(3):323--334, 2008.
\newblock \href {https://doi.org/10.1016/J.JCSS.2007.06.025}
  {\path{doi:10.1016/J.JCSS.2007.06.025}}.

\bibitem{oberwolfach2024}
Albert Atserias, Meena Mahajan, Jakob Nordström, and Alexander Razborov.
\newblock Proof complexity and beyond.
\newblock {\em Oberwolfach Report}, 2024.
\newblock Workshop held 24--29 March 2024.
\newblock \href {https://doi.org/10.14760/OWR-2024-15}
  {\path{doi:10.14760/OWR-2024-15}}.

\bibitem{DBLP:journals/siamcomp/AtseriasM13}
Albert Atserias and Elitza~N. Maneva.
\newblock {S}herali-{A}dams relaxations and indistinguishability in counting
  logics.
\newblock {\em {SIAM} J. Comput.}, 42(1):112--137, 2013.
\newblock \href {https://doi.org/10.1137/120867834}
  {\path{doi:10.1137/120867834}}.

\bibitem{Babai2016}
L{\'{a}}szl{\'{o}} Babai.
\newblock Graph isomorphism in quasipolynomial time [extended abstract].
\newblock In {\em 48th Annual {ACM} {SIGACT} Symposium on Theory of Computing,
  {STOC} 2016, Cambridge, MA, USA, June 18-21, 2016}, pages 684--697. {ACM},
  2016.
\newblock \href {https://doi.org/10.1145/2897518.2897542}
  {\path{doi:10.1145/2897518.2897542}}.

\bibitem{BeameBeckImpagliazzo2016}
Paul Beame, Chris Beck, and Russell Impagliazzo.
\newblock Time-space trade-offs in resolution: Superpolynomial lower bounds for
  superlinear space.
\newblock {\em {SIAM} J. Comput.}, 45(4):1612--1645, 2016.
\newblock \href {https://doi.org/10.1137/130914085}
  {\path{doi:10.1137/130914085}}.

\bibitem{BeckNT2013}
Chris Beck, Jakob Nordstr{\"{o}}m, and Bangsheng Tang.
\newblock Some trade-off results for polynomial calculus: extended abstract.
\newblock In {\em Symposium on Theory of Computing Conference, STOC'13, Palo
  Alto, CA, USA, June 1-4, 2013}, pages 813--822. {ACM}, 2013.
\newblock \href {https://doi.org/10.1145/2488608.2488711}
  {\path{doi:10.1145/2488608.2488711}}.

\bibitem{Ben-SassonIW2004}
Eli Ben{-}Sasson, Russell Impagliazzo, and Avi Wigderson.
\newblock Near optimal separation of tree-like and general resolution.
\newblock {\em Comb.}, 24(4):585--603, 2004.
\newblock \href {https://doi.org/10.1007/S00493-004-0036-5}
  {\path{doi:10.1007/S00493-004-0036-5}}.

\bibitem{Ben-SassonNordstrom2008}
Eli Ben{-}Sasson and Jakob Nordstr{\"{o}}m.
\newblock Short proofs may be spacious: An optimal separation of space and
  length in resolution.
\newblock In {\em 49th Annual {IEEE} Symposium on Foundations of Computer
  Science, {FOCS} 2008, October 25-28, 2008, Philadelphia, PA, {USA}}, pages
  709--718. {IEEE} Computer Society, 2008.
\newblock \href {https://doi.org/10.1109/FOCS.2008.42}
  {\path{doi:10.1109/FOCS.2008.42}}.

\bibitem{BenSassonNordstrom2011}
Eli Ben{-}Sasson and Jakob Nordstr{\"{o}}m.
\newblock Understanding space in proof complexity: Separations and trade-offs
  via substitutions.
\newblock In {\em Innovations in Computer Science - {ICS} 2011, Tsinghua
  University, Beijing, China, January 7-9, 2011. Proceedings}, pages 401--416.
  Tsinghua University Press, 2011.
\newblock URL:
  \url{http://conference.iiis.tsinghua.edu.cn/ICS2011/content/papers/3.html}.

\bibitem{DBLP:conf/stoc/Ben-SassonW99}
Eli Ben{-}Sasson and Avi Wigderson.
\newblock Short proofs are narrow - resolution made simple.
\newblock In {\em Proceedings of the Thirty-First Annual {ACM} Symposium on
  Theory of Computing, May 1-4, 1999, Atlanta, Georgia, {USA}}, pages 517--526.
  {ACM}, 1999.
\newblock \href {https://doi.org/10.1145/301250.301392}
  {\path{doi:10.1145/301250.301392}}.

\bibitem{Berkholz2012}
Christoph Berkholz.
\newblock On the complexity of finding narrow proofs.
\newblock In {\em 53rd Annual {IEEE} Symposium on Foundations of Computer
  Science, {FOCS} 2012, New Brunswick, NJ, USA, October 20-23, 2012}, pages
  351--360. {IEEE} Computer Society, 2012.
\newblock \href {https://doi.org/10.1109/FOCS.2012.48}
  {\path{doi:10.1109/FOCS.2012.48}}.

\bibitem{DBLP:conf/icalp/BerkholzG15}
Christoph Berkholz and Martin Grohe.
\newblock Limitations of algebraic approaches to graph isomorphism testing.
\newblock In {\em Automata, Languages, and Programming - 42nd International
  Colloquium, {ICALP} 2015, Kyoto, Japan, July 6-10, 2015, Proceedings, Part
  {I}}, volume 9134 of {\em Lecture Notes in Computer Science}, pages 155--166.
  Springer, 2015.
\newblock \href {https://doi.org/10.1007/978-3-662-47672-7\_13}
  {\path{doi:10.1007/978-3-662-47672-7\_13}}.

\bibitem{DBLP:conf/soda/BerkholzG17}
Christoph Berkholz and Martin Grohe.
\newblock Linear diophantine equations, group csps, and graph isomorphism.
\newblock In {\em Proceedings of the Twenty-Eighth Annual {ACM-SIAM} Symposium
  on Discrete Algorithms, {SODA} 2017, Barcelona, Spain, Hotel Porta Fira,
  January 16-19}, pages 327--339. {SIAM}, 2017.
\newblock \href {https://doi.org/10.1137/1.9781611974782.21}
  {\path{doi:10.1137/1.9781611974782.21}}.

\bibitem{BerkholzLichterVinallSmeeth2024Arxiv}
Christoph Berkholz, Moritz Lichter, and Harry Vinall{-}Smeeth.
\newblock Supercritical size-width tree-like resolution trade-offs for graph
  isomorphism.
\newblock {\em Computing Research Repository}, 2024.
\newblock arXiv preprint.
\newblock \href {https://doi.org/10.48550/ARXIV.2407.17947}
  {\path{doi:10.48550/ARXIV.2407.17947}}.

\bibitem{BerkholzNordstrom16}
Christoph Berkholz and Jakob Nordstr{\"{o}}m.
\newblock Near-optimal lower bounds on quantifier depth and
  {W}eisfeiler-{L}eman refinement steps.
\newblock In {\em 31st Annual {ACM/IEEE} Symposium on Logic in Computer
  Science, {LICS} 2016, New York, NY, USA, July 5-8, 2016}, pages 267--276,
  2016.
\newblock \href {https://doi.org/10.1145/2933575.2934560}
  {\path{doi:10.1145/2933575.2934560}}.

\bibitem{BerkholzNordstrom2020}
Christoph Berkholz and Jakob Nordstr{\"{o}}m.
\newblock Supercritical space-width trade-offs for resolution.
\newblock {\em {SIAM} J. Comput.}, 49(1):98--118, 2020.
\newblock \href {https://doi.org/10.1137/16M1109072}
  {\path{doi:10.1137/16M1109072}}.

\bibitem{CaiFuererImmerman1992}
Jin{-}yi Cai, Martin F{\"{u}}rer, and Neil Immerman.
\newblock An optimal lower bound on the number of variables for graph
  identification.
\newblock {\em Combinatorica}, 12(4):389--410, 1992.
\newblock \href {https://doi.org/10.1007/BF01305232}
  {\path{doi:10.1007/BF01305232}}.

\bibitem{DawarGraedelLichter22}
Anuj Dawar, Erich Grädel, and Moritz Lichter.
\newblock Limitations of the invertible-map equivalences.
\newblock {\em J. Log. Comput.}, 09 2022.
\newblock \href {https://doi.org/10.1093/logcom/exac058}
  {\path{doi:10.1093/logcom/exac058}}.

\bibitem{RFJN024}
Susanna~F. de~Rezende, Noah Fleming, Duri~Andrea Janett, Jakob Nordstr{\"{o}}m,
  and Shuo Pang.
\newblock Truly supercritical trade-offs for resolution, cutting planes,
  monotone circuits, and {W}eisfeiler-{L}eman.
\newblock {\em CoRR}, abs/2411.14267, 2024.
\newblock arXiv preprint.
\newblock \href {https://doi.org/10.48550/ARXIV.2411.14267}
  {\path{doi:10.48550/ARXIV.2411.14267}}.

\bibitem{Furer2001}
Martin F{\"{u}}rer.
\newblock {W}eisfeiler-{L}ehman refinement requires at least a linear number of
  iterations.
\newblock In {\em 28th International Colloquium on Automata, Languages, and
  Programming, {ICALP} 2001, Crete, Greece, July 8-12, 2001, Proceedings},
  volume 2076 of {\em Lecture Notes in Computer Science}, pages 322--333.
  Springer, 2001.
\newblock \href {https://doi.org/10.1007/3-540-48224-5\_27}
  {\path{doi:10.1007/3-540-48224-5\_27}}.

\bibitem{DBLP:conf/sat/GalesiT05}
Nicola Galesi and Neil Thapen.
\newblock Resolution and pebbling games.
\newblock In {\em Theory and Applications of Satisfiability Testing, 8th
  International Conference, {SAT} 2005, St. Andrews, UK, June 19-23, 2005,
  Proceedings}, volume 3569 of {\em Lecture Notes in Computer Science}, pages
  76--90. Springer, 2005.
\newblock \href {https://doi.org/10.1007/11499107\_6}
  {\path{doi:10.1007/11499107\_6}}.

\bibitem{GradelGPP2019}
Erich Gr{\"{a}}del, Martin Grohe, Benedikt Pago, and Wied Pakusa.
\newblock A finite-model-theoretic view on propositional proof complexity.
\newblock {\em Log. Methods Comput. Sci.}, 15(1), 2019.
\newblock \href {https://doi.org/10.23638/LMCS-15(1:4)2019}
  {\path{doi:10.23638/LMCS-15(1:4)2019}}.

\bibitem{Grohe2012}
Martin Grohe.
\newblock Fixed-point definability and polynomial time on graphs with excluded
  minors.
\newblock {\em J. {ACM}}, 59(5):27:1--27:64, 2012.
\newblock \href {https://doi.org/10.1145/2371656.2371662}
  {\path{doi:10.1145/2371656.2371662}}.

\bibitem{GroheLN23}
Martin Grohe, Moritz Lichter, and Daniel Neuen.
\newblock The iteration number of the {W}eisfeiler-{L}eman algorithm.
\newblock In {\em {LICS}}, pages 1--13, 2023.
\newblock \href {https://doi.org/10.1109/LICS56636.2023.10175741}
  {\path{doi:10.1109/LICS56636.2023.10175741}}.

\bibitem{DBLP:conf/focs/GroheLNS23}
Martin Grohe, Moritz Lichter, Daniel Neuen, and Pascal Schweitzer.
\newblock Compressing {CFI} graphs and lower bounds for the
  {W}eisfeiler-{L}eman refinements.
\newblock In {\em 64th {IEEE} Annual Symposium on Foundations of Computer
  Science, {FOCS} 2023, Santa Cruz, CA, USA, November 6-9, 2023}, pages
  798--809. {IEEE}, 2023.
\newblock \href {https://doi.org/10.1109/FOCS57990.2023.00052}
  {\path{doi:10.1109/FOCS57990.2023.00052}}.

\bibitem{GroheLichterNeuenSchweitzer2023Arxiv}
Martin Grohe, Moritz Lichter, Daniel Neuen, and Pascal Schweitzer.
\newblock Compressing {CFI} graphs and lower bounds for the
  {W}eisfeiler-{L}eman refinements.
\newblock {\em CoRR}, abs/2308.11970, 2023.
\newblock arXiv preprint.
\newblock \href {https://doi.org/10.48550/ARXIV.2308.11970}
  {\path{doi:10.48550/ARXIV.2308.11970}}.

\bibitem{DBLP:conf/bcc/GroheN21}
Martin Grohe and Daniel Neuen.
\newblock Recent advances on the graph isomorphism problem.
\newblock In Konrad~K. Dabrowski, Maximilien Gadouleau, Nicholas Georgiou,
  Matthew Johnson, George~B. Mertzios, and Dani{\"{e}}l Paulusma, editors, {\em
  Surveys in Combinatorics, 2021: Invited lectures from the 28th British
  Combinatorial Conference, Durham, UK, July 5-9, 2021}, pages 187--234.
  Cambridge University Press, 2021.
\newblock \href {https://doi.org/10.1017/9781009036214.006}
  {\path{doi:10.1017/9781009036214.006}}.

\bibitem{GroheNeuen2024}
Martin Grohe and Daniel Neuen.
\newblock Isomorphism for tournaments of small twin width.
\newblock In {\em 51st International Colloquium on Automata, Languages, and
  Programming, {ICALP} 2024, July 8-12, 2024, Tallinn, Estonia}, volume 297 of
  {\em LIPIcs}, pages 78:1--78:20. Schloss Dagstuhl - Leibniz-Zentrum f{\"{u}}r
  Informatik, 2024.
\newblock \href {https://doi.org/10.4230/LIPICS.ICALP.2024.78}
  {\path{doi:10.4230/LIPICS.ICALP.2024.78}}.

\bibitem{DBLP:journals/jsyml/GroheO15}
Martin Grohe and Martin Otto.
\newblock Pebble games and linear equations.
\newblock {\em J. Symb. Log.}, 80(3):797--844, 2015.
\newblock \href {https://doi.org/10.1017/JSL.2015.28}
  {\path{doi:10.1017/JSL.2015.28}}.

\bibitem{DBLP:journals/jcss/Immerman82}
Neil Immerman.
\newblock Upper and lower bounds for first order expressibility.
\newblock {\em J. Comput. Syst. Sci.}, 25(1):76--98, 1982.
\newblock \href {https://doi.org/10.1016/0022-0000(82)90011-3}
  {\path{doi:10.1016/0022-0000(82)90011-3}}.

\bibitem{DBLP:conf/coco/Karp72}
Richard~M. Karp.
\newblock Reducibility among combinatorial problems.
\newblock In Raymond~E. Miller and James~W. Thatcher, editors, {\em Proceedings
  of a symposium on the Complexity of Computer Computations, held March 20-22,
  1972, at the {IBM} Thomas J. Watson Research Center, Yorktown Heights, New
  York, {USA}}, The {IBM} Research Symposia Series, pages 85--103. Plenum
  Press, New York, 1972.
\newblock \href {https://doi.org/10.1007/978-1-4684-2001-2\_9}
  {\path{doi:10.1007/978-1-4684-2001-2\_9}}.

\bibitem{Lichter2023b}
Moritz Lichter.
\newblock Witnessed symmetric choice and interpretations in fixed-point logic
  with counting.
\newblock In {\em 50th International Colloquium on Automata, Languages, and
  Programming (ICALP 2023)}, volume 261 of {\em LIPIcs}, pages 133:1--133:20.
  Schloss Dagstuhl -- Leibniz-Zentrum f{\"u}r Informatik, 2023.
\newblock \href {https://doi.org/10.4230/LIPIcs.ICALP.2023.133}
  {\path{doi:10.4230/LIPIcs.ICALP.2023.133}}.

\bibitem{DBLP:conf/soda/ODonnellWWZ14}
Ryan O'Donnell, John Wright, Chenggang Wu, and Yuan Zhou.
\newblock Hardness of robust graph isomorphism, lasserre gaps, and asymmetry of
  random graphs.
\newblock In {\em Proceedings of the Twenty-Fifth Annual {ACM-SIAM} Symposium
  on Discrete Algorithms, {SODA} 2014, Portland, Oregon, USA, January 5-7,
  2014}, pages 1659--1677. {SIAM}, 2014.
\newblock \href {https://doi.org/10.1137/1.9781611973402.120}
  {\path{doi:10.1137/1.9781611973402.120}}.

\bibitem{Pago2023}
Benedikt Pago.
\newblock Finite model theory and proof complexity revisited: Distinguishing
  graphs in choiceless polynomial time and the extended polynomial calculus.
\newblock In {\em 31st {EACSL} Annual Conference on Computer Science Logic,
  {CSL} 2023, February 13-16, 2023, Warsaw, Poland}, volume 252 of {\em
  LIPIcs}, pages 31:1--31:19. Schloss Dagstuhl - Leibniz-Zentrum f{\"{u}}r
  Informatik, 2023.
\newblock \href {https://doi.org/10.4230/LIPICS.CSL.2023.31}
  {\path{doi:10.4230/LIPICS.CSL.2023.31}}.

\bibitem{PudlakImpagliazzo2000}
Pavel Pudl{\'{a}}k and Russell Impagliazzo.
\newblock A lower bound for {DLL} algorithms for \emph{k}-{SAT} (preliminary
  version).
\newblock In {\em Proceedings of the Eleventh Annual {ACM-SIAM} Symposium on
  Discrete Algorithms, January 9-11, 2000, San Francisco, CA, {USA}}, pages
  128--136. {ACM/SIAM}, 2000.
\newblock URL: \url{http://dl.acm.org/citation.cfm?id=338219.338244}.

\bibitem{DBLP:journals/jacm/Razborov16}
Alexander~A. Razborov.
\newblock A new kind of tradeoffs in propositional proof complexity.
\newblock {\em J. {ACM}}, 63(2):16:1--16:14, 2016.
\newblock \href {https://doi.org/10.1145/2858790} {\path{doi:10.1145/2858790}}.

\bibitem{SchweitzerSeebach21}
Pascal Schweitzer and Constantin Seebach.
\newblock Resolution with symmetry rule applied to linear equations.
\newblock In {\em 38th International Symposium on Theoretical Aspects of
  Computer Science, {STACS} 2021, March 16-19, 2021, Saarbr{\"{u}}cken, Germany
  (Virtual Conference)}, volume 187 of {\em LIPIcs}, pages 58:1--58:16. Schloss
  Dagstuhl - Leibniz-Zentrum f{\"{u}}r Informatik, 2021.
\newblock \href {https://doi.org/10.4230/LIPIcs.STACS.2021.58}
  {\path{doi:10.4230/LIPIcs.STACS.2021.58}}.

\bibitem{SeymourThomas93}
Paul~D. Seymour and Robin Thomas.
\newblock Graph searching and a min-max theorem for tree-width.
\newblock {\em J. Comb. Theory, Ser. {B}}, 58(1):22--33, 1993.
\newblock \href {https://doi.org/10.1006/jctb.1993.1027}
  {\path{doi:10.1006/jctb.1993.1027}}.

\bibitem{Thapen2016}
Neil Thapen.
\newblock A tradeoff between length and width in resolution.
\newblock {\em Theory Comput.}, 12(1):1--14, 2016.
\newblock \href {https://doi.org/10.4086/TOC.2016.V012A005}
  {\path{doi:10.4086/TOC.2016.V012A005}}.

\bibitem{Toran04}
Jacobo Tor{\'{a}}n.
\newblock On the hardness of graph isomorphism.
\newblock {\em {SIAM} J. Comput.}, 33(5):1093--1108, 2004.
\newblock \href {https://doi.org/10.1137/S009753970241096X}
  {\path{doi:10.1137/S009753970241096X}}.

\bibitem{DBLP:conf/sat/Toran13}
Jacobo Tor{\'{a}}n.
\newblock On the resolution complexity of graph non-isomorphism.
\newblock In {\em Theory and Applications of Satisfiability Testing - {SAT}
  2013 - 16th International Conference, Helsinki, Finland, July 8-12, 2013.
  Proceedings}, volume 7962 of {\em Lecture Notes in Computer Science}, pages
  52--66. Springer, 2013.
\newblock \href {https://doi.org/10.1007/978-3-642-39071-5\_6}
  {\path{doi:10.1007/978-3-642-39071-5\_6}}.

\bibitem{DBLP:conf/sat/ToranW23}
Jacobo Tor{\'{a}}n and Florian W{\"{o}}rz.
\newblock Cutting planes width and the complexity of graph isomorphism
  refutations.
\newblock In {\em 26th International Conference on Theory and Applications of
  Satisfiability Testing, {SAT} 2023, July 4-8, 2023, Alghero, Italy}, volume
  271 of {\em LIPIcs}, pages 26:1--26:20. Schloss Dagstuhl - Leibniz-Zentrum
  f{\"{u}}r Informatik, 2023.
\newblock \href {https://doi.org/10.4230/LIPICS.SAT.2023.26}
  {\path{doi:10.4230/LIPICS.SAT.2023.26}}.

\bibitem{DBLP:journals/tocl/ToranW23}
Jacobo Tor{\'{a}}n and Florian W{\"{o}}rz.
\newblock Number of variables for graph differentiation and the resolution of
  graph isomorphism formulas.
\newblock {\em {ACM} Trans. Comput. Log.}, 24(3):23:1--23:25, 2023.
\newblock \href {https://doi.org/10.1145/3580478} {\path{doi:10.1145/3580478}}.

\bibitem{Urquhart2011}
Alasdair Urquhart.
\newblock The depth of resolution proofs.
\newblock {\em Stud Logica}, 99(1-3):349--364, 2011.
\newblock \href {https://doi.org/10.1007/S11225-011-9356-9}
  {\path{doi:10.1007/S11225-011-9356-9}}.

\end{thebibliography}
        
        \newpage
\appendix

\section{Proof of Lemma~\ref{lem: game}} \label{a: prover_delayer}

\begin{lemma}[Lemma~\ref{lem: game} restated]
Let $k\ge 1$, let $\graphG, \graphH$ be colored graphs, and let $\pi$ be a $k$-narrow refutation of $\ISO$. Then Prover has a $(\lceil \log(|\pi|)  \rceil)$-point strategy in the $(k+1)$-narrow Prover-Delayer game on $\graphG,\graphH$. 
\end{lemma}

\begin{proof}
Let $k$, $\graphG, \graphH$ and $\pi$ be as in the statement of the lemma. The proof proceeds by showing that Prover can use $\pi$ to construct a winning strategy in the $(k+1)$-narrow Prover-Delayer game on $\graphG, \graphH$.
To do this, Prover traverses $\pi$ starting from the empty clause, which is its root. If in round $r$ Prover is at some clause of $\pi$, then in the next round Prover moves either to a child of this clause or to a leaf while maintaining a certain invariant. 
In order to state the invariant we require the following notion.
A partial assignment $\sigma$ \defining{contradicts} a clause $C \in \pi$ if
\begin{enumerate}
\item for every negated variable $\neg x_{u,v} \in C$, we have $\sigma(x_{u,v}) = 1$; and 
\item for every positive variable $x_{u,v} \in C$ either
\begin{enumerate}
\item $\sigma(x_{u,v}) = 0$ or
\item there is some $x_{u',v'} \in \dom(\sigma)$ with $\sigma(x_{u', v'}) = 1$ such that $(\neg x_{u,v} \vee \neg x_{u', v'} ) \in \ISO$.
\end{enumerate} 
\end{enumerate}
\noindent We show by induction of the number of rounds,
that Prover has a strategy such that after round~$t$ the following invariant holds:
\begin{description}
\item[(I)]  If Delayer has $p$ points,
then the partial assignment $\sigma_{t}$ contradicts a clause $C \in \pi$
and the subtree of $\pi$ rooted at $C$ has size at most $\lfloor |\pi| / 2^p \rfloor$. Moreover, $C$ is not a color clause of $\ISO$.
\end{description}
 
First, suppose that Prover can maintain (I) and that in round $t$ Delayer scores a point to take the total score to $\lceil \log(|\pi|) \rceil$.
We show that now the game ends.
Observe that by~(I), the assignment~$\sigma_{t}$ contradicts a clause $C$ such that the 
subtree of $\pi$ rooted at $C$ has size one. Therefore, $C \in \ISO$. Moreover, again by~(I), $C$ is either a bijection or an edge clause. Since such clauses contain only negated variables, $\sigma_{t}$ violates $C$.
Therefore, the game ends in round $t$.
It follows that if Prover maintains (I), then Delayer can score at most $\lceil \log(|\pi|) \rceil$ many points.
It remains to show that Prover indeed can maintain the invariant.

Before the first round, Delayer has zero points and $\sigma_0$ trivially violates the empty clause, which is the root of $\pi$. This acts as our base case.
So suppose that for some $t \ge 1$ the invariant holds at the end of round $t-1$. If $\sigma_{t-1}$ violates a leaf, then---as previously argued---the game ends. Otherwise, there is some $C \in \pi$ such that:
\begin{enumerate}
\item $\sigma_{t-1}$ contradicts $C$ but does not contradict any $C'$ lying in the subtree of $\pi$ rooted at $C$, and
\item the subtree of $\pi$ rooted at $C$ has size at most $\lfloor |\pi| / 2^p \rfloor$.
\end{enumerate} 
Prover begins by choosing $\sigma \subseteq \sigma_{t-1}$ to be the assignment including only the variables needed to contradict $C$. This implies $|\sigma| = w(C) \le k$, since $C \not \in \ISO$. We know that $C$ is derived by either the resolution or the narrow resolution rule; we deal with each of these possibilities in turn.
To simplify the case analysis, we assume, without loss of generality,
that no color clause is ever an assumption in an application of the
resolution rule: such steps can be replaced by narrow resolution
steps. We further assume, again without loss of generality, that every
narrow resolution step \emph{not} involving a color clause has exactly
three assumptions: if not we can replace such steps with resolution
steps.

\subparagraph{Case 1: Resolution.} Suppose $C$ is derived from $C'= A \vee x_{u,v}$ and $C''= B \vee \neg x_{u,v}$, where~$x_{u,v}$ is the resolved variable. Suppose further that $x_{u,v} \not \in \dom(\sigma)$.
Then Prover makes a resolution move and chooses $x_{u,v}$. Now we observe that since $\sigma$ contradicts $C$:
\begin{itemize}
\item $\sigma[x_{u,v} \mapsto 0]$ contradicts $C'$ and
\item $\sigma[x_{u,v} \mapsto 1]$ contradicts $C''$.
\end{itemize}
Recall that by assumption neither $C'$ nor $C''$ is a color clause. If Delayer makes a committal move, then $\sigma_t$ either contradicts $C'$ or $C''$. Moreover, the subtrees rooted at both these clauses are smaller than the one rooted at $C$ and so (I) is maintained. Otherwise, Delayer makes a point move. Then Prover can choose the value of $\sigma_t(x_{i,j})$, Therefore, Prover can in effect choose which of $C'$ and $C''$ is contradicted by $\sigma_t$. Prover chooses the one with the smaller subtree in~$\pi$. This is at most half the size of the subtree rooted at $C$ and so (I) is again maintained.

So suppose otherwise that $x_{u,v} \in \dom(\sigma)$. If $\sigma(x_{u,v}) = 0$, then $\sigma$ contradicts $C'$.
If $\sigma(x_{u,v}) = 1$, then $\sigma$ contradicts $C''$. But by the choice of $C$, neither of these situations occur.
Hence, $x_{u,v} \not \in \dom(\sigma)$. Therefore, Prover can maintain (I) if $C$ is derived by resolution.

\subparagraph{Case 2: Narrow Resolution.} If $C$ is derived by narrow resolution, then some clause of $\ISO$ is an assumption in the derivation of $C$. We need to distinguish two subcases.

\subparagraph{Case 2a: A color clause is an assumption.} Let $D$ be this assumption. Prover now queries~$D$. Let $x_{u,v} \in D \setminus \sigma^{-1}(0) $. If $x_{u,v}$ is resolved at this step, then there is some assumption~$C'$ of the form $(A \vee \neg x_{u,v})$. Since~$\sigma$ contradicts~$C$, the assignment $\sigma[x_{u,v} \mapsto 1]$ contradicts $C'$. If $x_{u,v}$ is not resolved at this step, then it occurs positively in $C$. Since $\sigma$ contradicts $C$ and as $x_{i,j} \not \in \sigma^{-1}(0)$ by assumption, there is some variable $x_{u', v'}$ such that $\sigma(x_{u', v'}) = 1$ and $C'' :=(\neg x_{u,v} \vee \neg x_{u', v'}) \in \ISO$. Therefore, $\sigma[x_{u,v} \mapsto 1]$ violates $C''$. 

From this it is easy to see that Prover can maintain the invariant similarly to the resolution case. In detail, suppose that Delayer makes a committal moves and chooses $x \in D \setminus \sigma^{-1}(0)$. Then,
$\sigma_{t} = \sigma[x \mapsto 1]$ either violates a bijection or an edge clause---so the game ends---or contradicts an assumption used to derive $C$. Either way, (I) is maintained. Suppose instead that Prover makes a point move and chooses $\{x,y\}$.
If, for some $z \in \{x,y\}$, the assignment $\sigma[z\mapsto1]$ violates a bijection or an edge clause of $\ISO$, then Prover chooses this $z$ and the game ends.
Otherwise, by the analysis above, $\sigma[x\mapsto1]$ and $\sigma[y\mapsto1]$ contradict distinct children of~$C$. Via their choice of $z$, Prover can decide which of these two children are contradicted and so---similarly to in the resolution case---can maintain (I). 

\subparagraph{Case 2b: No color clause is an assumption.} In this case, there is some assumption $D= (\neg x_{u,v} \vee \neg x_{u',v'})$ which is either a bijection or an edge clause. Moreover, there are three assumptions in total and so the other two must be of the form $(A \vee x_{u,v})$ and $(B \vee x_{u',v'})$. If $x_{u,v}$ occurs positively in $C$, then $\sigma(x_{u,v}) =0$ and so $\sigma$ contradicts $(A \vee x_{u,v})$. But this contradicts our choice of $C$, since $(A \vee x_{u,v})$ is in the subtree of $\pi$ rooted at $C$. Therefore, $x_{u,v} \not \in \sigma^{-1}(0)$. Moreover, if $\sigma(x_{u,v}) = 1$, then $\sigma$ contradicts $(B \vee x_{u',v'}) $ since $D = (\neg x_{u,v} \vee \neg x_{u',v'})$ is either a bijection or an edge clause. Again, by the choice of $C$, this case does not occur. Hence, $x_{u,v} \not \in \dom(\sigma)$. 
Therefore, Prover makes a resolution move and selects $x_{u,v}$. Observe that:
\begin{itemize}
\item  $\sigma [x_{u,v} \mapsto 0]$ contradicts $(A \vee x_{u,v})$ and
\item $\sigma [x_{u,v} \mapsto 1]$ contradicts $(B \vee x_{u',v'})$.
\end{itemize} 
Therefore, by essentially the same argument as when $C$ is derived by resolution, Prover can maintain (I).
\end{proof} 
\section{Proofs from Section~\ref{sec: Clique} \label{a: duplicator_to_delayer}}

\begin{lemma}[Lemma~\ref{lem: spoiler_win} restated]
Let $k \ge 3$ and $\graphG$ and $\graphH$ be colored graphs that do not have connected twins.
If $\graphG \notkequivr{k}{r} \graphH$, then $\graphXG \notkequivr{k}{r+1} \! \graphXH$. 
\end{lemma}

\begin{proof}
By Theorem~\ref{thm: logic_equiv},
it suffices to show that if Spoiler has a winning strategy in the $r$-round $k$-pebble game on $\graphG, \graphH$, then Spoiler as a winning strategy in the $(r+1)$-round $k$-pebble game on $\graphXG, \graphXH$. The following claim is crucial. 

\begin{claim} \label{claim: respect}
Let some $p,q\in[k]$ and $(\alpha, \beta)$ be a position in the $k$-pebble game on $\graphXG, \graphXH$ such that  $\{\alpha(p), \alpha(q) \} = \XG(v)$, for some $v \in \vertexG$, and $\{\beta(p), \beta(q) \} \neq \XH(u)$, for every $u \in \vertexH$. Then Spoiler has a winning strategy in one round from position $(\alpha, \beta)$.
\end{claim}

\begin{claimproof} 
Let $x:=\beta(p)$ and $y:=\beta(q)$. First note that if $x=y$, then the game ends immediately. Otherwise, $u:=\XHI(x) \neq \XHI(y)=:v$. Suppose $\set{u,v} \not \in \edgeH$. Then $\set{x,y} \not \in \edgeXH$ so---since there is an edge between $\alpha(p)$ and $\alpha(q)$---the game again ends immediately.
If $\set{u,v} \in \edgeH$, then by assumption $u$ and $v$ are not twins.
Let $w \in \vertexH$ be a witness for this, i.e., $w$ is either adjacent to $u$ and not $v$ or vice-versa.
In the next round, Spoiler picks the $o$-th pebble pair for some $o \not \in \{q, p\}$. This is possible since $k\ge 3$. Spoiler then places the $o$-th pebble for $\graphXH$ on $w_0 \in \vertexXH$. Since $\alpha(p)$ and $\alpha(q)$ are connected twins, every response of Duplicator is immediately losing.
\end{claimproof}

We associate a position $P = (\alpha, \beta)$  in the  $k$-pebble game on $\graphXG, \graphXH$, with a position $\XI(P):= (\gamma, \delta) $ in the $k$-pebble game on $\graphG, \graphH$ via $\gamma(i) := \XI(\alpha(i))$ and $\delta(i) := \XI(\beta(i))$ for all $i \in [k]$.
We show by induction on the number of rounds $t$ that if Spoiler can win in $t$ rounds from $\XI(P)$, then Spoiler can win from position $P$ in at most $t+1$ rounds. 

For the base case $t=0$, Spoiler immediately wins in position
$\XI(P)$ and we need to prove that Spoiler wins in $\leq 1$ round from
$P$.
First suppose there are distinct $p,q\in[k]$ such that
$\gamma(p) = \gamma(q)$ or $\delta(p) = \delta(q)$
but $\delta(p) \neq \delta(q)$ or $\gamma(p) \neq \gamma(q)$, respectively.
Then Spoiler can win from $P$ in at most one round by Claim~\ref{claim: respect}. Otherwise, there are distinct $p,q \in [k]$ such that $\{\gamma(p), \gamma(q)\} \in \edgeG$ if and only if $\{\delta(p), \delta(q)\} \not\in \edgeH$. But then $\{\alpha(p), \alpha(q)\} \in \edgeXG$ if and only if $\{\beta(p), \beta(q)\} \not\in \edgeXH$.
So $P$ is a winning position for Spoiler.

For the inductive step, suppose that Spoiler has a strategy to win in $t+1$ rounds from $\XI(P)$. Suppose that the next move of Spoiler according to this strategy is to pick up the $p$-th pebble pair and to place the $p$-th pebble for  $\graphG$ on $v$.
In the game on $\graphXG, \graphXH$, we stipulate that Spoiler also picks up the $p$-th pebble pair and place the $p$-th pebble for $\graphG$ on $v_0$. Duplicator responds by placing the $p$-th pebble for $\graphXH$ on some $u_i \in \vertexXH$ for $i \in \{0,1\}$.
Let $P'$ be the resulting position in the game on $\graphXG, \graphXH$.
In the game on $\graphG,\graphH$, we let Duplicator respond by
placing the $p$-th pebble on $u$. Let $Q$ be the resulting position in the game on  $\graphG,\graphH$.
Clearly, $Q = \XI(P')$.
Since we followed the winning strategy for Spoiler in the game on $\graphG,\graphH$,
Spoiler has a winning strategy in $t$ rounds from $Q$.
By the inductive hypothesis, Spoiler has also a winning strategy in $t+1$ rounds from $P'$.
Hence, Spoiler has a winning strategy from $P$ in the game on $\graphXG, \graphXH$ in $t+2$ rounds.
The case where Spoiler places a pebble on $\graphH$ in position $\XI(P)$ is analogous. 
\end{proof} 

\begin{lemma}[Lemma~\ref{lem:duplicator-blocking-to-delayer} restated]
        Let $\graphG$ and $\graphH$ be colored graphs
        and $k\geq 2$ an integer. If Duplicator has a winning strategy for the $r$-round $k$-pebble game with blocking on $\graphG, \graphH$, then Delayer has an $r$-point strategy in the $k$-narrow Prover-Delayer game on $\graphXG, \graphXH$.
\end{lemma}

\begin{proof}
The intuition behind Delayer's strategy for the $k$-narrow Prover-Delayer game played on $\graphXG,\graphXH$ is to 
simulate in the background positions of the $k$-pebble game with blocking on~$\graphG$ and~$\graphH$.
Whenever Duplicator marks a pebble pair placed on vertices $u$ and $v$ as $\regular$ in the game
(and there is not already a pebble pair on $u$ and $v$),
Delayer should score a point in the Prover-Delayer game since it `does not matter' whether we map $u_0$ to $v_0$ or to $v_1$.
But this is only the case if no existing variable assignments fix the image of $u_0$. For instance if $x_{u_1,v_1}$ is already assigned zero, then this forces $u_0$ to be mapped to $v_1$. However, whenever this is the case, we can show that Delayer already scored a point for this earlier assignment.

We formalize this idea.
Let $\sigma$ be a position in the $k$-narrow 
Prover-Delayer game on $\graphXG, \graphXH$.
A \defining{point variable in position $\sigma$} is a variable $x_{u,v} \in \inv{\sigma}(0)$ such that
for all variables $x_{u',v'} \in \dom(\sigma)$ with $\XI(u') = \XI(u)$ and $\XI(v') = \XI(v)$, we have $u=u'$ and $v=v'$.
Intuitively, point variables are those critical zero assignments---as in the previous example---that force Delayer to set a variable to one and for which a point was already scored. 

A position $(\alpha,\beta,c)$ in the $k$-pebble game with blocking on $\graphG$ and $\graphH$ is a \defining{$t$-round witness} for
a pair $(\sigma, V)$ of a position $\sigma$ in the $k$-narrow Prover-Delayer game and a set of point variables $V \subseteq \sigma^{-1}(0)$ in position $\sigma$ if
\begin{itemize}
        \item there is a variable $x_{u,v} \in \sigma^{-1}(1)$ if and only if
        there is a pebble pair $p \in [k]$ with $c(p) = \regular$, $\alpha(p) = \XI(u)$, and $\beta(p) = \XI(v)$,
        \item there is a variable  $x_{u,v} \in \sigma^{-1}(0)\setminus V$ if and only if
        there is a pebble pair $p \in [k]$ such that $\alpha(p) = \XI(u)$, $\beta(p) = \XI(v)$, and $c(p) = \blocking$, and
        \item Duplicator has a winning strategy for the $t$-round game starting in position $(\alpha, \beta,c)$. 
\end{itemize}

\begin{claim} \label{claim: witness_implies_points} For all $t \in \mathbb{N}$, $\ell \leq k$, all positions $\sigma$ in the $k$-narrow the Prover-Delayer
game on $\graphXG, \graphXH$, and all sets $V\subseteq \inv{\sigma}(0)$ of point variables in $\sigma$ of size $\ell$,
 if $(\sigma, V)$ has a $(t+\ell)$-round witness, 
then
Delayer has a $t$-point strategy from $\sigma$.
\end{claim}

\begin{claimproof}
We proceed by induction on $t$.
For the base $t=0$, note that trivially Delayer has a 0-point strategy from any position. So assume that the inductive hypothesis holds for some~$t$. Let $\ell \in [k]$.
We now do a further induction on~$\ell$.
Since the base and the inductive case are quite similar,
we will not split them but argue that the steps where we use the inductive hypothesis for~$\ell$ cannot occur if $\ell = 0$.
Let $\sigma$ be a position in the $k$-narrow Prover-Delayer game and $V\subseteq \inv{\sigma}(0)$ be a set of point variables in~$\sigma$ of size~$\ell$. Suppose that
$(\sigma,V)$ has a $(t+1+\ell)$-round witness $P$.
Prover chooses a subset $\sigma' \subseteq \sigma$ of size at most $k-1$. We pick up the corresponding pebble pairs in $P$, yielding the position $P'$. Clearly $P'$ is a $(t+1+\ell)$-round witness for $(\sigma',V \cap \dom(\sigma'))$.
First suppose that $V \not \subseteq \dom(\sigma')$. In this case
$\ell$ is nonzero and 
by the induction hypothesis for $\ell$ and since $|V \cap \dom(\sigma')|<\ell = |V|$, Delayer has a $(t+1)$-point strategy from $\sigma'$ and therefore from $\sigma$.
Now we can assume that $V \subseteq \dom(\sigma')$ so that $P'$ is a $(t+1+\ell)$-round witness for $(\sigma', V)$. We make a case distinction on the action of Prover.

\subparagraph{Case 1: Resolution move.} Prover chooses a variable $x_{u,v} \not \in \dom(\sigma)$. To determine Delayer's response, we suppose that from position $P'$ in the $k$-pebble game with blocking Spoiler makes a blocking move and
places a pebble pair on $\XI(u)$ in $\graphG$ and $\XI(v)$ in $\graphH$. Since $P'$ is a $(t+1+\ell)$-round witness for $(\sigma', V)$, it follows that Duplicator has a winning strategy in the $(t+1+\ell)$-round $k$-pebble game with blocking starting from position $P'$. We assume that Duplicator responds according to such a strategy. First suppose that Duplicator responds by marking this pebble pair with $\blocking$ yielding position $P''$. We let Delayer make a committal response and set $x_{u,v}$ to be zero yielding position $\sigma''$. It is easy to see, that $P''$ is a $(t+1+\ell)$-round witness for $(\sigma'', V)$.

By induction on the number of blocking rounds in which Duplicator responds by marking the pebble pair placed as $\blocking$,
we can repeat this argument until either Duplicator answers by marking the pebble pair placed as $\regular$, or Spoiler plays a regular move. 
Assume that, when this situation finally occurs, the $k$-pebble game with blocking is in position $Q$,
and the Prover-Delayer game is in position $\tau$. 
By the inductive argument, $Q$ is a $(t+1+\ell)$-round witness for $(\tau,V)$. By assumption, from $Q$ a new position $Q'$ is reached by placing a new pebble pair marked $\regular$ on $\XI(u)$ in $\graphG$ and on $\XI(v)$ in $\graphH$; note that it does not matter whether this happens during a blocking move or a regular move. Moreover, Duplicator has a winning strategy in the $(t+ \ell)$-round $k$ pebble game with blocking starting from position $Q'$. There are two possibilities to consider.
\begin{itemize}
        \item 
If there is a variable $x_{u', v'} \in V$ such that $\XI(u') = \XI(u)$ and $\XI(v') = \XI(v)$, then Delayer makes a committal response and sets $x_{u,v}$ to be one.
This yields position $\tau'$. Set $V' := V \setminus \set{x_{u, v}}$.
Hence, $|V'| = \ell -1$. Now, the position $Q'$ is a $(t+\ell)$-round witness for $(\tau',V')$. Note that in this case $\ell \neq 0$ so we may apply the induction hypothesis for $\ell$. Since $t+\ell = (t+1) + |V'|$, it follows that
Delayer has a $(t+1)$-point strategy from
$\tau'$ and hence from $\sigma$.

\item Otherwise, there is no such variable in $V$. In this  case, Delayer makes a point response and Prover assigns a value to $x_{u,v}$ resulting in position $\tau'$.
If $\tau'(x_{u,v})=1$, then we set $V' := V$.
The position $Q'$ is a $(t+\ell)$-round witness for $(\tau',V')$.
By the inductive hypothesis for $t$, Delayer has a $t$-point strategy from $\tau'$ and so has a $(t+1)$-strategy from $\sigma$.

If instead $\tau'(x_{u,v})=0$, then $x_{u,v}$ is a point variable.
Therefore we set $V' := V \cup \set{x_{u,v}}$. Hence, $|V'| = |V| + 1$ and $Q$ is a $(t+1+\ell)=(t+|V'|)$-round witness for $(\tau',V')$---note that here we consider $Q$ instead of $Q'$ because we add $x_{u,v}$ to the point variables and we hence want to `forget' the pebble pair placed in this round. 
Thus, by the inductive hypothesis for $t$, Delayer has a $t$-point strategy from position $\tau'$. Since Delayer scored a point in this round,
Delayer has a $(t+1)$-point strategy from position $\sigma$.
\end{itemize}

\subparagraph{Case 2: Narrow move.}
Prover chooses a color clause.
Assume there is some $u \in \vertexXG$ such that
this clause is $\bigvee_{v \in W_u} x_{u,v}$ where $W_u = \inv{\coloringH}(\coloringG(u))$.
The case  that $u$ is picked from $\vertexXH$ is symmetric.
To determine Delayer's response, we suppose that Spoiler makes a regular move from position $P'$ and places an unused pebble on $\XI(u)$ on $\graphG$. Again, since $P'$ is a $(t+1+\ell)$-round witness for $(\sigma', V)$, Duplicator has a winning strategy in the $(t+1+\ell)$-round $k$-pebble game with blocking starting from position $P'$. Suppose that Duplicator answer with a vertex $v \in \vertexH$ according to such a strategy and denote the resulting position in the $k$-pebble game with blocking by $P''$. We distinguish three cases:
\begin{itemize}
        \item Assume that $x_{u, v_0} \in (\sigma')^{-1}(0)$ or $x_{u, v_1} \in (\sigma')^{-1}(0)$. Since both cases are analogous,
        we assume $x_{u, v_0} \in (\sigma')^{-1}(0)$.
        We claim that in this case $x_{u, v_0} \in V$.
        To see this, suppose for a contradiction $x_{u,v_0}\notin V$.
        Then in position $P'$, there is a pair of blocking pebbles lying on $\XI(u)$ in $\graphG$ and on $v$ in $\graphH$.
        This implies that Spoiler wins in position $P''$, which is a contradiction
        to the fact that $P$ is a $(t+1+\ell)$-witness for $(\sigma,V)$
        and that we followed the strategy of Duplicator to obtain position $P''$.
        Hence, $x_{u, v_0} \in V$ and so $x_{u, v_0}$ is in particular a point variable in position $\sigma$.
        It follows that $x_{u,v_1} \not\in \dom(\sigma)$. In this case, Delayer makes a committal response and chooses $x_{u,v_1}$, yielding position $\sigma''$.
        We set $V'' :=  V \setminus \set{x_{u, v_0}}$.
        Then $P''$ is a $(t+\ell) = ((t+1)+|V''|)$-round witness for $(\sigma'',V'')$. In this case $\ell \neq 0$ so we may apply the inductive hypothesis for $\ell$. Therefore, Delayer has a $(t+1)$-point strategy from position $\sigma''$ and thus from position $\sigma$.
        \item 
        Assume that $x_{u, v_0} \in (\sigma')^{-1}(1)$ or $x_{u, v_1} \in (\sigma')^{-1}(1)$.
        We consider the case $x_{u, v_0} \in \sigma^{-1}(1)$, the other one is analogous.
        Delayer makes a committal response and chooses $x_{u, v_0}$.
        We see that in this case the Prover-Delayer game did not make any progress at all since $\sigma(x_{u, v_0}) = 1$.
        Hence, we may assume that this case does not occur.
        \item It remains to consider the case $\set{x_{u,v_0}, x_{u,v_1}} \cap \dom(\sigma) = \emptyset$.
        We set $V'' := V$.
        Delayer makes a point response and chooses $\set{x_{u,v_0}, x_{u,v_1}}$.
        Spoiler sets one of these two variables to one yielding position $\sigma''$.
        The position $P''$ is a $(t+\ell)=(t+|V''|)$-round witness for $(\sigma'',V'')$.
        By the inductive hypothesis for~$t$,
        Delayer has a $t$\nobreakdash-point strategy from position~$\sigma''$.
        Since a point was scored in this round,
        Delayer has a $(t+1)$\nobreakdash-point strategy from~$\sigma$. 
    	\claimqedhere
\end{itemize}
\end{claimproof}

Finally we show that the claim implies the lemma. The $k$-narrow Prover-Delayer game starts from the empty position. Moreover, $(\emptyset,\emptyset)$ clearly has an $r$-round witness since Duplicator has a winning strategy in the $r$-round $k$-pebble game with blocking on $\graphG$ and $\graphH$. Hence, the claim implies that Delayer has an $r$-point strategy from the initial position.
\end{proof} 
\section{Proofs from Section~\ref{sec:compressing-cfi}} \label{a: compress}

\begin{lemma}[Lemma~\ref{lem:precomp-comp-same-number-of-round} restated]
        Let $k\geq 3$, $r\in\nat$, let $\comp$ be a $\graphG$-compression, and
        let $f,g \, \colon \edgeG \to \FF_2$ be $\comp$\nobreakdash-compressible.
        \begin{enumerate}
                \item $\CFI{\graphG,f} \notkbequivr{k}{r} \CFI{\graphG,g}$
                implies $\precompCFI{\graphG,f}{\comp} \notkbequivr{k}{r} \precompCFI{\graphG,g}{\comp}$.
                \item  $\precompCFI{\graphG,f}{\comp} \notkbequivr{k}{r} \precompCFI{\graphG,g}{\comp}$
                implies $\compCFI{\graphG,f}{\comp} \notkbequivr{k}{r} \compCFI{\graphG,g}{\comp}$.
                \item $\compCFI{\graphG,f}{\comp} \notkbequivr{k}{r} \compCFI{\graphG,g}{\comp}$ implies
                $\precompCFI{\graphG,f}{\comp} \notkbequivr{k}{r+2} \precompCFI{\graphG,g}{\comp}$.
        \end{enumerate}
\end{lemma}

\begin{proof}
        The following proof is a rather straight-forward adaption of
        the proof of the corresponding statement for the bijective $k$-pebble game
        from~\cite{DBLP:conf/focs/GroheLNS23}.
        \begin{enumerate}
                \item Trivial since $\precompCFI{\graphG,f}{\comp}$ and  $\precompCFI{\graphG,g}{\comp}$ just extend $\CFI{\graphG,f}$ and $\CFI{\graphG,g}$, respectively, by an additional relation.
                \item 
        
         Assume that $\precompCFI{\graphG,f}{\comp} \notkbequivr{k}{r} \precompCFI{\graphG,g}{\comp}$,
        that is, Spoiler has a winning strategy in the $r$\nobreakdash-round $k$\nobreakdash-pebble game with blocking played on $\precompCFI{\graphG,f}{\comp}$ and $\precompCFI{\graphG,g}{\comp}$.
        We show that Spoiler also has a winning strategy in the $r$\nobreakdash-round game
        on $\compCFI{\graphG,f}{\comp}$ and $\compCFI{\graphG,g}{\comp}$.
        
        First note that we can always assume the Duplicator plays color preserving,
        which means that Duplicator always answers for regular pebbles
         with a vertex of the same color.
        Otherwise, Duplicator would lose immediately.

        Consider a position $(\alpha,\beta,c)$ of the game on $\compCFI{\graphG,f}{\comp}$ and $\compCFI{\graphG,g}{\comp}$.
        We say that a position $(\alpha',\beta',c')$ of the game on $\precompCFI{\graphG,f}{\comp}$ and $\precompCFI{\graphG,g}{\comp}$ is an \defining{$s$-round witness} for
        $(\alpha,\beta,c)$ if the following conditions are satisfied:
        \begin{enumerate}
        \makeatletter
        \renewcommand{\p@enumii}{}
        \renewcommand{\theenumii}{(\alph{enumii})}
\renewcommand{\labelenumii}{(\alph{enumii})}
        \setlength{\leftskip}{0.5em}
        \makeatother

\item $\dom(\alpha) = \dom(\beta) =\dom(c) = \dom(\alpha') = \dom(\beta') = \dom(c')$;
                \item \label{cond:same-type} $c(i) = c'(i)$ for all $i\in\dom(c)$;
                \item \label{cond:equiv} $\alpha'(i)/_\comp = \alpha(i)$ and $\beta'(i)/_\comp = \beta(i)$ for all $i\in\dom(\alpha)$; 
                \item \label{cond:color} $\alpha'(i)$ has the same color in $\precompCFI{\graphG,f}{\comp}$ as $\beta'(i)$ has in $\precompCFI{\graphG,g}{\comp}$ for every $i \in  \dom(\alpha)$ such that $c'(i) = \regular$;
                \item \label{cond:win} $(\alpha', \beta', c')$ is a winning position for Spoiler in the $s$-round game on $\precompCFI{\graphG,f}{\comp}$ and $\precompCFI{\graphG,g}{\comp}$.
        \end{enumerate}
        We first show that if $(\alpha', \beta', c')$ is a $0$\nobreakdash-round witness for $(\alpha,\beta,c)$,
        then Spoiler wins the game on $\compCFI{\graphG,f}{\comp}$ and $\compCFI{\graphG,g}{\comp}$.
        Because $(\alpha', \beta', c')$ is a winning position for Spoiler in the $0$\nobreakdash-round game on $\precompCFI{\graphG,f}{\comp}$ and $\precompCFI{\graphG,g}{\comp}$
        by Condition~\ref{cond:win},
        the map $\alpha'(i) \mapsto \beta'(i)$ for all $i$ with $c'(i) =\regular$
        is not a partial isomorphism
        or does not respect the blocking pebbles.
        We show that the mapping
        $\alpha'(i) \mapsto \beta'(i)$ for all $i$ with $c(i) =\regular$
        is not a partial isomorphism
        or does not respect the blocking pebbles.
        
        First note that $\alpha(i)$ has the same color
        as $\beta(i)$
        and $\alpha'(i)$ has the same color as $\beta'(i)$
        for every $i \in \dom(\alpha)$ with $c(i)=c'(i) = \regular$
        because Duplicator plays color-preserving and because of Condition~\ref{cond:color}. 
        
        Now assume that $\alpha'(i) \mapsto \beta'(i)$ does not respect the blocking pebbles, i.e. that there are $i,j\in[k]$ with $c'(i) = \regular$, $c'(j) =\blocking$, and
        $(\alpha'(i),\beta'(i)) = (\alpha'(j), \beta'(j))$.
        Then, by Condition~\ref{cond:equiv},
        we have that $(\alpha(i),\beta(i)) = (\alpha'(i)/_\comp, \beta'(i)/_\comp)
        = (\alpha'(j)/_\comp, \beta'(j)/_\comp) = (\alpha(j),\beta(j))$.
        Because we have $c(i) = \regular$ and $c(j) = \blocking$
        by Condition~\ref{cond:same-type},
        the position $(\alpha,\beta,c)$ also does not respect the blocking pebbles
        and hence Spoiler wins immediately.
        So we may assume that the mapping respects the blocking pebbles.
        
        Next, suppose that $\alpha'(i) \mapsto \beta'(i)$ for all $i$ with $c'(i) =\regular$
        is not a partial isomorphism. If there exists $i,j\in[k]$, $c'(i)=c(j)=\regular$ such that $\alpha'(i) \comp \alpha'(j)$ but $\beta'(i)\not\comp  \beta'(j)$, then
        by Condition~\ref{cond:equiv}, we have $\alpha(i)=\alpha(j)$ but $\beta(i)\neq \beta(j)$.
        Spoiler wins immediately.
        If instead $\alpha'(i) = \alpha'(j)$ but $\beta'(i) \neq \beta'(j)$,
        then---since Duplicator plays color-preserving---$\beta'(i)$ has the same color as $\beta'(j)$ (namely the one of $\alpha'(i)$)
        and $\beta'(i) \not\comp \beta'(j)$
        because distinct vertices of the same color are never $\comp$-\nobreakdash equivalent.
        Thus, $\alpha(i) = \alpha(j)$ but $\beta(i)\neq \beta(j)$ by Condition~\ref{cond:equiv}.
        Spoiler again wins immediately.
        
        Otherwise, the map does not respect the edge relation. So suppose that $\set{\alpha'(i),\alpha'(j)}$ is an edge in $\precompCFI{\graphG,f}{\comp}$
        but $\set{\beta'(i),\beta'(j)}$ is not an edge in $\precompCFI{\graphG,g}{\comp}$. Then $\set{\alpha(i),\alpha(j)}$ is an edge in $\compCFI{\graphG,f}{\comp}$.
        We prove that $\set{\beta(i),\beta(j)}$ is not an edge in $\compCFI{\graphG,g}{\comp}$ and hence that Spoiler wins the game on $\compCFI{\graphG,f}{\comp}$, $\compCFI{\graphG,g}{\comp}$.
        Because $\set{\alpha'(i),\alpha'(j)}$ is an edge,
        the origins of~$\alpha'(i)$ and~$\alpha'(j)$ are adjacent base vertices.
        Hence, the origins of~$\beta'(i)$ and~$\beta'(j)$ are also adjacent base vertices,
        because they have the same colors as~$\alpha'(i)$ and~$\alpha'(j)$, respectively.
       	Condition~\ref{itm:compression-order-agree} of a graph compression
       	implies that the set $\set{u,v}$ is not an edge in $\compCFI{\graphG,g}{\comp}$
       	(see~\cite[Lemma 11]{GroheLichterNeuenSchweitzer2023Arxiv}).
        
        The same arguments apply if $\set{\beta'(i),\beta'(j)}$ is an edge in $\precompCFI{\graphG,g}{\comp}$ but $\set{\alpha'(i),\alpha'(j)}$ is not an edge in $\precompCFI{\graphG,f}{\comp}$; this completes the proof that if $(\alpha', \beta', c')$ is a $0$\nobreakdash-round witness for $(\alpha,\beta,c)$, then Spoiler wins the game on $\compCFI{\graphG,f}{\comp}$ and $\compCFI{\graphG,g}{\comp}$. 
        
        It remains to prove, by induction on~$s\leq r$,
 that Spoiler has a strategy for the $r$\nobreakdash-round game on $\compCFI{\graphG,f}{\comp}$ and $\compCFI{\graphG,g}{\comp}$ such that the position in round~$s$ has an $(r-s)$\nobreakdash-round witness. 

The base case $s=0$ holds because the empty position in the game on $\precompCFI{\graphG,f}{\comp}$ and $\precompCFI{\graphG,g}{\comp}$ is an $r$\nobreakdash-round witness for the empty position in the game on  $\compCFI{\graphG,f}{\comp}$ and $\compCFI{\graphG,g}{\comp}$.

        For the inductive step, let $s<r$ and suppose that the position in round~$s$ has an ${(r-s)}$\nobreakdash-witness.
        We show by induction on $m$,
        that after $m$ blocking moves (but no regular ones),
        the resulting position still has an $(r-s)$\nobreakdash-witness.

        The case $m=0$ is trivial.
        So assume that the position $(\alpha,\beta,c)$ after $m$ moves
        still has an $(r-s)$-witness $(\alpha',\beta',c')$.
        Assume that, on the precompressed graphs $\precompCFI{\graphG,f}{\comp}$ and $\precompCFI{\graphG,g}{\comp}$, 
        Spoiler plays a blocking move.
        If, in the game on the precompressed CFI graphs, Spoiler removes a pair of pebbles, then Spoiler picks up the same pebble pair
        in the game on the compressed CFI graphs.
        We can remove the corresponding elements for the position and the witness relation remains intact.
        Now suppose Spoiler wants to place a pebble pair,
        say the $\ell$-th.
        Let $u \in V_{\precompCFI{\graphG,f}{\comp}}$ and $v \in V_{\precompCFI{\graphG,g}{\comp}}$ be the vertices on which Spoiler places the $\ell$-th pebble pair.
        In the game on the compressed CFI graphs $\compCFI{\graphG,f}{\comp}$ and $\compCFI{\graphG,g}{\comp}$,
        we stipulate that Spoiler plays a blocking move and
        places the $\ell$-th pebble pair on $u/_\comp$ and $v/_\comp$.

        If Duplicator now marks this pair with $\regular$, then we are done since there is nothing to prove.
        So assume Duplicator marks the pair with $\blocking$. Then the resulting position is $(\alpha'[\ell \mapsto u],\beta'[\ell \mapsto v], c'[\ell\mapsto \blocking])$. This is an $(r-s)$-witness
        for the position $(\alpha[\ell \mapsto u/_\comp],\beta[\ell \mapsto v/_\comp], c[\ell\mapsto \blocking])$ after $m+1$ blocking moves
        since Spoiler has a strategy to win on the precompressed CFI graphs. 
        
        Finally, we show that after a regular move
        the resulting position has an $(r-s-1)$-witness.
        Picking up a pebble pair from the precompressed CFI graphs
        is transferred to the compressed CFI graphs as  before.
        Now suppose Spoiler wants to place a pebble on the precompressed CFI graphs,
        say the $\ell$-th.
        Assume that Spoiler places the pebble on $\precompCFI{\graphG,f}{\comp}$
        (the case for $\precompCFI{\graphG,g}{\comp}$ is analogous).
        Let $u \in V_{\precompCFI{\graphG,f}{\comp}}$ be the vertex on which Spoiler places the pebble.
        Then in the game on the compressed CFI graphs $V_{\compCFI{\graphG,f}{\comp}}$ and $V_{\compCFI{\graphG,g}{\comp}}$,
        Spoiler places a pebble on $u/_\comp$.
        Assume that Duplicator answers with $v/_\comp$.
        Because we can assume that Duplicator plays color respecting,
        there is a unique vertex $w \in v/_\comp$ that has the same color as $u$.
        We let Duplicator respond with $w$ in the game on the the precompressed CFI graphs.
        Then the resulting position $(\alpha'[\ell \mapsto u],\beta'[\ell \mapsto w], c'[\ell\mapsto \regular])$,
        is an $(r-s-1)$-witness
        for the new position $(\alpha[\ell \mapsto u/_\comp],\beta[\ell \mapsto v/_\comp], c[\ell\mapsto \regular])$. 
        
        \item   
        Assume that $\compCFI{\graphG,f}{\comp} \notkbequivr{k}{r} \compCFI{\graphG,g}{\comp}$.
        We turn a Spoiler winning strategy in the $r$\nobreakdash-round  $k$\nobreakdash-pebble game with blocking played on $\compCFI{\graphG,f}{\comp}$ and $\compCFI{\graphG,g}{\comp}$
        into a Spoiler winning strategy in the $(r+2)$\nobreakdash-round 
        $k$\nobreakdash-pebble game with blocking played on $\precompCFI{\graphG,f}{\comp}$ and $\precompCFI{\graphG,g}{\comp}$.
        
        A position $(\alpha',\beta',c')$ in the game played on $\compCFI{\graphG,f}{\comp}$ and $\compCFI{\graphG,g}{\comp}$
        is an \defining{$s$-witness} for  a position
        $(\alpha,\beta,c)$ in the game played on $\precompCFI{\graphG,f}{\comp}$ and $\precompCFI{\graphG,g}{\comp}$ if
        \begin{enumerate}
        \makeatletter
        \renewcommand{\p@enumii}{}
        \renewcommand{\theenumii}{(\alph{enumii})}
\renewcommand{\labelenumii}{(\alph{enumii})}
        \setlength{\leftskip}{0.5em}
        \makeatother
                \item $\dom(\alpha) = \dom(\beta) =\dom(c) = \dom(\alpha') = \dom(\beta') = \dom(c')$;
        
                \item $c(i) = c'(i)$ for all $i\in \dom(c)$;
                \item \label{cond:class}$\alpha'(i) = \alpha(i)/_\comp$ and $\beta'(i) = \beta(i)/_\comp$
                for all $i \in \dom(\alpha)$;
                \item Spoiler has a winning strategy in the next $s$  rounds in position $(\alpha,\beta',c')$. 
        \end{enumerate}

        We first show that if $(\alpha,\beta,c)$ has a $0$-witness $(\alpha',\beta',c')$,
        then Spoiler wins the game on $\precompCFI{\graphG,f}{\comp}$ and $\precompCFI{\graphG,g}{\comp}$ in at most $2$ additional rounds.
        Since $(\alpha',\beta',c')$ is a $0$-witness, Spoiler wins in this position.
        First, suppose the position does not respect the blocking pebbles.
        So there are $i,j\in[k]$ with $c'(i) = \regular$, $c'(j) = \blocking$, and
        $(\alpha'(i),\beta'(i)) = (\alpha'(j), \beta'(j))$.
        Then, by Condition~\ref{cond:class},
        we have that $\alpha(i) \comp \alpha(j)$ and  $\beta(i) \comp \beta(j)$.
        Spoiler picks up any pebble pair apart from the $i$-th and $j$-th pair,
        say the $\ell$-th one (such a pair exists because $k \geq 3$).
        Spoiler plays a regular move and picks $\alpha(j)$.
        If Duplicator answers with $\beta(j)$,
        Duplicator loses since the resulting position does not respect the blocking pebble pair $j$.
        So assume otherwise that Duplicator answers with a different vertex $w$.
        The vertex $w$ has to be the same color as $\alpha(j)$.
        However, there is exactly one vertex $w \comp \beta(i)$
        that has the same color as $\alpha(j)$, namely $\beta(j)$.
        Hence, $w \not\comp \beta(i)$ but $\alpha(i) \comp\alpha(j)$.
        This means that the resulting position is not a partial isomorphism
        and Spoiler wins.
        
        So consider the case where the position $(\alpha',\beta')$ in the game on $\compCFI{\graphG,f}{\comp}$ and $\compCFI{\graphG,g}{\comp}$ does not induce a partial isomorphism (without blocking). 
        For all $i \in [k]$ with $c(i)=c'(i) = \regular$,
        it holds that $\alpha(i)$ and $\beta(i)$ have the same color---since Duplicator plays in a color-preserving manner---and thus $\alpha'(i) = \alpha(i)/_\comp$ and~$\beta'(i) = \beta(i)/_\comp$ have the same color, too.
        If, for some regular pebble pairs $i$ and $j$, we have $\alpha(i)/_\comp=\alpha(j)/_\comp$ but $\beta(i)/_\comp\neq \beta(j)/_\comp$,
        then  $\alpha(i) \comp \alpha(j)$ but
        $\beta(i) \not\comp \beta(j)$ and Spoiler wins immediately.
        Suppose $\set{\alpha(i)/_\comp, \alpha(j)/_\comp}$ is an edge in $\compCFI{\graphG,f}{\comp}$
        but $\set{\beta(i)/_\comp, \beta(j)/_\comp}$ is not an edge in $\compCFI{\graphG,g}{\comp}$.
On the one hand, there are vertices $u_i\comp \alpha(i)$ and $u_j\comp \alpha(j)$
        such that $\set{u_i,u_j}$ is an edge in $\precompCFI{\graphG,f}{\comp}$.
        On the other hand, for every $v_i\comp \beta(i)$ and $v_j\comp \beta(j)$,
        the set $\set{v_i,v_j}$ is not an edge in $\precompCFI{\graphG,f}{\comp}$.
        Spoiler picks up a pebble pair different from the $i$-th and $j$-th,
        say the $\ell$-th
        (such a pair exists because $k \geq 3$).
        Spoiler plays a regular move and places a pebble on~$u_i$.
        Then Duplicator answers with some vertex $v_i$.
        If $v_i \not \comp \beta(i)$, then Spoiler wins.
        Otherwise, Spoiler picks up the $i$-th pebble pair.
        Spoiler plays another regular round and
        places one pebble on~$u_j$.
        Duplicator answers with some vertex~$v_j$.
        If $v_j \not \comp \beta(j)$,
        then Spoiler wins again.
        Otherwise, as already argued above, $\set{v_i,v_j}$ is not an edge, but $\set{u_i,u_j}$ is.
        Thus, Spoiler wins after $2$ additional rounds.
        
        The same arguments apply in the case that $\set{\beta(i)/_\comp, \beta(j)/_\comp}$  is an edge in $\compCFI{\graphG,g}{\comp}$
        but  $\set{\alpha(i)/_\comp, \alpha(j)/_\comp}$ is not an edge in  $\compCFI{\graphG,f}{\comp}$.
        
        We now prove by induction on~$s\leq r$ that Spoiler has a winning strategy
        in the ${(r+2)}$\nobreakdash-round game on $\precompCFI{\graphG,f}{\comp}$ and $\precompCFI{\graphG,g}{\comp}$ such that the position reached in round~$s$ 
        has an $(r-s)$-witness.
        Clearly, the initial position has an $r$\nobreakdash-witness because
        Spoiler wins the $r$\nobreakdash-round game $\compCFI{\graphG,f}{\comp}$ and $\compCFI{\graphG,g}{\comp}$. Assume $s < r$ and that, by the induction hypothesis,
        the current position $(\alpha,\beta,c)$
        of the game on $\precompCFI{\graphG,f}{\comp}$ and $\precompCFI{\graphG,g}{\comp}$
        has an $(r-s)$\nobreakdash-witness.
        If Spoiler removes a pair of pebbles, then we can update the position accordingly and the resulting position is still witnessed.
        
        First, consider the case that Spoiler---according to their winning strategy on the compressed CFI graphs---plays a regular move
        in the $(r-s)$-witnessing position $(\alpha', \beta', c')$
        on $\compCFI{\graphG,f}{\comp}$ and $\compCFI{\graphG,g}{\comp}$.
        Spoiler picks up the same pebble pair in the game on
        the precompressed CFI graphs  $\precompCFI{\graphG,f}{\comp}$ and $\precompCFI{\graphG,g}{\comp}$
        and plays a regular move, too.
        Assume Spoiler places the pebble on $u/_\comp \in V_{\compCFI{\graphG,f}{\comp}}$ (the case for $u/_\comp \in V_{\compCFI{\graphG,g}{\comp}}$ is analogous).
        Then, in the game on the precompressed CFI graphs,
        we let Spoiler place the pebble on $u$ (which is not unique)
        and Duplicator answers with some vertex $v \in V_{\precompCFI{\graphG,g}{\comp}}$. 
        Since Spoiler had a winning strategy for $r-s$ rounds from position $(\alpha', \beta', c')$,
        Spoiler has a winning strategy for $r-(s+1)$ rounds from position
        $(\alpha'[\ell \mapsto u/_\comp], \beta'[\ell \mapsto v_\comp], c'[\ell \mapsto \regular])$.
        Hence, on the compressed CFI graphs, the resulting position $(\alpha[\ell \mapsto u], \beta[\ell \mapsto v], c[\ell\mapsto \regular])$
        has an $(r-(s+1))$-witness.
        
        Second, consider the case that Spoiler plays a blocking move according to their strategy.
        We follow the strategy as in the previous case.
        If Spoiler places pebbles on  $u/_\comp \in V_{\compCFI{\graphG,f}{\comp}}$
        and  $v/_\comp \in V_{\compCFI{\graphG,f}{\comp}}$ for the compressed CFI graphs,
        then we let Spoiler place pebbles on  $u \in V_{\precompCFI{\graphG,f}{\comp}}$
        and  $v \in V_{\precompCFI{\graphG,f}{\comp}}$ for the precompressed CFI graphs.
        If Duplicator decides to place regular or blocking pebbles on $u$ and $v$, then we consider the position in which Duplicator places
        regular or blocking, respectively, pebbles on $u/_\comp$ and $v/_\comp$.
        An induction on the number of blocking moves
        shows that the resulting position has an $(r-s-1)$-witness.
\qedhere
        \end{enumerate}
\end{proof}
 
\begin{lemma}[Lemma~\ref{lem:robber-to-duplictor-blocking} restated] 
        Let $\comp$ be a $\graphG$-compression and 
        suppose $f,g \, \colon \edgeG \to \FF_2$ only twist a single edge $e$.
        If the robber, initially placed on the edge $e$, has a strategy for the first $r$ rounds in the compressed and blocking $k$-Cops and Robber game
        on $\graphG$ and $\comp$, then
        $\precompCFI{\graphG,f}{\comp} \kbequivr{k}{r} \precompCFI{\graphG,g}{\comp}$.
\end{lemma}

\begin{proof}
        We show that Duplicator has a winning strategy in the $r$-round $k$-pebble game with blocking. Duplicator maintains a function $g' \, \colon \edgeG \to \FF_2$, an edge $e' \in \edgeG$,
        and an isomorphism $\phi  \, \colon \precompCFI{\graphG,g}{\comp} \to \precompCFI{\graphG,g'}{\comp}$ 
        such that after $s\leq r$ rounds in position $(\alpha, \beta, c)$ the following properties hold.
        \begin{enumerate}
                \item  \label{cond:iso} The isomorphism $\phi$ satisfies
                \begin{itemize} 
                \item $\phi(\beta(i)) = \alpha(i)$ for every $i\in \dom(\alpha)$ such that $c(i) = \regular$, and
                \item   $\phi(\beta(i)) \neq \alpha(i)$ for every $i\in \dom(\alpha)$ such that $c(i) = \blocking$.
                \end{itemize}
                \item \label{cond:one-twist} Only the edge $e'$ is twisted by $f$ and $g'$.
                \item \label{cond:not-caught} At most one endpoint of $e'$ is the origin of a vertex on which a pebble in placed by $\alpha$. (Recall that the origin of a vertex $w = (u, \tup{a})$ in the (precompressed) CFI graph is the vertex $u$ in the base graph $\graphG$).
                \item \label{cond:not-lose} The robber has a winning strategy in the $(r-s)$-round compressed and blocking $k$-Cops and Robber game
                starting in the following position:
                \begin{itemize}
                        \item The robber is placed on $e'$.
                        \item For every $i$ with $c(i) = \regular$, a cop is placed on
                        the $\comp$-equivalence class of the origin of $\alpha(i)$. 
                        \item For every $i$ with $c(i) = \blocking$
                        and for which $\alpha(i)$ and $\phi(\beta(i))$ (and so $\beta(i)$) have the same origin $x$,
                        the following roadblock is placed on $x/_\comp$. Let $\alpha(i) = (x,\tup{a})$ and $\phi(\beta(i)) = (x, \tup{b})$ and $x$ be of degree $d$. We place the roadblock $\setcond{i \in [d]} { a_i \neq b_i}$
                        on $x/_\comp$,
                        which by the definition of CFI graphs is a set of even size.
                \end{itemize}
        \end{enumerate}
        
        \noindent We first argue that maintaining the invariant for $r$ rounds implies the statement of the lemma.
        By Conditions~\ref{cond:one-twist} and~\ref{cond:not-caught}, at the end of round $s$ for all $s \le r$,
        the functions $f$ and $g'$
        twist exactly one edge $e'$
        and at most one endpoint of $e'$ is the origin of a vertex on which a pebble is placed.
        The edges $\precompCFI{\graphG,f}{\comp}$ and $\precompCFI{\graphG,g'}{\comp}$ only differ over the twisted connection. Over all other connections
      there is an edge between two vertices of $\precompCFI{\graphG,f}{\comp}$  if and only if there is one in $\precompCFI{\graphG,g'}{\comp}$.
        Since pebbles are placed on at most endpoint of the twisted connected, 
        by applying Condition~\ref{cond:iso}, we see that the pebbles induce a partial isomorphism with blocking between $\precompCFI{\graphG,f}{\comp}$ and $\precompCFI{\graphG,g}{\comp}$.
        Therefore, Duplicator survives round $s$ for all $s \le r$ and so $\precompCFI{\graphG,f}{\comp} \kbequivr{k}{r} \precompCFI{\graphG,g}{\comp}$.
        
        We now show that Duplicator has a strategy to maintain the invariant.
        Because initially no pebbles are placed, the invariant clearly holds for $g' := g$ and $e' := e$.
        Assume, by the inductive hypothesis, that after $s<r$ many rounds the invariant holds and that in round $s+1$ Spoiler picks up the $\ell$-th pebble pair.
        If at the end of round $s$, the pair was placed on vertices of different colors (in which case $c(\ell) = \blocking$), we do nothing.
        Otherwise, the pair was placed on vertices of the same origin.
        In the Cops and Robber game, the Cops Player picks up the corresponding cop, if $c(\ell)=\regular$, or roadblock, if $c(\ell) =\blocking$, according to the invariant.
        Clearly, the invariant is maintained.
        
        First consider the case where Spoiler makes a regular move
        and places a pebble on $u \in V_{\precompCFI{\graphG,f}{\comp}}$
        (the case where $u \in V_{\precompCFI{\graphG,g}{\comp}}$ is analogous).
        Duplicator determines the destination for the pebble on $\precompCFI{\graphG,g}{\comp}$ as follows.
        In the compressed and blocking $k$-Cops and Robber game, we suppose the Cops Player performs a cop move where they place a cop  on the $\comp$-class of the origin of $u$.
        Let $T$ be the $\comp$-compressible $\graphG$-twisting, with which the robber
        moves from $e'$ to an edge $e''$ according to the robber's strategy.
        This compressible $\graphG$-twisting gives rise to an isomorphism $\psi \colon \precompCFI{\graphG,g'}{\comp} \to \precompCFI{\graphG,g''}{\comp}$~\cite[Lemma~12]{DBLP:conf/focs/GroheLNS23} such that $e'$ and $e''$ are precisely the edges twisted by $g'$ and $g''$.
        Duplicator places the pebble on $\inv{\phi}(\inv{\psi}(u))$.
        
        We show that $g''$, $e''$, and $\psi \circ \phi$ satisfy the invariant.
        Conditions~\ref{cond:iso} and~\ref{cond:one-twist} of the invariant hold because
        of the choice of $T$:
        We still have $\phi(\beta(i)) = \alpha(i)$ for all $i \in \dom(\alpha)$ for which $c(i) = \regular$ because $T$ avoids the cop on the origin $x$ of $\beta(i)$ and thus the induced isomorphism $\psi$ is the identity on the gadget of $x$~\cite[Lemma 13]{DBLP:conf/focs/GroheLNS23}.
        In particular, $\psi(\phi(\beta(i))) = \phi(\beta(i))$, which $\psi(\phi(\beta(i))) = \alpha(i)$ by the inductive hypothesis.
        We also still have $\psi(\phi(\beta(i))) \neq \alpha(i)$ for all $i \in \dom(\alpha)$ for which $c(i) = \blocking$ because $T$ avoids the roadblock $\setcond{(x,y_i)} { a_i \neq b_i}$, where $\alpha(i) = (x,\tup{a})$ and $\phi(\beta(i)) = (x, \tup{b})$ and $y_i$ is the $i$-th neighbor of $x$, on the origin $x$ of $\beta(i)$. This implies that $\psi(\phi(\beta(i))) \neq \alpha(i)$ by the definition of the induced isomorphism $\psi$ (see again~\cite{DBLP:conf/focs/GroheLNS23}).
        Condition~\ref{cond:not-caught} is satisfied because the robber is not caught in the game.
For Condition~\ref{cond:not-lose}, observe that since followed the strategy of the robber, the robber has a  strategy for the remaining $(r-s-1)$ rounds in the game. Moreover, since the roadblocks in the game are updated using the compressible $\graphG$-twisting $T$, they 
        game are exactly the ones required in the game position by Condition~\ref{cond:not-lose} of the invariant.
      Thus Condition~\ref{cond:not-lose} holds.
        We finally update $g' \leftarrow g''$, $e' \leftarrow e''$, and $\phi \leftarrow \psi \circ \phi$.
        
        It remains to consider the case where Spoiler makes a blocking move
        and places a pebble pair on  $u \in V_{\precompCFI{\graphG,f}{\comp}}$
        and $v \in V_{\precompCFI{\graphG,g}{\comp}}$.
       	If $u$ and $v$ are not of the same color (and thus have different origins),
        Duplicator marks the pebble pair $\blocking$ and the invariant clearly holds.
      	Otherwise, let the common origin of $u$ and $v$ be $x$
      	and let $u=(x,\tup{a})$ and $\phi(v)=(x, \tup{b})$.
      	Let \[N:=\setcond{i \in [d]} { a_i \neq b_i}.\]
      	If $N = \emptyset$, then we suppose the Cops Player makes a cop move and
      	announce a cop for $x/_\comp$.
      	The robber moves with a $\comp$-compressible $\graphG$-twisting $T$
      	and we update $e'$, $g'$, and $\phi$ as in the case of a cop move.
      	If $T$ uses $N$, then we mark the pebble pair $\regular$.
      	Otherwise, $T$ avoids $N$ and we mark the pebble pair $\blocking$.
      	We show that the invariant is satisfied:
      	If $T$ uses $N$,
      	then the isomorphism $\psi$, as defined in the regular move,
      	is the identity on the gadget of $x$.
      	Hence, we have $\phi(v) = u$ also for the updated $\phi$
      	and 
      	Condition~\ref{cond:iso} is true.
      	In the Cops and Robber game, a cop is placed on $x/_\comp$ and since we followed the strategy of the Robber, Condition~\ref{cond:not-lose} holds.
      	If $T$ otherwise avoids $N$,
      	$\psi$ is not the identity for all vertices in the gadget for $x$.
      	Hence, because we have $\tup{a} = \tup{b}$,
      	we have $\psi(v) \neq u$ and thus $\phi(\psi(v)) \neq u$
      	and 
      	Conditions~\ref{cond:iso} is true.
      	Also, a roadblock is placed in the cops and robber game,
      	which is exactly the one as in Condition~\ref{cond:not-lose}.
      	
      	If otherwise $N\neq \emptyset$,
      	we suppose the Cops Player makes a blocking move
      	and announces the roadblock $N$
      	for $x/_\comp$.
      	Robber answers with $\comp$-compressible twisting
      	$T$ with which the robber moves
      	from the edge $e'$ to the edge $e''$.
      	We update $e'$, $g'$, and $\phi$ as in the case of a regular move.
      	If $T$ uses $N$,
      	we will have that $u = \phi(v)$ for the updated isomorphism $\phi$.
      	For the same reasons as in the 
      	 regular move case,
      	this ensures that Condition~\ref{cond:iso} is satisfied.
      	Similarly, the remaining invariants hold for the same reasons as the 
      	regular move case.
      	
      	If $T$ avoids $N$,
      	then $u \neq \phi(v)$ for the updated isomorphism $\phi$.
        The invariant holds for the same reason as before:
        Conditions~\ref{cond:iso} and~\ref{cond:one-twist} hold because
        of the choice of $g'$ and $e'$ and the properties of $T$,
        Condition~\ref{cond:not-caught} holds because the robber is not caught in the game, and Condition~\ref{cond:not-lose} holds because
        we followed the strategy of the robber (and again the roadblocks are updated after the robber moved).
        Now it is Spoiler's turn to make the next move,
        for which we proceed as before.
        An induction on the number of blocking moves in round $s+1$
        shows that after the round, the invariant is satisfied.
\end{proof} 
\section{Proofs and Material from Section~\ref{sec:super-linear-lower-bound}}
\label{a: super-linear-lower-bound}

\subparagraph{Discussion of the Generalized Construction.}
The original construction considers the case $t=1$
and uses a cylindrical grid of length $J/2$~\cite{DBLP:conf/focs/GroheLNS23}.
The generalized construction adds $f(k)$ columns to the begin and the end of the
cylindrical grid but avoids the $f(k) < j,j' \leq J-f(k)$ condition~\cite{RFJN024}.
Hence, the definition presented here can be seen as a variant of the one of~\cite{RFJN024}, where we turn some non-singleton equivalence classes into singleton ones and cut off some columns at both ends.
The former change only makes it easier for the robber and  the latter change
is not a restriction since there are still $f(k)$ columns containing only singleton classes.
Moreover, the generalized construction puts the twists at the vertices
and not the edges. This is unimportant since both variants of CFI graphs are isomorphic~\cite{DawarGraedelLichter22}.
For simplicity, we use $f(k)=4k$ instead of $f(k,t) = 2(k+t)$ so that $J$ does not depend on $t$.

\subparagraph{Formal Definitions of Periodic Paths and $t$-Critical Sets.}
	We now provide the formal definitions of periodic paths and $t$-critical sets
	adapted from~\cite{RFJN024,DBLP:conf/focs/GroheLNS23},
	to which we refer for further explanations.
	We continue to work in the setting of Section~\ref{sec:super-linear-lower-bound}.
	We may call $\comp^t$\nobreakdash-equivalence classes just $\comp^t$-classes.
	We will also use sets of $\comp^t$-classes~$W$ ti mean the set of all vertices contained in classes in~$W$, so for the vertex set $\bigcup_{C\in W}C$,
	e.g., a $\grid$-twisting~$T$ avoids a set of $\comp^t$-classes~$W$
	if~$T$ avoids $\bigcup_{C\in W}C$.

	\begin{definition}[Periodic Path]
		A path $(u_1,\dots,u_m)$ in $\grid$ is \defining{$\ell$-periodic} if $u_1$ is in the left and $u_m$ is in right end, and for
		all $i < m$ and $v \in \vertexC$ in the same row as $u_i$ such that 
		the distance of $v$ and $u_i$ is divisible by $\ell$
		and $u_i$ and $v$ are not on the ends of $\grid$,
		there is a $j < m$ such that $u_j = v$, and
		$u_{i+1}$ and $u_{j+1}$ are in the same row
		and their distance is divisible by $\ell$.
	\end{definition}
	
	\noindent Intuitively, an $\ell$-periodic path is composed out of a path of length $\ell$,
	which is copied for all repetitions of $\ell$ columns.
	We continue with the formal definition of $t$-critical sets.

	\begin{definition}[Separating Sets]
		Let $t \in [k-1]$, $I \subseteq [k]$ and $q = \gcd \setcond{f(k)p_i\cdot \ldots \cdot  p_{i+t}}{i \in I}$. A set $W\subseteq \vertexC$ is \defining{$(I,t)$-separating}
		if there is no $q$-periodic path only using vertices of rows $I$ such that the induced twisting avoids 
		$\setcond{(i,j)}{(i,j') \in W, j'-j \text{ is divisible by }q}$.
	\end{definition}

	\begin{definition}[Virtual Cordon]
		A set $S \subseteq \vertexC$ 
		is a \defining{vertical separator}
		if in  $\grid-S$ 
		the left end is not connected to the right end of $\grid$.
		For $t \in [k-1]$, a \defining{$t$-virtual cordon} for a set  $W \subseteq \vertexC$ is a vertical separator $S$ such that
		\begin{enumerate}
			\item for every row $i \in [k]$, 
			the number of vertices in $S$ lying in row $i$
			is at most the number of $\comp^t$-classes,
			of which $W$ contains vertices in row $i$, and
			\item for every row $i \in [k]$ for which $W$ contains only vertices of one $\comp^t$-class $c_i$,
			we have $S\cap (\set{i}\times [J]) \subseteq c_i$,
			that is, $S$ contains only vertices of $c_i$ in row $i$.
		\end{enumerate}
		A \defining{minimal} $t$-virtual cordon is an inclusion-wise minimal $t$-virtual cordon.
		We call \confORfull{}{the set} $W$ \defining{$t$\nobreakdash-critical} if there exists a $t$-virtual cordon for $W$ and $W$ is $(I,t)$-separating for all $I \subseteq [k]$ of size at most $t+1$.
	\end{definition}
	
	\noindent This means, for rows in which~$W$ contains vertices of exactly one $\comp^t$-class~$C$,
	a $t$-virtual cordon~$S$ for~$W$ contains exactly one vertex of the class $C$.

\subparagraph{Cops do not Benefit From Roadblocks.}

We now give a detailed proof of Theorem~\ref{thm:compression-blocking-summary}, which follows by showing a lower bound on the round number of the compressed and blocking Cops and Robbers game played on $\grid$ and $\comp^t$. To do this we first need two auxiliary lemmas. The first says that if there are not too many cops placed, then the robber can always move from one end of the grid to the other. 

\begin{lemma}[Lemma~\ref{lem:no-separator-with-too-many-roadblocks} restated]
	Let $t \in [k-1]$ and $c \leq \frac{2}{5}k-1$ be integers.
	Consider the compressed and blocking $(k + c)$-Cops and Robber
	game on $\grid$ and $\comp^t$
	and assume that cops are placed in at most $c$ rows.
	Then there is a $t$\nobreakdash-end-to-end twisting
	that avoids all cop-occupied $\comp^t$\nobreakdash-equivalence classes
	and avoids all roadblocks.
\end{lemma}
\def\ignorepathfigure{}
\begin{proof}
	
	We can assume that cops are placed in exactly $c$ rows.
	Otherwise, we convert roadblocks into cops
	until cops are placed in exactly $c$ rows.
	As already argued, converting roadblocks to cops only
	makes it harder to find the desired $t$-end-to-end twisting
	(or to move the robber in general).
	If this is not possible, then there is a row not containing cops
	and roadblocks, which means that the path straight through
	this row is the desired $t$-end-to-end twisting.
	
	Let $I\subseteq [k]$ be the set of rows in which no cop is placed.
	We call these rows \defining{cop-free}.
	Clearly, there are $(k - c)$ cop-free rows.
	A cop-free row $i \in I$ is \defining{lonely}
	if $i-1 \notin I$ and $i+1 \notin I$
	(recall that we use addition on rows such that
	the $(k+1)$-th row is the $1$st one).
	A cop-free row is \defining{socialized} if it is not lonely.
	Let $\ell$ be the number of lonely rows.
	Hence, there are $s := k - c - \ell$ socialized rows.

	A \defining{horizontal} roadblock is a roadblock
	that blocks the use of exactly the two horizontal edges of a vertex or $\comp^t$-class, the one to the left and the one to the right.
	A roadblock \defining{belongs} to a row $i \in I$
	if 
	\begin{enumerate}[(a)]
		\item it is a horizontal roadblock contained in row $i$, or
		\item it is not contained in row $i$
		but blocks an edge incident to a vertex in row $i$
		and a horizontal edge in another row
		(and hence a roadblock can belong to at most one row).
	\end{enumerate}
	
	\noindent If there is a cop-free row
	that does not contain a horizontal roadblock,
	then the path straight through this row
	is a desired $t$-end-to-end twisting. 
	Hence, for every cop-free row $i$, 
	there is a horizontal roadblock that belongs to row $i$
	unless the desired $t$-end-to-end twisting exists.
	
	Let $i\in I$ be a socialized row
	and assume that $i+1 \in I$ (the case that $i-1 \in I$
	is analogous by considering $i-1$ instead of $i$).
	We show that,
	unless the desired $t$-end-to-end twisting exists,
	there have to be two non-horizontal roadblocks in addition to the horizontal roadblocks from before,
	one that belongs to row $i$ blocking an edge incident to a vertex in row $i+1$
	and one that belongs to row $i+1$ and blocks an edge incident to a vertex in row $i$.
	Consider the two paths:
		\begin{align*}
		\pi_1&= \big((i,0),(i,1),(i+1,1),(i+1,2),(i,2)\big)\\
		\pi_2&= \big((i+1,0),(i+1,1),(i,1),(i,2),(i+1,2)\big),
	\end{align*}
	Each of them induces a $2$-periodic path by  copying~$\pi_1$ and~$\pi_2$, respectively, to every second column because~$\pi_1$ starts and ends in row~$i$ and~$\pi_2$ starts and ends in row $i+1$ (the $2$-periodic paths by are drawn in Figure~\ref{fig:paths}).
	These $2$\nobreakdash-periodic paths induce $t$\nobreakdash-end-to-end twistings
	by Lemma~\ref{lem:periodic-paths-to-twistings}, since $2$ is a divisor of $f(k)$. Clearly, the induced twistings
 avoid every possible horizontal roadblock.
 Moreover, for every vertex $v$, the incident edges of $v$ used by the two induced twistings differ from one another.
	This means that a roadblock that is not avoided by~$\pi_1$
	is avoided by~$\pi_2$ and vice versa.
	Hence, there have to be at least two additional non-horizontal roadblocks
	in rows~$i$ and $i+1$
	unless there is a desired $t$\nobreakdash-end-to-end twisting.
	Every roadblock which blocks the path~$\pi_1$ or~$\pi_2$
	belongs to row~$i$ or $i+1$, respectively.
	Hence,  for each socialized row there are at least two non-horizontal roadblocks belonging to it unless an $t$-end-to-end twisting exists.
	
So suppose for a contradiction that no $t$-end-to-end twisting exists. We now count the number of roadblocks.
For every cop-free row $i$, we have shown that up to three additional roadblocks belonging to row $i$ have to exist:
	one horizontal roadblock and two non-horizontal roadblocks. The horizontal roadblock always has to exist to prevent the path straight through row $i$ from being a $t$-end-to-end twisting. Whether the non-horizontal roadblocks have to exist depends on whether there are cops in rows $i-1$ and $i+1$: we require
	a non-horizontal one blocking an edge incident to row $i+1$ (if this row is also cop-free)
	and a non-horizontal one blocking an edge incident to row $i-1$ (if this row is also cop-free).
	A row containing a cop can reduces the number of roadblocks needed in some other row by one for at most two cop-free rows. If row $i$ contains a cop,
	then the required number of non-horizontal roadblock in rows $i-1$ and $i+1$ is reduced by one
	(if they are cop-free).
	Since $(k-c)$ rows do not contain a cop, there have to be at least
	$3(k-c) - 2c$ roadblocks.
	Since $c$ rows contain a cop and as we consider the compressed and blocking $(c+k)$-Cops and Robber game,
	there are at most $k$ roadblocks.
	This implies 
	\begin{align*}
		3(k-c)-2c &\leq k  \qquad\text { and thus} \\
		\frac{2}{5}k &\leq c 
	\end{align*}
	This is a contradiction to the assumption that
	$c \leq \frac{2}{5}k - 1$.
	Hence, the desired $t$-end-to-end twisting must exist.
\end{proof}

\begin{lemma}
	\label{lem:separator-move-when-some-are-fixed}
	Let $t\in [k-1]$ and $W_1, \dots, W_m$ be sets of at most $(k+t)$ many $\comp^t$-equivalence classes
	such that $|W_i\cup W_{i+1}| \leq k+t$ for all $i < m$.
	Let $I\subseteq [k]$ be a set of rows,
	of which all~$W_i$ contain exactly one and the same class.
	Also assume that all $W_i$ are $t$-critical.    
	Then all minimal $t$-semi-separators for each $W_i$
	coincide on the rows in $I$.
	In particular,  if $I$ is nonempty, the \defining{diameter} of all vertex minimal $t$-semi-separators together
	is at most $2(k+t)$.
	This means that there is a $j$ such that
	all these $t$-semi-separators 
	are contained within column $j, \dots, j + 2(k+t)-1$.
\end{lemma}
\begin{proof}
	We prove, for every $j < m$, that the statement holds for the sets $W_1,\dots,W_j$ by induction on $j$.
	For $j = 1$, the statement obviously holds.
	For the inductive step,
	assume that all minimal $t$-semi-separators for each $W_i$ with $i \leq j$ agree on the rows in $I$.
	Now consider the set $W = W_j \cup W_{j+1}$.
	By assumption, $W$ has size at most $k+t$.
	By~\cite[Proposition 4.9]{RFJN024},
	all minimal $t$-semi-separators for $W$ agree on the rows in $I$. 
	Combined with the inductive hypothesis,
	this implies that all minimal $t$-semi-separators for each $W_i$ with $i \leq j+1$
	agree on the rows in $I$,
	since every minimal $t$-semi-separator for $W_{i+1}$
	is also one for $W$~\cite{RFJN024}.
	
	Now assume that $I$ is nonempty.
	So all minimal $t$-semi-separators for each $W_i$ with $i \in [m]$
	share at least one vertex. 
	Since the $W_i$ are of size at most $k+t$,
	every $t$-semi-separator for them has size at most $k+t$.
	Hence, their individual diameter is at most $k+t$~\cite{RFJN024}.
	Since all these $t$-semi-separators share a vertex,
	their diameter together is at most $2(k+t)$.
\end{proof}

\noindent We now use the two preceding lemmas to prove our round number lower bound.

\begin{lemma}[Lemma~\ref{lem:compression-lower-bound-blocking} restated]
	Let $1\leq t \leq \frac{2}{5}k -1$ be an integer.
	Then the robber, initially placed on the left or right end of~$\grid$,
	has a strategy for the first $\Omega(J)$
	rounds in the compressed and blocking $(k+t)$-Cops and Robber game
	on $\grid$ and $\comp^t$.
\end{lemma}
\begin{proof}
We use letters $P$ and $Q$ for positions in the compressed and blocking $(k+t)$-Cops and Robber game.
	For such a position $P$, we write $\hat{P}$ for the position
	in which every roadblock is turned into a cop,
	that is, if there is a roadblock for a vertex $u$ in $P$,
	then we put a cop on $u$ in $\hat{P}$.
	Existing cops are copied.
	We call such a position $\hat{P}$ $t$-critical,
	if the set of cop-occupied classes in $\hat{P}$ is $t$-critical.
	Similarly, a $t$-virtual cordon for $\hat{P}$
	is a $t$-virtual cordon for the set of cop-occupied classes in $\hat{P}$.
	
	As already argued for Lemma~\ref{lem:no-separator-with-too-many-roadblocks},
	replacing roadblocks with cops only makes the game  harder for the robber.
	So, if the robber has a strategy for the next $r$ rounds from position $\hat{P}$,
	then the robber also has such a strategy from position $P$.
	Moreover, $\hat{P}$ can also be seen as a position
	in the non-blocking game.
	
	\begin{claim}
		\label{clm:only-constant-rows-per-round}
		Let $P$ be a position in the game such that $\hat{P}$ is $t$-critical
		and the robber 
		has distance at least $r$ to every 
		minimal $t$-virtual cordon for~$\hat{P}$.
		Then the robber has a strategy in the compressed and blocking $(k+t)$-Cops and Robber game in position $P$
		for the next round, resulting in a position $Q$,
		such that the robber has distance at least $r-O(k)$ to every minimal $t$-virtual cordon for $\hat{Q}$.
	\end{claim}
	\begin{claimproof}
		We prove the statement by induction on the number of moves in the current round as follows:
		Let $P=Q_0, Q_1, \dots ,Q_m$ be the sequence of positions during the first $m$ moves of the current round.
		Also let $Q_0^-, \dots, Q_{m-1}^-$ be the intermediate positions in which a cop or roadblock was picked up.
		This means, that the game changes from position $Q_i$ to $Q_i^-$ to $Q_{i+1}$ for every $0 \leq i < m$.
		We show that the robber has a strategy such that all positions do not contain empty roadblocks and either
		\begin{enumerate}
			\item all the positions $\hat{Q}_i^-$ are $t$-critical and
			every minimal $t$-virtual cordon for $\hat{Q}_m$ has distance at least $r-O(k)$ to the robber, or
			\item for the minimal $\ell \in [m]$ such that $\hat{Q}_i^-$
			is $t$-critical for all $\ell\leq i \leq m-1$, the robber has distance $\Omega(J)$ to
			the minimal $t$-virtual cordons for $\hat{Q}_i$
			for each $\ell\leq i \leq m$.
		\end{enumerate}
		For the base case $m=0$,
		the inductive hypothesis clearly holds.
		So assume that after $m$ necessarily blocking moves with roadblock responses (otherwise the round already ended)
		the game is in position $Q_m$
		and the inductive hypothesis is satisfied for $m$.
		Now a cop or roadblock is picked up resulting in
		position $Q_m^-$.
		We make the following case distinction:
		\begin{enumerate}[(a)]
			\item If $\hat{Q}_m^-$ is not $t$-critical,
			then we show that if the next position $\hat{Q}_{m+1}$
			 is $t$-critical, then
			the distance of the robber to every $t$-virtual cordon in $\hat{Q}_{m+1}$ is $\Omega(J)$. 
			
			This is straightforward to see if the Cops Player makes a cop move.
			We just follow the strategy in the non-blocking game in position $\hat{Q}_m^-$ to ensure this distance~\cite[Proof of Theorem 4.1]{RFJN024}.
			The robber moves to the end with larger distance to all the $t$-virtual cordons. Since different minimal $t$-virtual cordons can have distance at most $O(k)$~\cite{RFJN024},
			one end has distance at least $J/2 - O(k)$ to these $t$-virtual cordons.
			 Otherwise, the Cops Player makes a blocking move.
			Let $N$ be the announced roadblock on a $\comp^t$\nobreakdash-equivalence class $C$. Moreover, imagine that in the non-blocking game a cop is announced for $C$; then the robber responds with some appropriate $\comp^t$\nobreakdash-compressible twisting---which we denote by $T$---according to the robber's strategy. Call the resulting position in the non-blocking game $Q$,
			in which the robber again has distance $O(J)$ to all $t$-virtual cordons.
			The robber moves using $T$ also in the blocking game.
			If $T$ uses $N$ a cop is placed on $C$
			and the resulting position is $Q_{m+1}$.
			 But observe that $\hat{Q}_{m+1}$ is identical to $Q$.
			Otherwise, $T$ avoids $N$. Then a roadblock is placed and 
			again, the resulting position $Q_{m+1}$ is such that $\hat{Q}_{m+1}$ is 
			identical to $Q$.

			\item Now assume that $\hat{Q}_m^-$ is $t$-critical and, moreover, that
			 in position $Q_m^-$ there are cops in at most $\frac{2}{5}k-1$ rows.
			Then there is a $\comp^t$-compressible end-to-end twisting $T$
			with which the robber can move to the other end of $\grid$
			by Lemma~\ref{lem:no-separator-with-too-many-roadblocks}.
			If the Cops Player makes a cop move,
			the robber can use $T$ to move to the end with larger distance to all minimal $t$-virtual cordons (which might appear after the cop is placed).
			The robber may need to avoid cops placed in the ends.
			This is always possible because all minimal $t$-virtual cordons are not in the end where the robber is located.
			Note that there are only singleton equivalence classes in both ends so inside the ends the robber can move easily.
			From now we silently assume that, after possibly moving sides,
			the robber also avoid these non-dangerous cops inside ends locally.
			This means that all twisting we use to move the robber get modified for this in-end avoiding of cops. 
			
			Otherwise the Cops Player makes a blocking move. In this case
			the robber first checks whether the robber they should switch ends, i.e. whether the robber is at the end with larger distance to all minimal $t$-virtual cordons.
			If switching is not required, 
			define $T'$ to be the trivial twisting with which the robber does not move at all.
			If switching ends is required, let $T' := T$.
			The robber moves using twisting $T'$.

			\item Otherwise, $\hat{Q}_m^-$ is $t$-critical and there are at least $\frac{2}{5}k$ rows in which cops are placed in $Q_m^-$. In this case the robber stays in the end where th robber is currently located.
			If the Cops Player makes a cop move,
			then the robber does not move (apart from possibly to avoid in-end cops).
			If the Cops Player announces a roadblock $N$ on a $\comp^t$\nobreakdash-equivalence class $C$, then the robber similarly only moves to avoid in-end cops and roadblocks.
			
			If the resulting position $\hat{Q}_{m+1}$ is not $t$-critical, we are done.
			Otherwise, let $\ell \in \set{0,\dots,m}$ be the minimal number, such that $\hat{Q}_i^-$
			is $t$-critical for all $\ell\leq i \leq m$.
			Let $W$ be a minimal $t$-virtual cordon for $\hat{Q}_\ell^-$.
			Every other minimal  $t$-virtual cordon for $\hat{Q}_\ell^-$
			has distance at most $O(k)$ to $W$~\cite{RFJN024}.
			
			Let $W'$ be a  minimal $t$-virtual cordon for $\hat{Q}_m^-$.
			Let $I$ be the set of rows,
			in which there is exactly one cop in both $Q_\ell^-$ and in $Q_{m+1}^-$.
			We claim that $|I|\geq 1$. To see this note that
			no new cops are placed when roadblocks are placed,
			that there are at most $k + (\frac{2}{5}k-1)$ cops in $Q_\ell^-$,
			and that there are cops in at least $\frac{2}{5}k$ rows in $Q_{m+1}^-$.

			The positions $\hat{Q}_\ell^-, \dots, \hat{Q}_m^-, \hat{Q}_{m+1}$ satisfy the prerequisites of Lemma~\ref{lem:separator-move-when-some-are-fixed}:
			all of them are $t$-critical and in each step exactly one cop is moved (and for $\hat{Q}_m^-$ and $\hat{Q}_{m+1}$ one cop is added).
			Hence, $W'$ and $W$ together have diameter at most $2(k+t) = O(k)$.
			If $\ell = 0$, then the robber has distance at least $r-O(k)$
			to every minimal $t$-virtual cordon for $\hat{Q}_{m+1}$.
			Otherwise, the robber has distance at least $\Omega(J) - O(k) = \Omega(J)$
			to every minimal $t$-virtual cordon for $\hat{Q}_{m+1}$.
			\claimqedhere
		\end{enumerate}
		
	\end{claimproof}

	\noindent Claim~\ref{clm:only-constant-rows-per-round} shows that in each round where the position is $t$-critical,
	independent of the number of roadblock responses,
	the cops can move a minimal virtual cordon by at most $O(k)$ many columns towards the robber
	without providing the robber the opportunity to switch  between the ends.
	If the cops allow the robber to switch ends, we essentially can start the game from the beginning again
	because the robber can enforce a distance of $\Omega(J)$ to all minimal $t$-virtual cordons.
	This means that we can follow the strategy for the robber used in the non-blocking game as outlined before~\cite{RFJN024},
	where after each round we convert all roadblocks to cops.
	If the position is not critical, it does not matter where the robber is.
	If the position becomes critical, the robber moves to the end with larger
	distance to the minimal virtual cordons.
	If the position is critical,
	then by Claim~\ref{clm:only-constant-rows-per-round}, this distance of the robber to the nearest $t$-virtual cordon decreases by only $O(k)$ per round.
	Hence this strategy can be maintained for $\Omega(J/O(k))$ rounds.
	Together with the observation that a lower bound for $\hat{P}$
	in the blocking game implies a lower bound for~$P$ in the blocking game,
	this implies a $\Omega(J/O(k))=\Omega(J)$ lower bound for the robber in the
	blocking game since $k$ is considered a constant.
\end{proof} 

\begin{theorem}[Theorem~\ref{thm:compression-blocking-summary} restated]
	For all integers $k\geq 3$, $1\leq t \leq \frac{2}{5}k-1$, and $n\in \nat$,
	there are  
	two colored graphs~$\graphG$ and~$\graphH$ of order $\Theta(n)$ and color class size~$8$
	such that
	\begin{enumerate}
		\item $\graphG \notkequiv{k+1} \graphH$, that is, Spoiler wins the $(k+1)$-pebble game on $\graphG, \graphH$,
		and 
		\item $\graphG \kbequivr{k+t}{\Omega(n^{k/(t+1)})} \graphH $,
		that is, Duplicator has a strategy for the first $\Omega(n^{k/(t+1)})$
		rounds in the $(k+t)$-pebble game with blocking on $\graphG, \graphH$.
	\end{enumerate}
\end{theorem}
\begin{proof}
	Let $k \geq 3$, $1\leq t \leq  \frac{2}{5}k-1$, and $n \in \nat$.
	If $n$ is sufficiently large,
	for $w = \lceil\sqrt[t+1]{n} \rceil$ the required pairwise coprime numbers $p_1,\dots, p_k$ exist~\cite{DBLP:conf/focs/GroheLNS23}.
	Let $J = f(k) p_1\cdot\ldots \cdot p_k$
	and $\mathcal{C}$ be the $k\times J$ cylindrical grid.
	Also, let $\comp^t$ be the compression defined earlier with respect
	to $p_1,\dots, p_k$.
	By Lemma~\ref{lem:compression-lower-bound-blocking},
	the robber has a strategy for the first $\Omega( J) = \Omega(w^k) = \Omega(n^{k/(t+1)})$
	rounds in the compressed and blocking $(k+t)$-Cops and Robber game
	on $\mathcal{C}$ and $\comp^t$, when the robber is placed on the left or right end. 
	
	Now let $e$ be an edge in the left end,
	$f,g \, \colon \edgeC \to \FF_2$ only twist $e$,
	and set $\graphG :=\compCFI{\mathcal{C},f}{\comp^t}$ and $\graphH :=\compCFI{\mathcal{C},f}{\comp^t}$.
	By Lemmas~\ref{lem:precomp-comp-same-number-of-round}
	and~\ref{lem:robber-to-duplictor-blocking},
	Duplicator has a strategy for the first  $\Omega(n^{k/(t+1)})$
	rounds of the $(k+t)$-pebble game with blocking
	on $\graphG$ and $\graphH$.
	It is known that Spoiler wins the $(k+1)$-pebble game
	on CFI graphs over cylindrical grids and thus over the compressed ones
	by Lemma~\ref{lem:precomp-comp-same-number-of-round}.
	By Lemma~\ref{lem:compression-size},
	the graphs $\graphG$ and $\graphH$ have size $\Theta(n)$
	and are of color class size $8$.
\end{proof}
         
\end{document}